\documentclass[9pt,twocolumn,twoside]{pnas-new}

\templatetype{pnasresearcharticle} 
\usepackage{stmaryrd}
\usepackage{booktabs}
\usepackage{graphicx}
\usepackage{subcaption}
\usepackage{float}
\usepackage{amsthm}
\usepackage[capitalise]{cleveref}
\usepackage[resetlabels,labeled]{multibib}
\newcites{SI}{References}

\usepackage{nomencl}
\makenomenclature

\begin{document}

\title{Resilience of the Electric Grid through Trustable IoT-Coordinated Assets}


\author[a,1]{Vineet J. Nair}
\author[b]{Priyank Srivastava}
\author[c]{Venkatesh Venkataramanan}
\author[d]{Partha S. Sarker}
\author[d]{Anurag Srivastava}
\author[e]{Laurentiu D. Marinovici}
\author[f]{Jun Zha}
\author[g]{Christopher Irwin}
\author[h]{Prateek Mittal}
\author[i]{John Williams}
\author[f]{Jayant Kumar}
\author[h,1]{H. Vincent Poor}
\author[a]{Anuradha M. Annaswamy}
\affil[a]{Department of Mechanical Engineering, MIT, Cambridge, MA 02139}
\affil[b]{Department of Electrical Engineering, Indian Institute of Technology Delhi, India}
\affil[c]{National Renewable Energy Laboratory, Golden, CO}
\affil[d]{West Virginia University, Morgantown, WV}
\affil[e]{Pacific Northwest National Laboratory, Richland, WA}
\affil[f]{Larsen \& Turburo Digital Energy Services, San Francisco, CA}
\affil[g]{Office of Electricity, Department of Energy, Washington, DC}
\affil[i]{Department of Civil and Environmental Engineering, MIT, Cambridge, MA 02139}
\affil[h]{Princeton University}

\leadauthor{Nair}

\significancestatement{This paper provides a framework for achieving power grid resilience against cyber-physical attacks through the coordination of resources at the grid edge that are trustable and resilient. It is proposed that such coordination can be enabled through a suite of market operators suitably located in a distribution grid. We use this coordination to validate the mitigation of attacks of different levels of severity, with attack magnitudes that range from 5 to 40\% of the total peak load. Both grid-connected and islanded cases are studied. In all cases, we show that grid resilience can be obtained through a combination of locally available flexible assets and reconfiguration of the grid topology.}


\authorcontributions{A.M.A, V.N., P.S., V.V., and H.V.P. designed research; A.M.A., V.N., P.S., and V.V. performed research; V.N., P.S., V.V., and P.S.S. developed the analytical tools, V.N., L.M., J.Z., and V.V. performed validation; 
A.M.A., V.N., P.S., V.V., and A.S. wrote the paper; C.I., P.M., J.W., and J.K. provided critical inputs and valuable guidance.}
\authordeclaration{The authors declare no conflicts of interest.\\

}

\correspondingauthor{\textsuperscript{1}To whom correspondence should be addressed. E-mail: jvineet9@mit.edu, poor@princeton.edu}


\begin{abstract}
The electricity grid has evolved from a physical system to a cyber-physical system with digital devices that perform measurement, control, communication, computation, and actuation. The increased penetration of distributed energy resources (DERs) including renewable generation, flexible loads, and storage provides extraordinary opportunities for improvements in efficiency and sustainability. However, they can introduce new vulnerabilities in the form of cyberattacks, which can cause significant challenges in ensuring grid resilience. 
We propose a framework in this paper for achieving grid resilience through suitably coordinated assets including a network of Internet of Things (IoT) devices. A local electricity market is proposed to identify trustable assets and carry out this coordination. Situational Awareness (SA) of locally available DERs with the ability to inject power or reduce consumption is enabled by the market, together with a monitoring procedure for their trustability and commitment. With this SA, we show that a variety of cyberattacks can be mitigated using local trustable resources without stressing the bulk grid. Multiple demonstrations are carried out using a high-fidelity co-simulation platform, real-time hardware-in-the-loop validation, and a utility-friendly simulator.
\end{abstract}

\dates{This manuscript was compiled on \today}
\doi{\url{www.pnas.org/cgi/doi/10.1073/pnas.XXXXXXXXXX}}

\maketitle
\thispagestyle{firststyle}
\ifthenelse{\boolean{shortarticle}}{\ifthenelse{\boolean{singlecolumn}}{\abscontentformatted}{\abscontent}}{}

\firstpage[15]{2}

\dropcap{T}he electricity grid is going through a rapid transformation in an effort toward deep decarbonization. Large synchronous generators powered by fossil fuels such as oil, natural gas, and coal are being phased out in favor of solar and wind-based generation. While the latter enables the necessary move towards a reduced carbon footprint, it brings two major challenges in ensuring reliable and resilient delivery of electricity to the end-user. The first of these is the temporal signature of these renewables  – the amount of generation varies with time, both in terms of intermittency and uncertainty. The second is that these are distributed and large in number. A strong enabler of the scale of the DERs is IoT, which denotes a network of physical devices such as water heaters (WHs), air-conditioners, and electric vehicles (EVs), as they enable automated and fast operation of various loads. And their pervasiveness brings in complexities of heterogeneity, decentralization, and scale. In order to ensure the reliability of the grid despite these challenges, a precise coordination of these DERs, both in space and time, has to be carried out. In particular, power balance of generation and consumption has to be ensured at all locations and at each instant. These challenges are being overcome using a pervasive cyber layer that senses, communicates, coordinates, and enables the requisite power injection and consumption throughout the grid.

In addition to reliability, an essential property of the electricity grid is resilience \cite{NAP2017}. This central property, which denotes the ability of the grid to withstand and recover quickly to supply critical loads following a major disruption, such as an outage, a natural calamity, a cyberattack, or a cascading failure, is paramount, even with higher penetration of DERs. In this context of ensuring resilience, the very transformations that enable deep decarbonization, including the development of cyber-grid infrastructure, adoption of IoT devices, use of dynamic renewable energy sources, and increased electrification of transportation, could also introduce new vulnerabilities. Cyberattacks can disclose, deceive, or disrupt crucial information, thereby causing significant damage, ranging from small outages to brownouts and blackouts.  Recent reports \cite{Chernovite,nguyen2020electric,whitehead2017ukraine,NAP26704}  
indicate the ubiquity, ease, and scale of cyberattacks on sensitive industrial environments including supervisory control and data acquisition (SCADA), operational technology (OT), and industrial control systems (ICS), underscoring the importance of ensuring resilience to such adversaries. 

By and large, most of the information for power grid operations flows through utility-controlled communication networks which are more reliable and resilient than commercial networks, and utilize commercial telecommunications services for other informational needs such as accessing the internet and communicating with customers. Such a tight separation is challenged by the increased information flow which becomes necessary with a stronger presence of a cyber-layer, which in turn is necessitated due to increased coordination and automation at the grid edge. What have remained as tight closed systems thus far, may have to relax their boundaries, introducing complexities in the underlying communication. While air gaps and protections will always be important and included, imperfect protections are inevitable as complexity increases. With increased penetration of instrumentation and automation, motors and generators may be manipulated by adversaries to open and close at will. Another point to be noted is that with increased complexities due to intermittent and uncertain generation and consumption, utilities alone cannot cater to all needs; public and private partnerships may be necessary. It is therefore extremely important to design an appropriate cyberinfrastructure that ensures that the lights stay on, despite increased communication, which may be between disparate stakeholders. The focus of this paper is on such a distributed decision-making framework. 

Given the size and complexity of the problem of cyberattacks, providing a complete resilience framework for the entire power grid is a tremendously difficult task. In this paper, we propose a first step, of providing SA to the grid operators in a distribution grid, with SA corresponding to the knowledge of local DERs in terms of their location and the amount of power that they are able to provide, as well as a resilience score (RS) that the operators can make use of to provide resilience. We argue that this first step, of providing SA, is enabled through a local electricity market (LEM) structure that consists of operators at different voltage levels in a distribution grid. This market structure is proposed to be local, across the distribution grid, electrically co-located with primary and secondary circuits, with operators scheduling all DERs at the corresponding nodes in a given region. The market will also include IoT-coordinated assets (ICAs), with the assumption that each ICA will have computing capability and the ability to exchange information. The overall framework, EUREICA (Efficient, Ultra-REsilient, IoT-Coordinated Assets), is the innovation in the proposed cyberinfrastructure, and will be shown to lead to SA made available to operators placed hierarchically at various locations, thereby providing an important first step in ensuring resilience.

LEMs have been addressed in several studies including \cite{nair2022hierarchical,taft,chen2018next,bjarghov2021developments,sousa2019peer,lembook}, with real-field implementations beginning to be reported \cite{nyutilitydive,luth2020distributional}, all of which show the feasibility of a local market structure, and its advantages compared to alternate solutions that are designed to encourage full participation of DERs \cite{Haider2021ReinventingMarkets,NAP26704}. The LEM structure that we propose in this paper builds on that in \cite{nair2022hierarchical}. The resilience of the electricity grid to cyberattacks has been explored in a very large number of studies (see \cite{nguyen2020electric,DIBAJI2019394,brummitt2012suppressing,li2017networked} and references therein), with new results appearing continuously. Broadly, these approaches can be categorized into detection and isolation of the attack \cite{li2022detection}, prevention of the attack, and resilience in the presence of attacks. For large-scale attacks such as those described in \cite{whitehead2017ukraine,madiot2, eder2017cyber}, these methods are inadequate; it may be near-impossible to identify the attacker but rather that an attack has occurred. Prevention of the attack can be enabled through varying levels of access and authorization \cite{he2016cyber} and monitoring, isolation, and protection at the component level \cite{whitehead2017ukraine}. However, as the scale, location, and number of IoT devices in particular, and DERs in general grow, it becomes exceedingly difficult to completely prevent attacks. Ensuring resilience, especially in the face of large-scale attacks, for a large-scale system such as the electricity grid, is exceedingly difficult; current literature has either focused on systems at a small scale or with low levels of renewables. The EUREICA framework that we propose will provide SA that detects that an attack has occurred, and with this SA, deploys trustable ICAs in order to mitigate the impact of the attack, and ensure grid resilience through a distributed decision-making strategy.

The distributed decision-making in EUREICA is enabled through an LEM. The same market structure \cite{nair2022hierarchical}, which has been shown to lead to grid reliability \cite{nair2024resilience} and provide grid services such as voltage support \cite{nair2023local} in addition to overall power balance, is demonstrated in this paper to ensure grid resilience against cyberattacks using local trustable DERs. In particular, the results will show that local resilience is attainable through SA of locally available ICAs that have the ability to inject power or reduce consumption as well as a procedure for monitoring their trustability and commitment. The demonstrations are carried out using a variety of platforms such as (i) GridLAB-D which enables the simulation of distribution grids with high fidelity, (i) the advanced research on integrated energy systems (ARIES) platform that includes a real-time digital simulator (RTDS) and enables hardware-in-the-loop (HIL) validation, and (iii) General Electric's advanced distribution management system (ADMS) \cite{amiri2021updated}, distribution operations training simulator (DOTS), and DER integration middleware (DERIM).

The problem statement is laid out in \cref{sec:problem-statement}, and the local market structure together with the SA and Resilience score (RS) it will facilitate, is described in \cref{sec:retail-markets}. Various use cases that constitute the various attack surfaces on the grid are described in \cref{sec:usecases}. The main results are presented in \cref{sec:results}, followed by a summary in \cref{sec:conclusion}.

\section{Problem Statement}\label{sec:problem-statement}
A typical path of electricity delivery to end users traverses generation, transmission, and distribution. Distribution substations connect to the transmission system and gradually step down the voltage from 44kV or higher to 33kV (denoted as a primary network), then down to 11.2/4.6kV (denoted as a secondary network), and further down to 110V or 220V, depending on the specific region in the world. While the 20th century witnessed distribution systems operating as simple distribution lines as vehicles for sharing the electricity from transmission networks, today’s distribution systems are increasingly becoming heavily integrated with distributed energy resources, that correspond to resources that are located closer to the load, including renewable generation, some of which may be behind the meter \cite{NAP26704}, batteries, and flexible consumption units. This in turn is causing distribution systems to become more independent, and to be required to take on increased responsibilities of services such as grid reliability and grid resilience.  Other examples of DERs are distributed photovoltaics (DPVs) like rooftop solar systems, combined heat and power plants, electric vehicles, and diesel generators. DERs vary in size, from DPV systems that range between 1-1000kW in size to larger ground-mounted solar farms that range to several MW. With technological advances in power electronics and associated smart inverters as well as protection systems, fewer restrictions are being placed on the size and locations of the DERs, providing an opportunity for them to play stronger and more central roles in grid reliability and resilience \cite{horowitz2019overview}.

Over the past years, DERs have been shown to be increasingly useful in providing key grid services such as volt-var control \cite{VVC}. The central idea in these explorations is that key information is exchanged, in a distributed manner, between suitable individual components in the primary and secondary networks, coordinated both in space and time, thereby allowing local control over power injection and reduction of load at key locations and instants. Such a correct operation of the complete distribution network is predicated on this key information reaching the recipients in a secure manner. This sets the stage for malicious attacks that can disconnect and disrupt the overall grid by impairing key components. 

Several attacks on power systems have been recently reported \cite{zetter18,kennedy17,lee16,zeller11,madiot,huang19,madiot2,whitehead2017ukraine} on the central control systems, key nodes in the distribution grid, or at the devices at the end-user level. Those at the device end, denoted as MadIoT (Manipulation of Demand via IoT) attacks, correspond to a botnet at a secondary network node that causes the corresponding load to change abruptly. If this node corresponds to a high-wattage device, and the attack is coordinated through malware that simultaneously corrupts a large number of these devices, an argument can be made that it can cause frequency instabilities, line failures, and subsequently a severe disruption on the overall power grid. Building on the results in \cite{madiot,huang19}, the results in \cite{madiot2} show that even with realistic load profiles, a strategically coordinated attack can show a better success rate than in \cite{madiot,huang19} requiring fewer compromised IoT devices without triggering well-established protection systems. The well-known attack studied in \cite{whitehead2017ukraine} on the other hand is at the central control system level, which was a well-planned strategic attack that led to a power outage affecting 250,000 customers over a significant period of time. The question we address in this paper is: \textit{How can we use a cyberinfrastructure with IoT-Coordinated Assets (ICA) to support grid resilience against cyber-attacks?}

The specific approach that we propose to circumvent the anomalous scenario consists of two steps: (1) Enable improved visibility over the grid and net power injections available at various nodes through a hierarchical market structure with operators at the primary network and secondary network nodes; (2) Enable the market operators to determine an RS computed through monitoring of various features of the communication network. Steps (1) and (2) together provide SA to the grid operators (as shown in \cref{eq:sa}). Our central thesis is that through this SA, operators can determine that an attack has occurred and take appropriate steps to mitigate the impact of the attack in a timely manner. The system operators and resilience managers are suitably co-located with the electrical assets so as to respond quickly through a distributed decision-making framework. The framework therefore avoids the computational pitfalls of a centralized architecture while still underpinned by a substrate of communication, sensing, and actuation. The overall solution is also well-placed to integrate with the existing grid operational and market structures, helping accelerate its adoption in the field. 

\begin{table*}[]
\centering
\caption{Summary of attack scenarios and use-cases, LA = load alteration attack, DG = distributed generator attack.}
\begin{tabular}{@{}ccccccc@{}}
\toprule
\textbf{Attack Number} & \textbf{Attack type} & \textbf{Attack surface} & \textbf{Grid connection} & \textbf{Power flow model} & \textbf{Grid model}  & \textbf{Scale of attack [kW]} \\ 
\midrule
1a                   & LA                 & PMA               & Grid-connected           & Current injection         & Unbalanced, 3-phase  & 36 \\ 
1b                    & DG                   & PMA               & Grid-connected           & Current injection         & Unbalanced, 3-phase & 45 \\
1c                    & DG                   & SMA                     & Grid-connected           & Current injection         & Unbalanced, 3-phase  & 157 \\ 
2a                    & DG                   & PMA               & Grid-connected           & Branch flow               & Balanced, single-phase & 261 \\ 
2b                    & DG                   & PMA               & Grid-connected           & Branch flow               & Balanced, single-phase & 650 \\ 
3                  & DG                   & PMA                    & Islanded                 & Current injection         & Unbalanced, 3-phase & 2500 \\
\bottomrule
\end{tabular}
\label{tab:attacks}
\end{table*}

\section{Our approach: Local Electricity Markets}\label{sec:retail-markets}
\begin{figure}[htb]
\centering
\includegraphics[width=0.8\columnwidth]{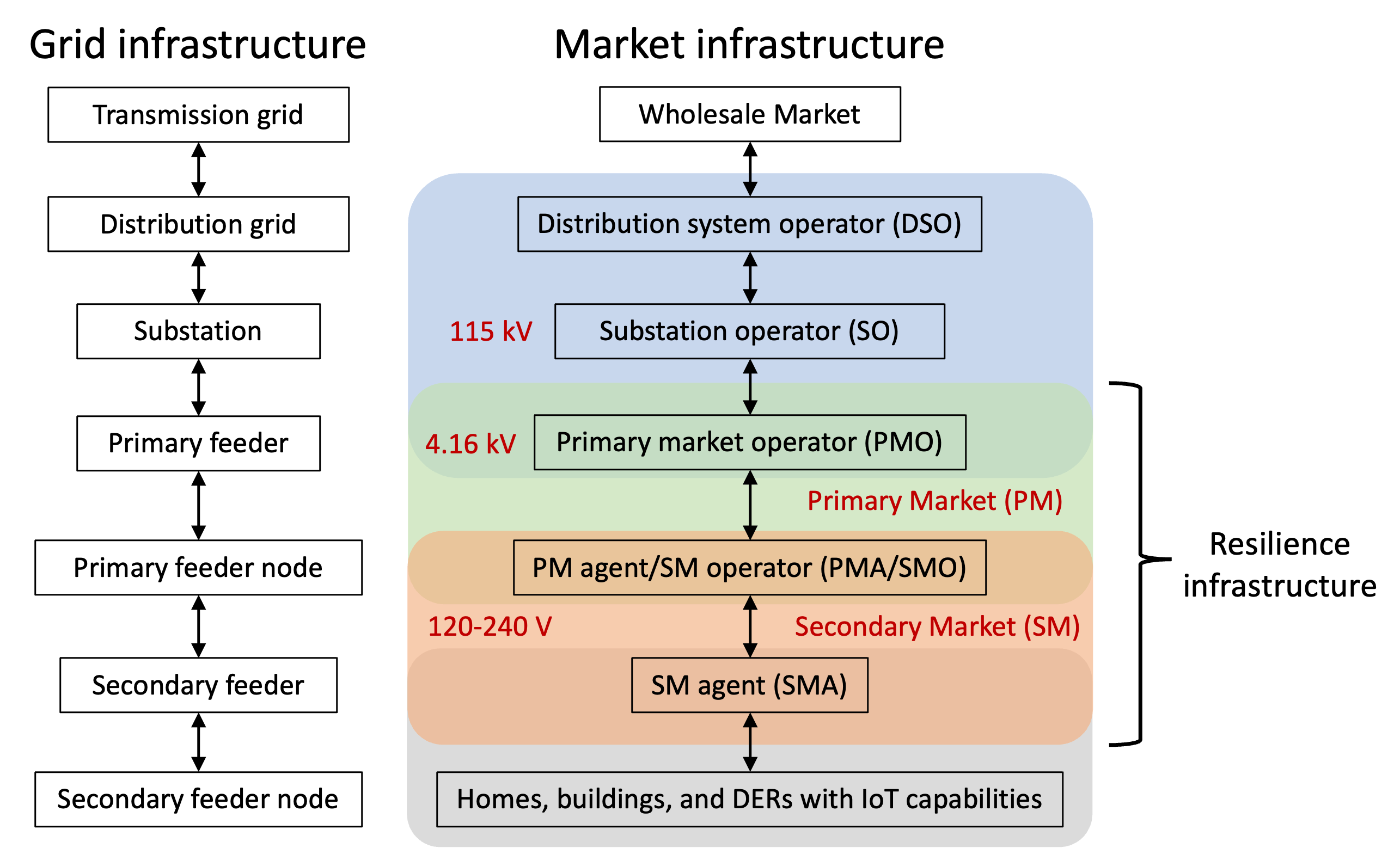}
\caption{A Hierarchical LEM for a Distribution Grid. The resilience infrastructure utilizes the dual market layer consisting of PM-SM. 
\label{fig:mkt_agents}}
\end{figure}

\begin{figure}[htb]
\centering
\includegraphics[width=0.95\columnwidth]{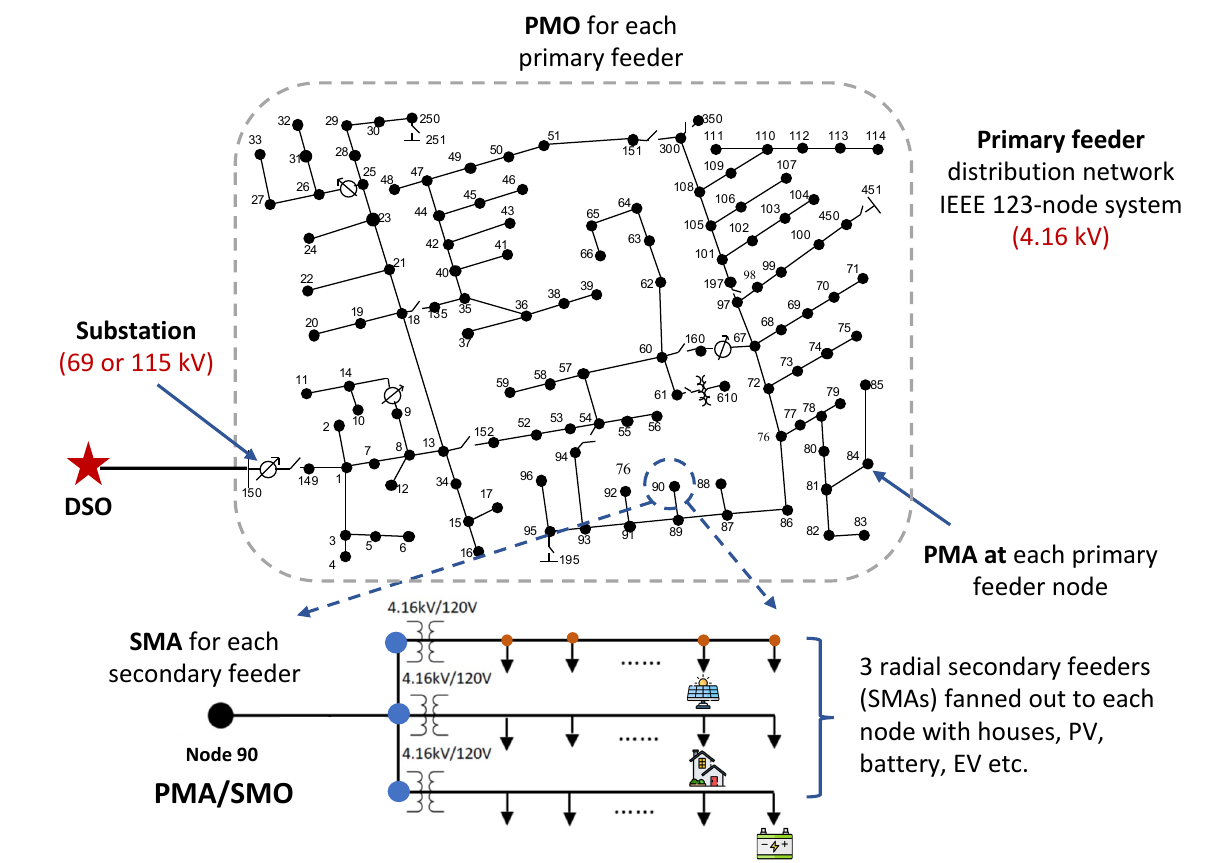}
\caption{LEM co-located with distribution grid. This shows a primary and secondary feeder distribution network based on the modified IEEE-123 node test case.\label{fig:network_schematic}}
\end{figure}

In order to provide visibility into a distribution grid, we propose an LEM that is hierarchical (see \cref{fig:mkt_agents}) in nature and electrically collocated with a radial network. The starting point for the overall LEM is a distribution system operator (DSO) that oversees several substations in the distribution grid with multiple primary and secondary markets downstream and acts as their representative in its transactions with the wholesale electricity market (WEM). The substation connects the distribution grid to the high voltage transmission grid at the point of common coupling (PCC) (node 150 in \cref{fig:network_schematic}). The dual-layer downstream of the substation, consisting of a primary market (PM) and a secondary market (SM), is the core of the resilience infrastructure of this paper. The PM consists of Primary market operators (PMOs) and Primary market agents (PMAs). The PMA at each of the primary nodes either own a DER at a primary feeder node or are aggregators representing DERs at the secondary feeder level and below.  In the latter case, the PMA plays a second role as an SM operator (SMO) and coordinates with SM agents (SMAs). The PMO, PMAs/SMOs, and the SMAs are located at the coupling between the substation and the primary feeder, primary feeder nodes, and secondary feeder nodes, respectively (see \cref{fig:network_schematic}). The PM and SM operate at medium- and low-voltage levels, respectively. The DSO supervises the entire distribution grid - our proposal could be viewed as an expansion of the current responsibilities of a DSO, which comprise grid maintenance and grid reliability, to include market oversight and regulation as well. In this sense, the role of the DSO would be analogous to that of existing independent system operators for transmission grids \cite{hogan1998independent}. SI \cref{sec:lem_example} illustrates a possible implementation of the dual-layer LEM for the city of Boston, MA, USA.

Formally, we define SA at an operator $x$ as the tuple
\begin{align}\label{eq:sa}
    \text{SA}_x=\{\text{ICA}_x,\text{RS}_x\},
\end{align}
where ICA$_x$ stands for IoT-coordinated assets and denotes the generation and/or consumption flexibilities of DERs under the purview of agent $x \in \{\text{SMA}, \text{SMO}\}$, and RS$_x$ denotes their resilience scores, to be defined in \cref{sec:commit}. We will show that $\text{RS}_x$ can be determined based on the asset's market performance and security against possible attacks.

We now show how the LEM made of the PM-SM layers will allow the computation of SA. The operation of a distribution grid is challenging due to its scale, complex topology, and presence of various active DER assets and fixed load nodes. We separate this complex task by having the PM focus on grid-specific costs and constraints while the SM focuses on consumer-centric costs and constraints. We assume that PM and SM clear once every 5 minutes and 1 minute respectively. The main reason for this separation of timescales is that the SM typically needs to monitor fewer assets than the PM, and is closer to DER devices (such as rooftop solar and batteries) and therefore may need to operate at a faster timescale than a PM. The starting point for both markets is the submission of bids by the corresponding agents. Bids for the SM are submitted by the SMAs exogenously, whereas bids for the PM are computed by the SMOs via the SM. We describe the operation of the SM before going into the details of the PM.

\subsection{Secondary market} 
The operation of the SM consists of three sequential stages: bidding, clearing, and monitoring. We denote $\mathcal{N}$ to be the set of all SMOs in the network and $\mathcal{N}_i$ to be the set of all SMAs under a given SMO $i \in \mathcal{N}$.

\subsubsection{SM bidding}
During the bidding phase, each SMA $j \in \mathcal{N}_i$ submits  a bid $\mathcal{B}_j^{iS}$ defined as
$$
   \mathcal{B}_j^{iS} =\{P_j^{i0},Q_j^{i0},\underline{P}_j^i,\underline{Q}_j^i,\overline{P}_j^i,\overline{Q}_j^i,\beta_j^{iP},\beta_j^{iQ}\}. \label{eq:sm_bid}
$$
$P_j^{i0}$ and $Q_j^{i0}$ denote the baseline active and reactive injections of SMA $j$, along with the upward ($\overline{P}^i_j,\overline{Q}^i_j$) and downward flexibility ($\underline{P}^i_j,\underline{Q}_j^i$). $\beta_j^{iP}$ and $\beta_j^{iQ}$ denote the disutility parameters associated with providing active and reactive power flexibility, respectively. It should be noted that Bid ${B}_j^{iS}$ requires SMA $j$ to have a realistic estimate of its energy profile for the next 1 minute. Since it is not always trivial to predict future power availability, agents deploy a decentralized federated learning (FL)-based framework \cite{venkataramanan2022forecast} to determine their bids. Using FL helps ensure that the privacy of the participating agents is preserved and the computational aspects of the prediction algorithm scale well as the number of agents increases. Further details on the FL implementation can be found in SI \cref{sec:fl}. SI \cref{fig:SMbids} summarizes details of the overall LEM.

\subsubsection{SM clearing\label{sec:sm_clear}}
Once the SMO $i$ has received bids from the participating SMAs, it clears the market with active and reactive power injection setpoints ($P^{i*}_j,Q_j^{i*}$) and the corresponding retail tariffs ($\mu_j^{iP*},\mu_j^{iQ*}$). 
In addition, the SMO also solves for the optimal flexibility ranges ($\delta P_j^{i*},\delta Q_j^{i*}$) for $j \in \mathcal{N}_i$.
The SMO clears the markets with the following objectives: (O1)  maximization of aggregate resilience $f_i^1$, (O2) minimization of the net cost to the SMO, $ f_i^2$, (O3) maximization of total flexibility $f_i^3$ that the SMO can extract from all its SMAs and (O4) minimization of the disutility of the SMAs $f_i^4$, arising from flexibility provision (see SI \cref{sec:sm_details} for all details). This gives rise to a multiobjective constrained optimization problem:
\begin{subequations}
\label{eq:opt}
\begin{align}
& \min_{\textbf{y}^S_i} f_i^S=\{f_i^1, f_i^2, f_i^3, f_i^4\}^\top \\
& \text{s.t.} \underline{P}^i_j + \delta P^i_j \leq P^i_j \leq \overline{P}^i_j - \delta P^i_j \; \forall j \in \mathcal{N}_i, \; \forall \; \text{constraints} \nonumber \\ 
& \underline{Q}^i_j + \delta Q^i_j\leq Q^i_j \leq \overline{Q}^i_j - \delta Q^i_j \nonumber \\ 
& \delta P^i_j, \; \delta Q^i_j \geq 0, 0 \leq \mu_j^{iP} \leq \overline{\mu}^{iP}, 0 \leq \mu_j^{iP} \leq \overline{\mu}^{iQ}  \; \nonumber  \\ 
& \sum_{t_p \in \mathcal{T}_p}\sum_{t_s \in \mathcal{T}_s} \sum_{j \in \mathcal{N}_i} \mu_j^{iP}(t) P_j^i(t) \Delta t_s \leq \sum_{t_p \in \mathcal{T}_p} \mu^{iP^*}(\hat{t}_p) P_i^*(\hat{t}_p) \Delta t_p  \nonumber \\
& \sum_{t_p \in \mathcal{T}_p}\sum_{t_s \in \mathcal{T}_s} \sum_{j \in \mathcal{N}_i} \mu_j^{iQ}(t) Q_j^i(t) \Delta t_s \leq \sum_{t_p\in \mathcal{T}_p} \mu^{iQ^*}(\hat{t}_p) Q_i^*(\hat{t}_p) \Delta t_p  \nonumber \\ 
& \sum_{j \in \mathcal{N}_i} P_j^i(t_s) = P^{i*}(\hat{t}_p), \quad \sum_{j \in \mathcal{N}_i} Q_j^i(t_s) = Q^{i*}(\hat{t}_p) \; \forall t_s \in \mathcal{T}_s \nonumber 
\end{align}
\end{subequations}
The constraints include capacity limits and operational bounds on SMA injections (including flexibilities), budget balance constraints, price ceilings, and lossless power balance \cite{nair2022hierarchical}. $\mathcal{T}_p, \mathcal{T}_s$ denote the set of all PM clearing timesteps, and secondary timesteps per PM clearing, respectively. $\hat{t}_p$ refers to the most recent PM timestep before each SM clearing. Note that, here we do not account for all the power physics, these will be considered in the PM. The decision variables consist of the P and Q injection setpoints as well as retail tariffs for each SMA i.e. $\textbf{y}^S_i = \{\textbf{y}^{iS}_j\} \; \forall j \in \mathcal{N}_i$ where $\textbf{y}^{iS}_j = [P_j^i, Q_j^i, \delta P_j^i, \delta Q_j^i, \mu^{iP}_j, \mu^{iQ}_j]$. We note from the choice of $f^1$ that the solution of \cref{eq:opt} requires the resilience scores $\text{RS}^i_j$. This is assumed to be communicated by the secondary resilience manager (SRM) to the SMA, the details of the SRM are addressed in the next section.

In general, the optimization problem in \cref{eq:opt} has multiple solutions known as Pareto points, with each solution prioritizing different objectives. 
However, since the objective functions have different units, instead of finding the Pareto solutions, we use a hierarchical ranked approach proposed in \cite{nair2023local} where the SMO optimizes one objective at a time in descending order of importance. While optimizing the subsequent objective functions, additional constraints on the degradation of prior objectives are added to the optimization problem (see \cite{nair2022hierarchical} for details). The cleared market schedules $\textbf{y}^{S*}_i$ are sent by the SMO to their corresponding SMAs, as well as to their SRM.

\subsubsection{SM monitoring and resilience scores}\label{sec:commit}
The final stage in the SM is monitoring. During the market operation, the responses of each SMA $j$ to the market schedules, in terms of its actual DER injections $\hat P_j^i$ and $\hat Q_j^i$ are suitably monitored by its corresponding SRM.
\begin{figure}[htb]
\centering
\includegraphics[width=0.7\linewidth]{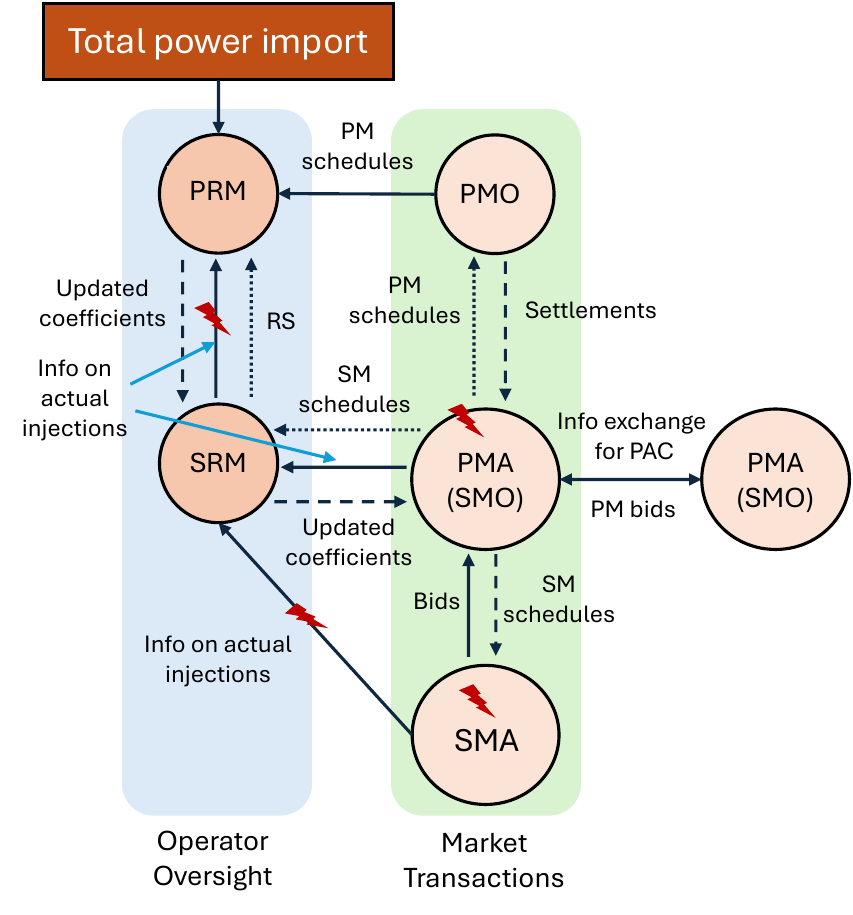}
\caption{Sequence of communication steps and events leading to SA with an LEM. The red arrows indicate the entities and communication links that would be affected by an attack. A more detailed diagram can be found in SI \cref{fig:detailed_comm}.}
\label{fig:prmsrm}
\end{figure}
In addition to the market operators, we also propose the addition of two new entities, which we denote as the primary resilience manager (PRM) and the SRM, both of which provide grid functionalities, with the PRM located at the primary circuit level and the SRM at the secondary level, as shown in \cref{fig:prmsrm}. With the market clearing providing the first step of awareness in the form of power available at each of the nodes at the secondary and primary level, the PRM and SRM monitor the actual injections, determine corresponding scores of commitment, trustability, and resilience (to be defined below), and communicate them using protected channels to the PRM. Not only do these entities enable a separation between grid-specific decision-making from market-specific decisions, but they also provide a pathway for mitigating the impact of any attacks that can occur through the addition of local resources, as will be shown in the following sections.

In this monitoring stage, the SRM assigns each SMA an RS that is updated constantly based on its performance in the market and susceptibility to being compromised. The RS is a weighted combination of its commitment score (CS) and trustability score (TS). Formally, for an agent $j$
\begin{align*}
    \text{RS}_j= \alpha \text{CS}_j + (1-\alpha) \text{TS}_j,
\end{align*}
where $\alpha \in (0,1)$ is a parameter chosen by the SRM. 
The CS and TS are defined below.
\begin{itemize}
    \item \textit{Commitment score (CS)}.
    The CS of an agent measures its reliability in executing its cleared schedules and is updated at every SM clearing instance.
    The first step in updating $CS_j$ for each agent $j \in \mathcal{N}_i$ is the computation of any relative deviation between the cleared schedule and its executed value over the past market period. A moving average is then computed to account for the past performance. Finally, a min-max normalization across all the SMAs is performed to keep $CS_j \in [0,1]$ for all $j$ (see SI \cref{sec:commit_app} for further details).
    
    \item \textit{IoT Trustability score (TS)}.
     The TS captures the possibility of the agents (or the devices underneath them) being compromised. TSs are computed using an FL-based anomaly detector and like the CS, past values are used again to compute a weighted moving average. However, unlike the CS, which solely depends on power injections, the TS is a cyber-power metric \cite{Sarker2023ResiliencyIoTs} that also takes into account the associated cyber information, e.g., packet length, arrival time, and communication protocols, etc. (see SI \cref{sec:ts_resmetric} for further details).
\end{itemize}
In summary, the overall SM operation allows the computation of schedules $\textbf{y}^{S*}_i =[P_j^{i*},Q_j^{i*},\delta P_j^{i*},\delta Q_j^{i*},\mu_j^{iP*},\mu_j^{iQ^*}] \; \text{and} \; \text{RS}^i_j \; \forall \;\text{SMAs} \; j$, all of which provide $\text{SA}^i_j$ for the SRMs corresponding to all SMAs $j$ at primary node $i$. Similar measures of the resilience of large-scale networks to attacks can be found in \cite{schneider2011mitigation}.

\subsection{Primary market\label{sec:primary-market}}
PM transactions happen between the PMO and the PMAs. Similar to the SM, the operation of the PM also consists of bidding, clearing, and monitoring.

\subsubsection{PM bidding}
Here, we describe the link between the PM and SM. As noted previously, the PM is cleared every 5 minutes while the SM operates more frequently at 1-minute intervals. Before each PM clearing, the SMOs (or PMAs) aggregate the schedules and cleared flexibilities of all their SMAs resulting from the most recent prior SM clearing (at the lower level) to submit their flexibility bid to the upper-level PM. All market bidding and clearing for both the SM and PM are based on forecasts (assuming perfect foresight) and for the very next period. The complete bid submitted by each SMO $j \in \mathcal{N}$ into the PM $\mathcal{B}_i^{P}$ defined as:
\begin{align}
   \mathcal{B}_i^{P} =\{P_i^{0},Q_i^{0},\underline{P}_i,\underline{Q}_i,\overline{P}_i,\overline{Q}_i,\alpha_i^P, \alpha_i^Q, \beta_i^P,\beta_i^Q\}. \label{eq:pm_bid}
\end{align}
This includes the nominal power available, the lower and upper bounds, net generation cost coefficients, and flexibility disutility parameters (see details in SI \cref{sec:pm_bidding}). We note that standalone PMAs such as a large industrial facility, a community solar farm, or an EV charging station, may also be present. In this case, the PMA would directly bid into the PM on its own instead of aggregating over SMAs.

\subsubsection{PM clearing}
At each PM clearing instance, an optimal power flow (OPF) problem is solved to optimize the PMO's objective while satisfying all grid physics and network power flow constraints. For simplicity, in this work, we consider the cost functions of all the PMAs (or SMOs) to be quadratic. The objective function 
utilized is a weighted linear combination of (i) maximization of social welfare, (ii) minimization of total generation costs, and (iii) minimization of electrical line losses (see SI \cref{sec:pm_obj} for details of these functions). The total cost includes paying the locational marginal price (LMP) $\lambda$ for importing power from the transmission grid at the PCC, as well as the payments to local generator PMAs that provide net positive injections into the PM. We divide by suitable base values to convert all quantities to per unit (between 0 and 1 p.u.). Thus, it is reasonable to combine all the terms into a single objective function using a simple weighted sum. 

With the objective function thus defined, the constraints are determined by the choice of the power flow model used to describe the system. Since the original alternating current OPF (ACOPF) is inherently nonconvex and NP-hard, we need to convexify the problem to make it more tractable. In this study, we considered two different approaches for this convexification. The first is a branch flow (BF) model or nonlinear \textit{DistFlow} \cite{Molzahn2019AEquations} based on a second-order conic program (SOCP) convex relaxation - this is a simpler implementation that is valid for radial and balanced networks. The second is a linear current injection (CI) model \cite{Ferro2020ANetworks} based on a McCormick envelope convex relaxation that is more generally applicable to unbalanced and meshed grids common in distribution systems (in addition to radial, balanced), although this adds some overhead due to certain pre-processing steps needed. 

We deployed both of these models for different use cases considered in this paper, as shown in \cref{tab:attacks}. Further details of the BF and CI approaches are provided in SI \cref{sec:bf} and SI \cref{sec:ci}, respectively. The exact set of decision variables $\textbf{y}^P_i$ for each PMA $i$ differs slightly depending on the OPF model used. Both models solve for the nodal power injections and voltages. However, the BF model only considers branch currents while the CI model also considers nodal current injections. BF also models all variables as only having a single phase while the CI models these as three-phase, complex phasor quantities. For simplicity, we have only included the single-phase formulations in the paper thus far. However, these can easily be extended to the complex three-phase representation by simply modifying all variables to 3-dimensional complex vectors instead of scalars. A three-phase extension of the SM optimization is given in SI \cref{sec:sm_ci}.

\subsubsection{Distributed optimization for PM clearing}
We employ a distributed proximal atomic coordination (PAC) algorithm \cite{Romvary2022AOptimization} to solve the OPF using peer-to-peer communication between the agents. This approach is preferred over traditional centralized optimization solvers since the number of nodes (and hence the number of PMAs) in a primary feeder could be arbitrarily large. It also helps preserve data privacy since each PMA only needs to exchange limited information with its immediate neighbors. A distributed approach also enables the PMAs to clear the market independently of the PMO, alleviates the communication burden, and reduces latencies since PMAs do not need to send all their data to a centralized entity, thus allowing for scalability. This is achieved by a process called atomization wherein the overall global optimization problem is decomposed into several local optimization problems called atoms for each PMA. 
The constraints can also similarly be decoupled. However, certain network constraints also depend on other PMAs' variables. To deal with this we include additional coupling or consensus constraints to ensure consistency. We also used an enhanced variant of PAC known as NST-PAC that employs Nesterov (NST) acceleration and has enhanced privacy features by further masking the variables exchanged between atoms (i.e., the PMAs) \cite{Ferro2022AHubs}. After a sufficient number of iterations, both the PAC and NST-PAC algorithms provably converge to globally optimal and feasible solutions $\textbf{y}^{P*}= \{\textbf{y}^{P*}_i\}$ for each of the PMAs. Further details on PAC and NST-PAC are provided in SI \cref{sec:dist_opt}. These cleared market schedules are communicated by the PMAs to their respective SRMs as well as to the PMO.

\subsubsection{PM prices}
Using distributed optimization, we obtain the electricity prices for the PM from the solutions that the dual variables $\mu_i^{P*},\mu_i^{Q*}$ converge to. As indicated in \cite{Haider2021ReinventingMarkets}, this can be viewed as a distributed locational marginal price (d-LMP) and reflects the local value that the market agents are providing through local generation and flexible consumption.

\subsubsection{PM monitoring and resilience scores} 
During the actual market operation, the injections $\hat P_i$ and $\hat Q_i$ from the DERs at PMA $j$ are monitored by their SRM. These could be either from standalone PMAs or aggregated information from all the SMAs at a given PMA. The SRM also assembles resilience scores $\text{RS}_i$ for each PMA $i$. This is done through aggregation (via a weighted average) of $\text{RS}^i_j \; \forall j \in \mathcal{N}_i$. The RSs for standalone PMAs can also be directly computed at the SRM using their monitored injections. $\textbf{y}^P_i$ and $\text{RS}_i$ thus provide complete SA at each PMA node $i$. All SRMs send this information to the PRM so that the PRM has complete SA of all PMAs. This SA can then be used to redispatch the ICAs in both the PM and SM to mitigate the impact of various attacks. Further details on the mitigation strategy can be found in SI \cref{sec:attack_mitig}.

\subsection{Reconfiguration paths}
The final tool that we use in our proposed EUREICA framework is grid reconfiguration in the wake of islanding which can occur if an attack, fault or natural disaster causes an entire section of the grid to be disconnected from the main grid. In such cases, an algorithm that determines a self-sustaining operation of the islanded system, which is enabled by reconfiguration paths with suitable switch settings, is essential. We propose a reconfiguration algorithm (see SI \cref{sec:reconfig_alg} for details) that considers power flow feasibility, available distributed generators (DGs), critical load, as well as RS information, to determine switching actions to restore specific sections of the distribution feeder. Reconfiguration paths will be determined based on the available amount of generation and the amount of critical load to be supplied, which is obtained through the SA provided by the EUREICA framework. In addition, the TSs are used at the secondary feeder level to intelligently disconnect non-critical loads, thus enabling the maximum restoration of critical loads. Once the feasible paths are determined for the optimal selection of loads, the RSs for all feasible paths are computed, and the most resilient path is implemented in the system.

\section{Use cases\label{sec:usecases}} 
In this section, we present use cases that illustrate how SA can be leveraged to ensure grid resilience in a distribution grid with a high penetration of DERs. We consider four different attack scenarios, all of which are motivated by the two large-scale attacks in  \cite{whitehead2017ukraine,madiot} on power grids. Disruptive attacks are assumed to occur in the form of (a) a sudden loss of generation, and/or (b) a sudden increase in load, at multiple vulnerable locations. All use cases are simulated using an IEEE 123-node test feeder (see SI \cref{sec:ieee123}) modified to have a high DER penetration (see SI \cref{sec:network_pnnl_high_der}); extensions to more realistic and larger networks \cite{meyur2022ensembles} can be implemented similarly.

\subsection{Attack 1}
In this attack, it assumed that a small percentage of generation or load resources at either the primary or secondary feeder level are compromised. In particular, it is assumed that these units are offline due to either an outage, natural calamity, or malicious cyber-attacker using elevated privileges to disconnect the units. In addition to the generation shortfall, it is assumed that the communication link between the market operators (PMO/SMO) and the resilience managers (PRM/SRM) is also affected by a denial of service (DoS) attack, which compromises the availability of a resource (see \cite{spower} for an attack which occurred on an sPower installation in Utah). Attack 1 draws inspiration from \cite{whitehead2017ukraine}, where a malicious attacker used (i) elevated and unauthorized access to disconnect several resources, and (ii) severed communication links, to hamper operator visibility and response. While these attacks occurred at the transmission level, it is feasible that a similar impact can be had by targeting distribution grid entities, especially with the larger attack surface provided by grid-edge devices. Independently, it is possible that IoT load devices such as heating, ventilation, and air conditioning (HVAC) devices, WHs, EV chargers, or refrigerators may be attacked as well as noted in \cite{madiot2}. Elements of both of these types of attacks are explored here in two different cases, 1a and 1b. 

\subsubsection{Case 1a}
In this case, the grid is assumed to be subjected to a sudden increase in load at the primary feeder level level (SMO or PMA) due to malicious agents. There are several large loads connected to the primary feeder such as commercial buildings or industries, and a malicious agent can manipulate the loads in these entities to affect the grid. Typically, the grid would rely on the margin provided by grid inertia to mitigate the effect of a sudden load increase. However, in a case where the grid’s resources are stretched, such as a cold snap or similar natural hazards, it is imperative that the grid-edge IoT resources be tapped to mitigate this condition. Examples of this scenario are already seen in operations, such as requests from grid operators in Alaska, Texas, and others in response to cold snaps. The operators requested customers to reduce their power consumption to support large critical loads such as chillers in hospitals. Furthermore, increased DER penetration will also lead to a loss of inertia, currently provided largely by large coal and gas plants. Case 1(b) details the performance of the proposed framework from this generation shortfall, even when the PRM does not have complete observability in the system.

\subsubsection{Case 1b}
Here, several generating resources are assumed to be unavailable at the primary feeder level (i.e. SMO or PMA). There are several scenarios that motivate this - for example, in the case of several cloudy days in a row (affecting wind power), or unforeseen maintenance on generating units, the grid operates at a lower margin than under normal conditions. There is also the case of a malicious actor disconnecting generation resources. The grid experiences a supply shortfall, and in combination with the DoS attack, the system operator (PRM) loses observability.  

\subsubsection{Case 1c} 
In this case, the grid is subjected to a sudden increase in load and/or corruption of distributed generation from the IoT devices, in a coordinated fashion directly at the secondary feeder. DER IoT devices will soon be operated via cloud-based service mechanisms that allow them to be controlled remotely. Thus, a sufficiently motivated malicious actor could gain control of a large number of these to suddenly reduce generation or increase load in a coordinated fashion. We simulated a case in which a large number of DGs (such as solar PV smart inverters) are attacked at the SMA level.

\subsection{Attack 2}
A larger-scale attack is assumed to occur at the distribution grid level in the form of several DGs being corrupted, causing them to go offline. The scale of this attack is assumed to be such that the impact is felt even in the transmission grid. We will explore how SA by the PRM and SRM helps mitigate this impact. Similar to attack 1.0, this use case combines elements of both \cite{madiot2} and \cite{whitehead2017ukraine}. The similarity to the latter is that the corruption is inserted in the form of outages of large DGs, while that to the former is that it introduces oscillations at the transmission level. For this purpose, we will utilize the well-known Kundur 2-area test system used to understand the transient and dynamic transmission-level impacts \cite{kundur2007power}. In particular, we will assume that there is an outage in one of the two areas (Area 2) that is load-rich, which introduces additional stress on the tie-line connecting the 2 areas (see SI \cref{fig:kundur} for a diagram of the 2-area system). 

\subsection{Attack 3}\label{sec:attack3_desc}
The substation transformer is located at node 150, which is connected to the main transmission grid under normal operating conditions. However, under this attack, the distribution grid is islanded from the main grid at node 150. This could be due to a multitude of factors – such as wildlife tripping the transmission line from the substation to the distribution system, or a cyber-attack (i.e., integrity or disruption attack) that trips the circuit breaker from the main grid. With the increased SA introduced through our framework, we will demonstrate that the distribution system loads can be picked up in a coordinated fashion.

\section{Results and discussion\label{sec:results}}
In this section, we focus on a few specific attack scenarios 1a, 2b, and 3 described in \cref{tab:attacks}, along with some results for attacks 1b and 1c. The remaining scenarios are elaborated in SI \cref{sec:mkt_other_attacks} and SI \cref{sec:valid_other_attacks}. With the numerical simulation setup described in \cref{sec:sim_setup}, we present details of how each of these attacks is mitigated using the proposed EUREICA framework. Note that for all attacks except attack 2.0, we use the mitigation strategy described in SI \cref{sec:costcoeff2}. For attacks 2a and 2b, we use the algorithm in SI \cref{sec:costcoeff1} instead. In addition to market simulations, we also validated our results using high-fidelity software at the Pacific Northwest National Lab (PNNL), Larsen \& Turburo Digital Energy Services (LTDES), and the National Renewable Energy Lab (NREL). Technical details for each validation platform can be found in SI \cref{sec:validation}. For each attack, we focus on a specific subset of results to highlight the most pertinent findings and insights. Complete details including the remaining market simulation results for all attack scenarios can be found in SI \cref{sec:mkt_other_attacks}, along with the complete set of validation results in SI \cref{sec:valid_other_attacks}.

\subsection{Mitigation of attack 1a\label{sec:attack1a_mit_mkt}} 
We note that in attack 1a, loads are compromised leading to an increase in the power import from the bulk grid. It is also assumed that the communication from all SRM to the PRM is disrupted, while the communication from the PRM to the SRM remains intact. That is, the PRM loses observability but is still able to communicate the redispatch of the new coefficients to the SRM. We do not consider the case when such observability is not lost, a discussion of which is beyond the scope of this paper. With the redispatch, the PM-SM framework identifies all of the new trustable PMAs (through the SA computations described in SI \cref{sec:attack_mitig}), which will provide the injections needed to fully mitigate the attack, and the overall power balance is thus met at all points in the distribution grid. 

\begin{figure}[htb]
\centering
\begin{subfigure}[t]{0.49\columnwidth}
    \includegraphics[width=\columnwidth]{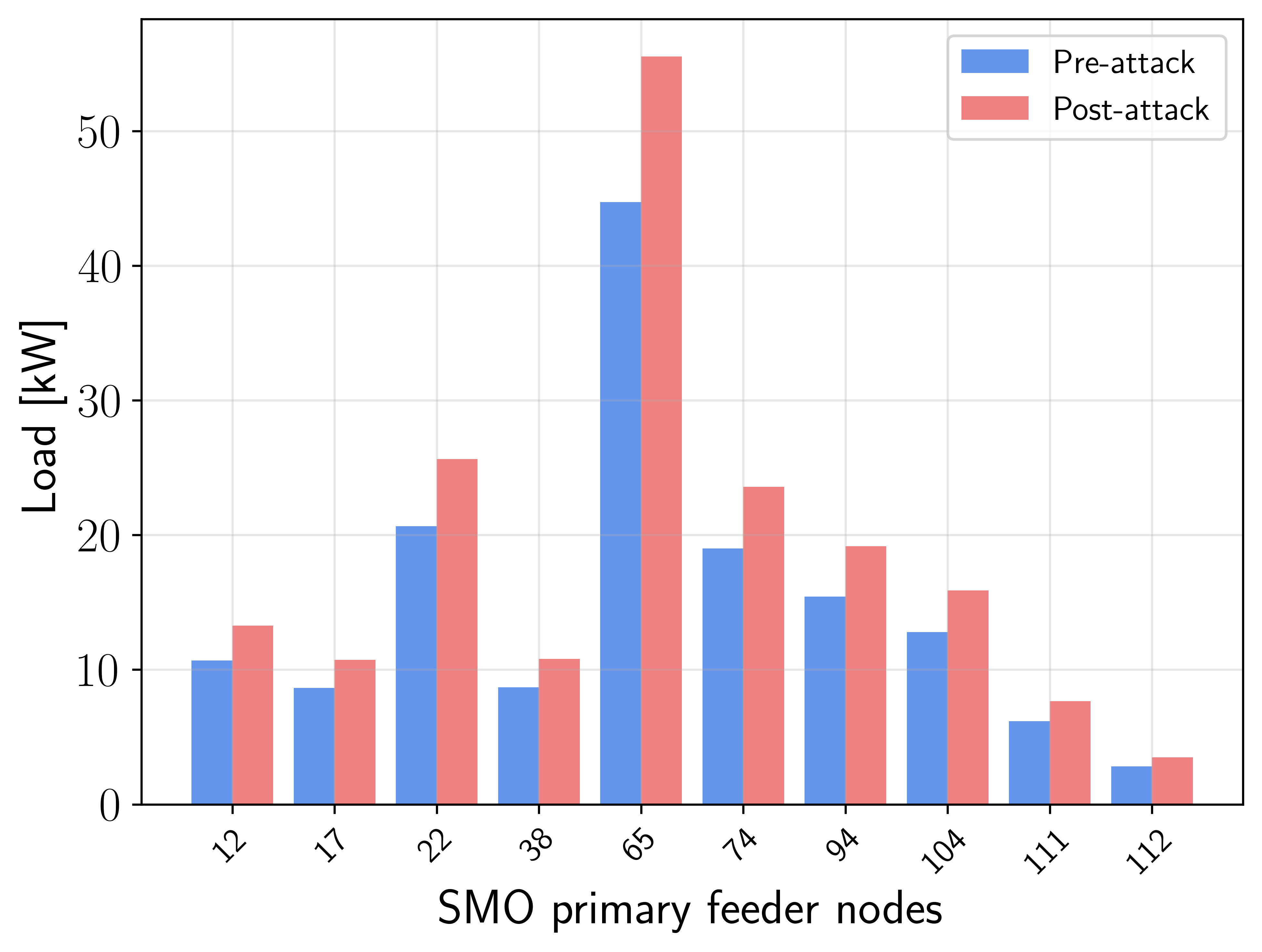}
    \hfill
\caption{Net load at attacked nodes. \label{fig:attack1a_load}}
\end{subfigure}
\begin{subfigure}[t]{0.49\columnwidth}
    \includegraphics[width=\columnwidth]{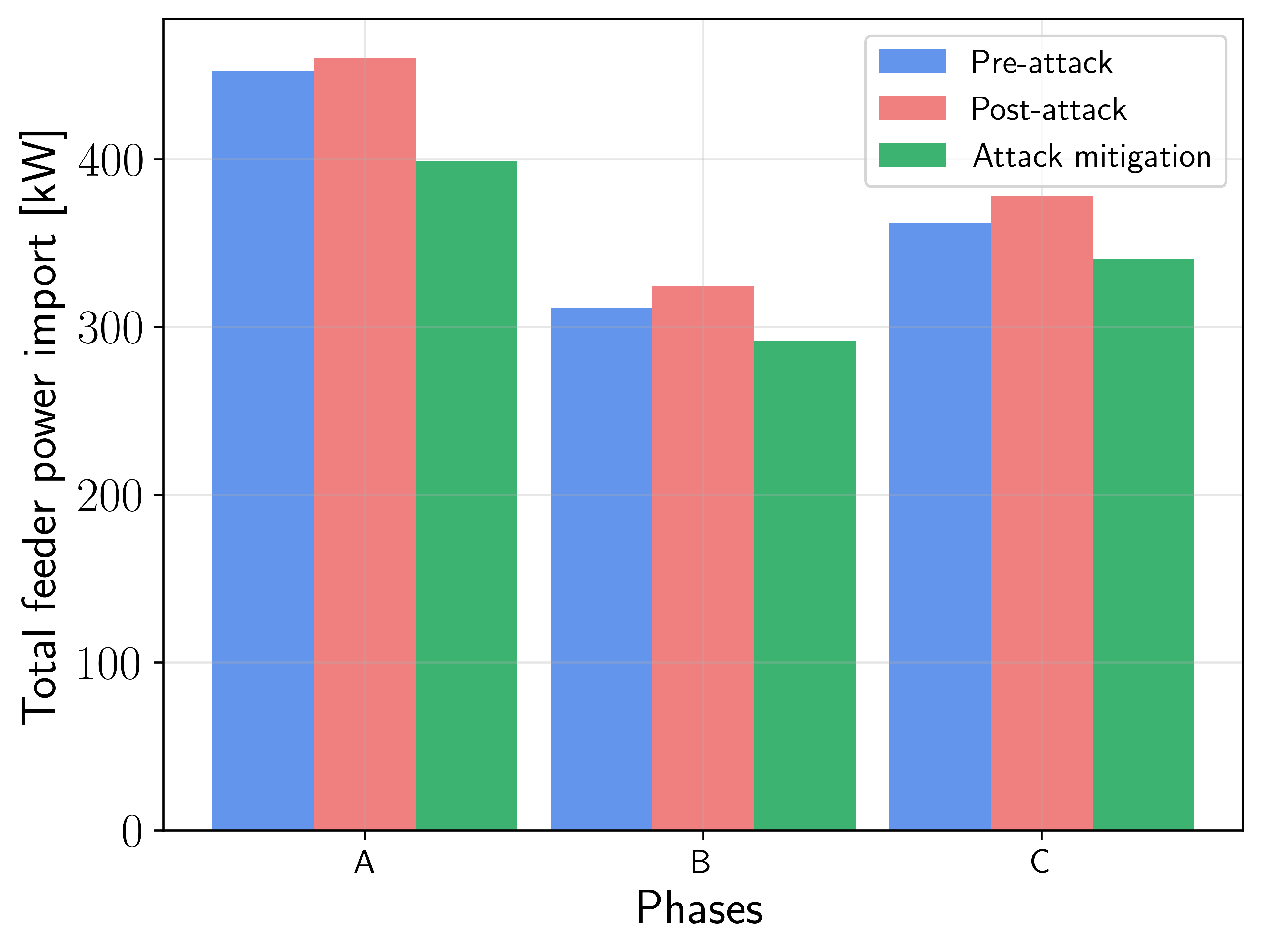}
\hfill
\caption{Feeder power import from main grid.\label{fig:attack1a_import}}
\end{subfigure}
\caption{Effect of attack 1a and mitigation.}
\end{figure}

\begin{figure}[htb]
\centering
\includegraphics[width=0.7\columnwidth]{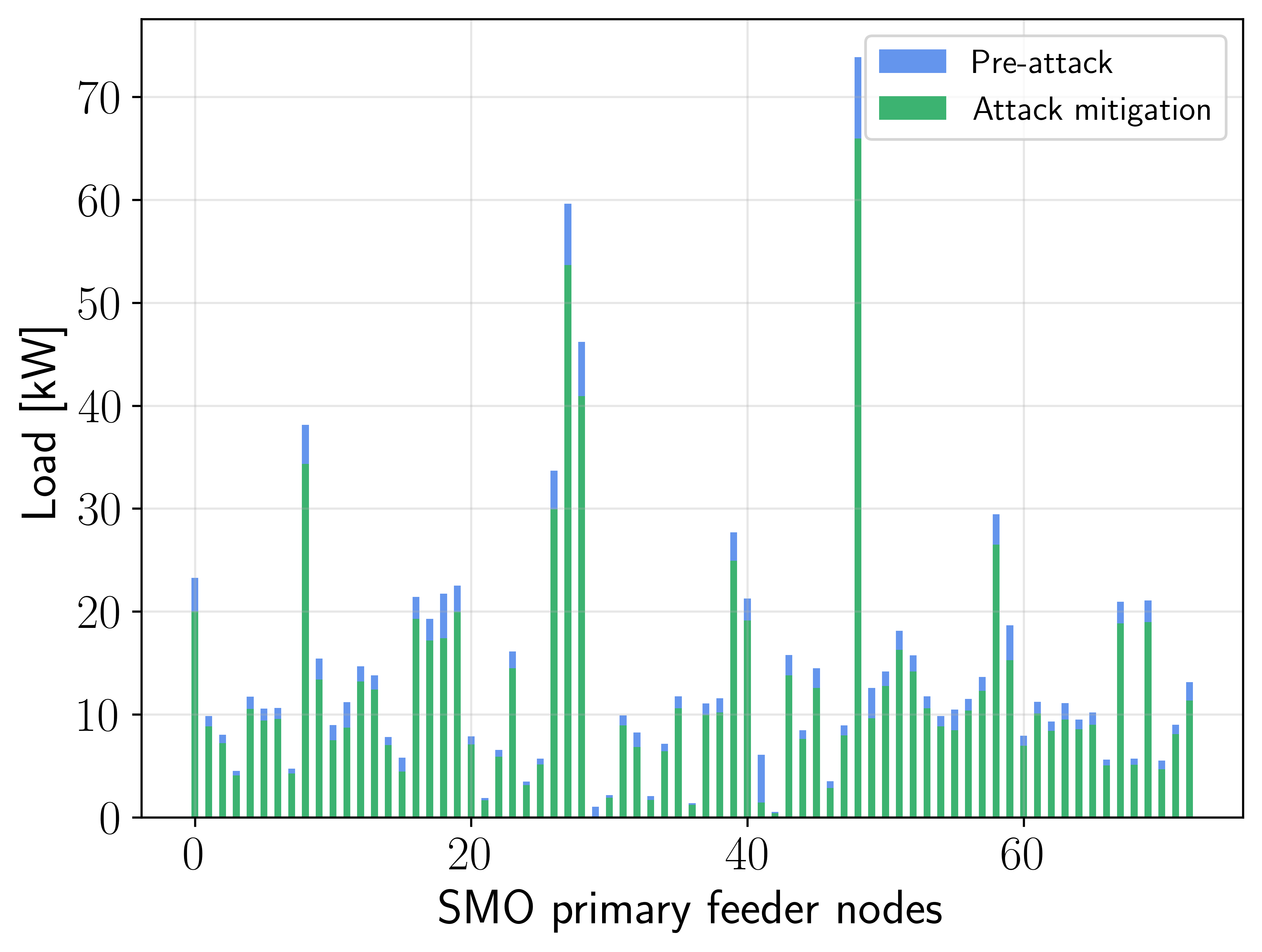}
\caption{Curtailment of flexible loads for attack 1a mitigation. \label{fig:attack1a_curtail}}
\end{figure}

\begin{figure}[htb]
\centering
\includegraphics[width=0.55\columnwidth]{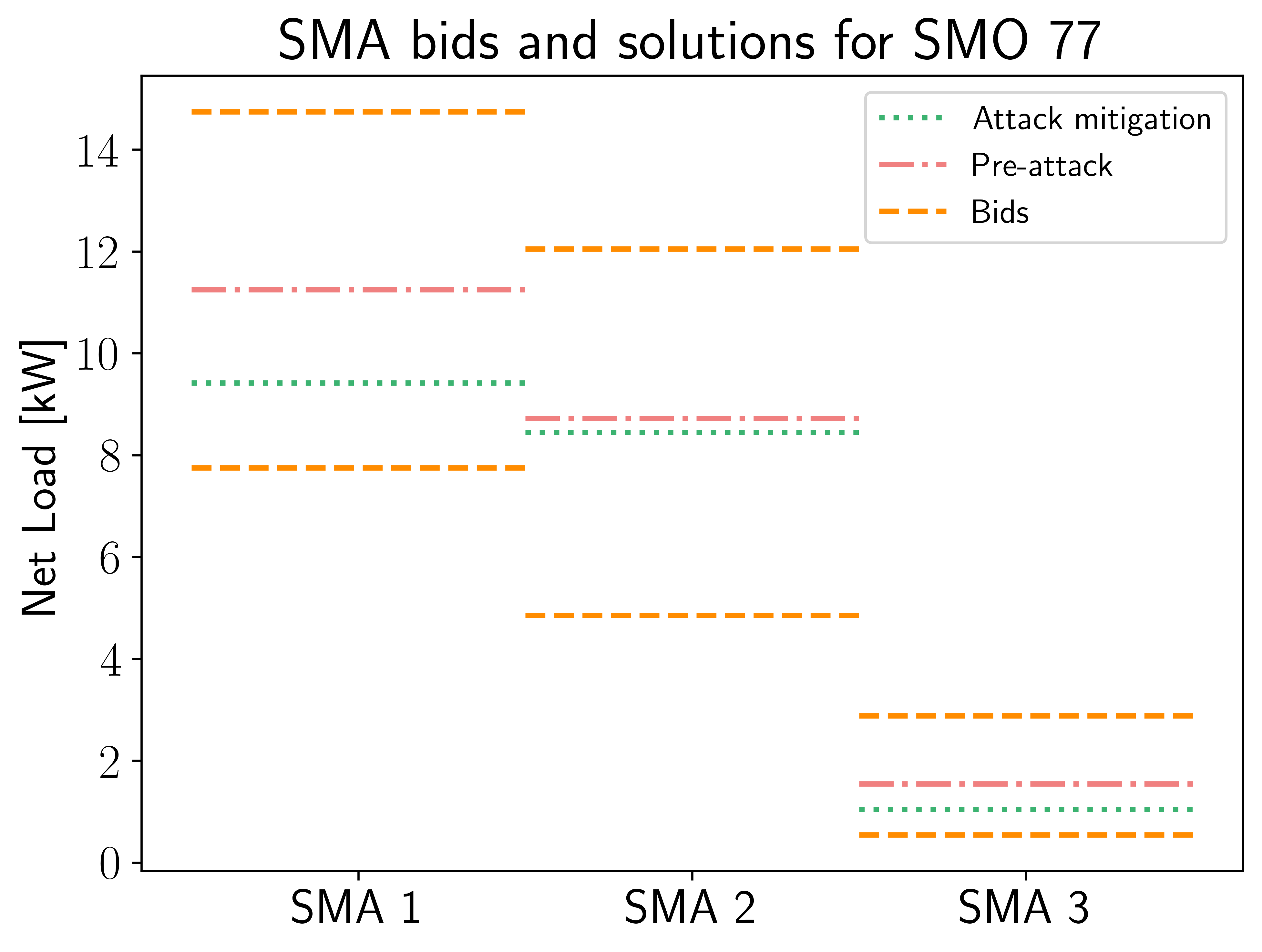}
\caption{Dis-aggregation of setpoint changes (from the PM) for SMO at node 77 across its 3 SMAs (in the SM) on phase B, after attack 1a mitigation.\label{fig:attack1a_icas}}
\end{figure}

The steps in mitigation are as follows: 10 SMO nodes are attacked, resulting in a total increase in load (generation shortfall) of 36 kW for the entire feeder as seen in \cref{fig:attack1a_load}. A large number of flexible load nodes across the entire feeder help with mitigation by curtailment and shifting as in \cref{fig:attack1a_curtail}. Flexible load curtailment at individual SMO nodes ranges from a minimum of 0.55 kW to a maximum of 7.8 kW reduction per primary feeder node - using a combination of resources like HVAC, WHs, batteries, and EVs to reduce the net load. There is a 123 kW decrease in power import after mitigation as seen in \cref{fig:attack1a_import}. The new SMO setpoints from the PM redispatch are then disaggregated amongst their SMAs during the following SM redispatch, with an example for SMO 77 shown in \cref{fig:attack1a_icas}.

\begin{figure}[htb]
\centering
\includegraphics[width=0.8\columnwidth]{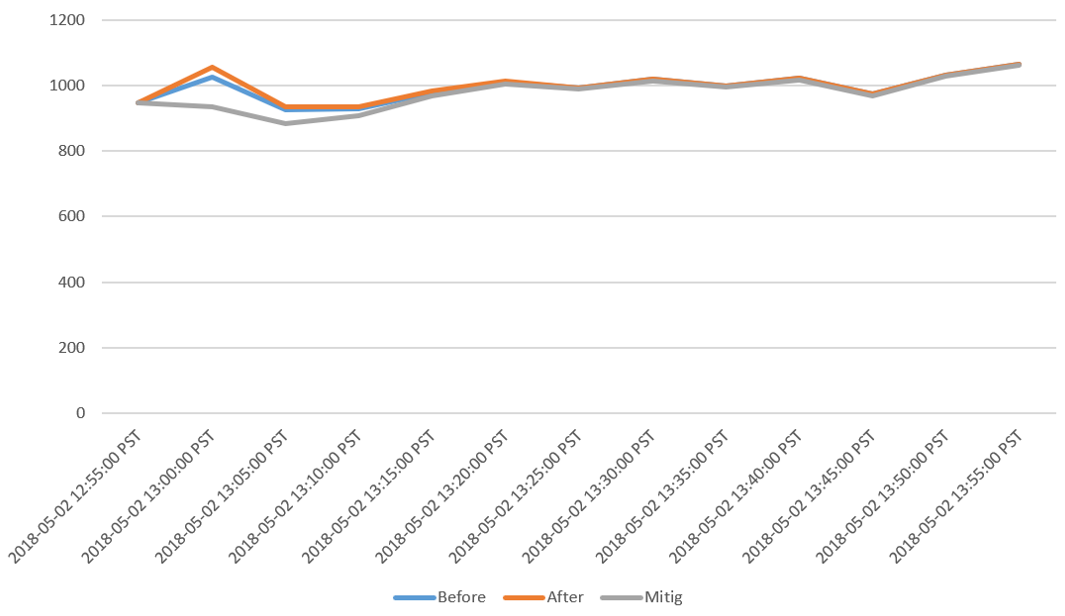}
\caption{LTDES validation of attack 1a in the DERIM-ADMS platform, showing total power import at the substation around the attack time at 13:00.} \label{figure18-ltdes}
\end{figure}

The outputs from the PM-SM market framework were sent to the DERIM interface using which the effect on the total net load at the substation feeder head could be determined with the DERIM-ADMS-DOTS software platform (see SI \cref{fig11-ltdes} for an overview of the validation process). It is clear from \cref{figure18-ltdes} that without the intervention of EUREICA, the impact of the attack is a 37 kW jump in the feeder demand; in contrast with EUREICA, the feeder demand is cut by 94 kW. Moving further ahead from the attack timestep, the feeder net load eventually approaches back to the same value as if there hadn't been an attack. See SI \cref{sec:attack1a_ltdes} for further details on this validation. See SI \cref{sec:attack1a_pnnl} and SI \cref{sec:attack1a_nrel} for the other validation results using the HELICS and ARIES platforms, respectively.

\subsection{Attack 1b validation based on resilience\label{sec:attack1b_rs_mit}}
\begin{figure}[htb]
\centering
\begin{subfigure}[t]{0.49\columnwidth}
    \includegraphics[width=\columnwidth]{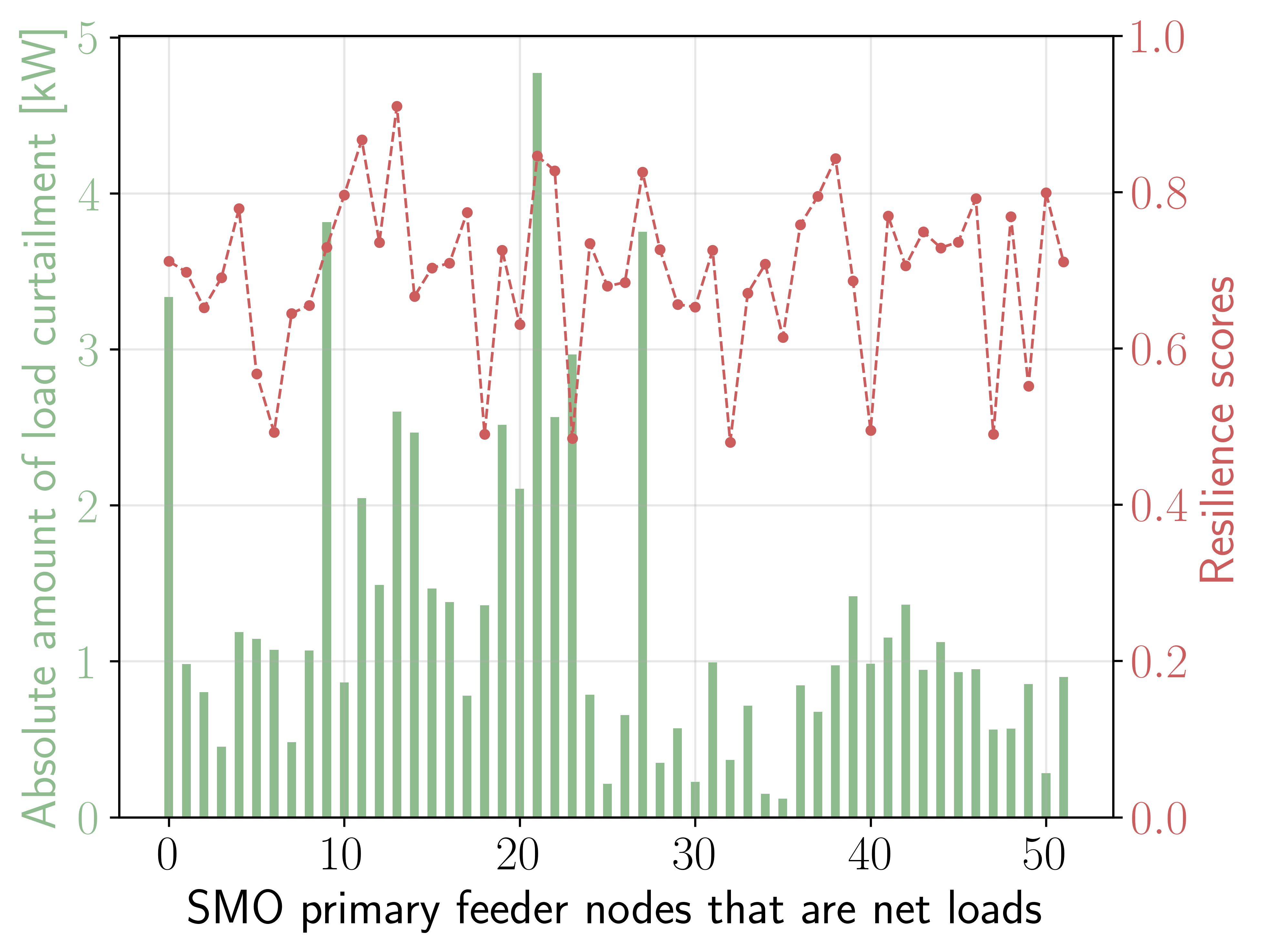}
    \caption{Absolute curtailment and RS.  \label{fig:attack1b_rs_absolute}}
\end{subfigure}
\begin{subfigure}[t]{0.49\columnwidth}
    \includegraphics[width=\columnwidth]{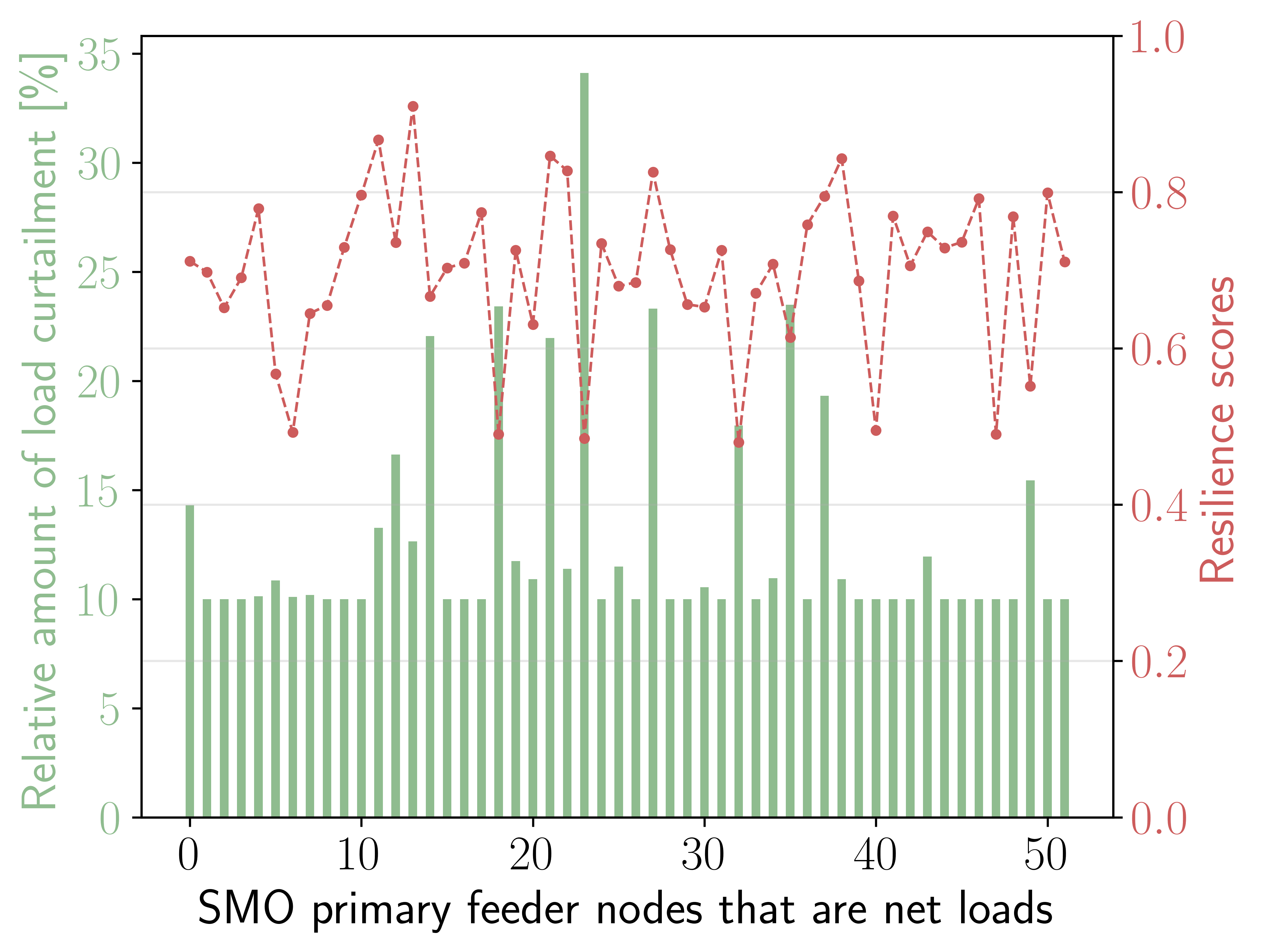}
    \caption{Relative curtailment and RS.  \label{fig:attack1b_rs_relative}}
\end{subfigure}
\caption{Distribution of absolute and relative amounts of load curtailment across the flexible net load SMOs, along with their corresponding resilience scores.\label{fig:attack1b_smo_rs_curtail}}
\end{figure}

Here, we briefly highlight the effects of resilience scores on the mitigation of attack 1b, where a number of DGs are attacked. Full details on this attack can be found in SI \cref{sec:attack1b}. Here, we focus only on how the RSs of SMOs and SMAs influence which resources are utilized to mitigate the attack. The RSs of the flexible SMOs are plotted against their absolute and relative levels of net load curtailment in \cref{fig:attack1b_rs_absolute} and \cref{fig:attack1b_rs_relative}, respectively. We see that in relative terms, the curtailment is generally distributed evenly to ensure that no single SMO is disproportionately affected. However, if the PMO does need to utilize more flexibility from certain SMOs, it generally calls upon more reliable ones with higher RSs. The absolute amounts of curtailment vary for each SMO based on their baseline load. This also holds while dis-aggregating SMO setpoints at the SM level, where the SMO allocates greater flexibility to SMAs with higher RSs, as seen in \cref{fig:attack1a_icas}. See SI \cref{sec:attack1b_ltdes} and SI \cref{sec:attack1b_pnnl} for validation results of this attack using DERIM-ADMS-DOTS and HELICS, respectively.

\subsection{Contributions of SM and PM to attack 1c mitigation}
\begin{figure}[htb]
\centering
\includegraphics[width=0.8\columnwidth]{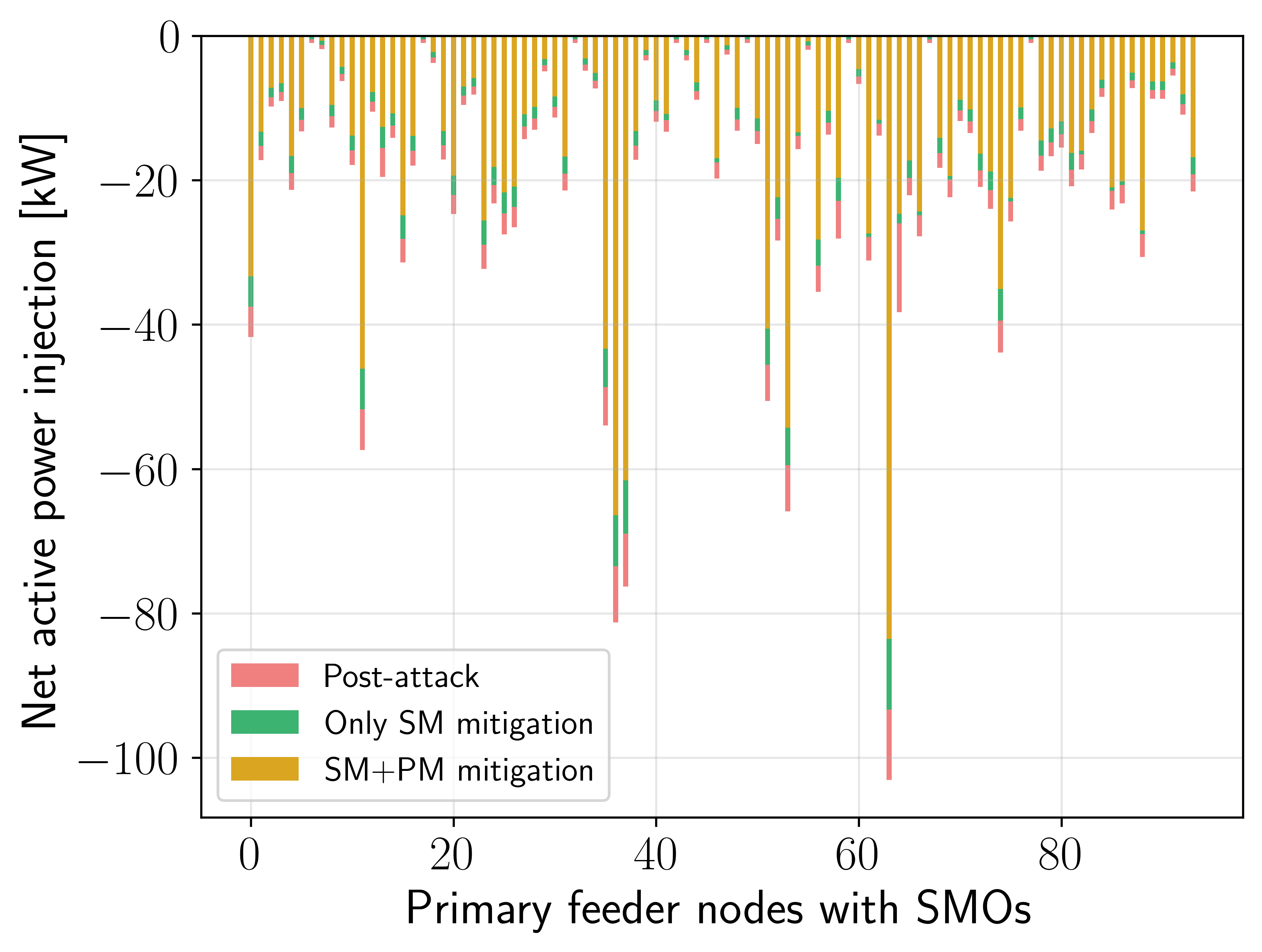}
\caption{Contributions of SM and PM flexibility for attack 1c mitigation. \label{fig:attack1c_sm_pm}}
\end{figure}

Attack 1c is a more distributed attack where individual SMAs are attacked directly. Here, we show how both the SM and PM flexibility are needed to fully mitigate the attack. \cref{fig:attack1c_sm_pm} shows the contributions of the SM and PM toward attack mitigation. We see that for most of the SMO nodes, both the SM and PM flexibility play a significant role in reducing the net load compared to the post-attack case. At the SM level, we utilize the available upward flexibility of any SMAs with remaining online DGs and use the downward flexibility of all net load SMAs. At the PM level, we utilize the downward load flexibility of the SMOs (which are all net loads after the attack). Further details on this attack can be found in SI \cref{sec:attack1c}. 

\subsection{Mitigation of attack 2}
Here, we describe the mitigation of two attacks at the primary feeder level that are relatively broader in scope, one is a medium-scale and the second is a large-scale attack. Both are disruption attacks where the attacker shuts down one or more of the large DGs in the network. We only consider a single primary market time step to study the effects of an instantaneous attack. Mitigation can use P dispatch from batteries, P and Q curtailment from flexible loads, limited P dispatch from PV, Q support from smart inverters (connected to PV and batteries), as well as conventional dispatchable fossil fuel sources like diesel generators. We only present results for the large-scale attack here, details on the medium-scale attack can be found in SI \cref{sec:attack2a}. 

\subsubsection{Large-scale attack 2b\label{sec:attack_2b}}
\begin{figure}[htb]
\centering
\includegraphics[width=\columnwidth]{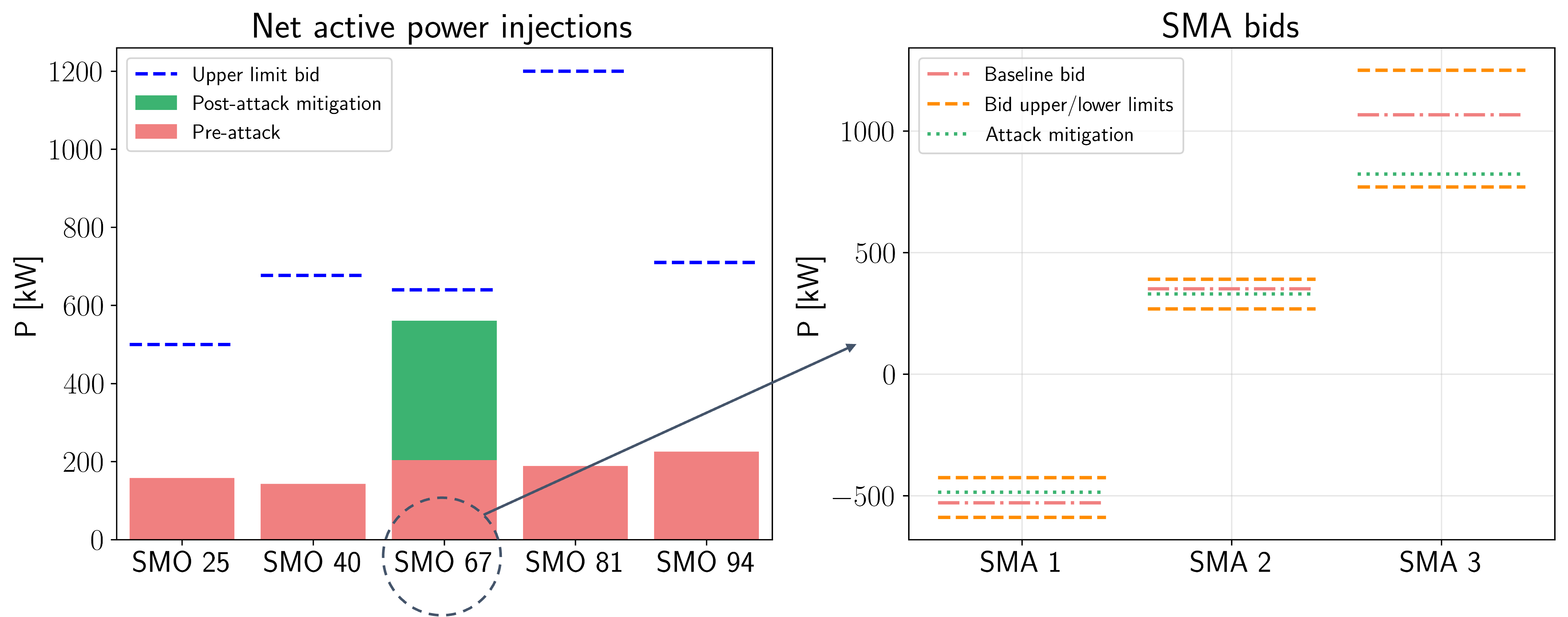}
\caption{Mitigation of large-scale attack 2b. \label{fig:attack_large}}
\end{figure}

Here, we adopted a top-down approach in emulating an attack and started with a Kundur 2-area transmission model, with the attack occurring in Area 2 (see SI \cref{fig:kundur}) which consists of a load of 1767 MW. Noting that Area 2 can be broken down into 552 IEEE-123 feeders, each with approximately 3.2 MW, we assume that an attack that compromises about 650 kW of generation, occurs in each of these 552 feeders. This in turn corresponds to an overall shortfall of 359 MW at the transmission level. We introduced this 650 kW shortfall in the form of a generation loss at four nodes 25, 40, 81, and 94, in each of the 552 primary feeders. The only remaining SMO with significant generation capability is at node 67. With the same procedure as outlined in the previous scenarios, the use of our proposed EUREICA framework leads to the results in \cref{fig:attack_large}. In order to mitigate the attack, we need to leverage the upward generation flexibility of the remaining SMO 67 to increase its output injection after attack mitigation, while the net injections for all the other four attacked SMOs drop to zero as seen in the left plot. The right plot shows the new SMA schedules resulting from the revised SM clearing.

\begin{figure}[htb]
\centering
\includegraphics[width=0.7\columnwidth]{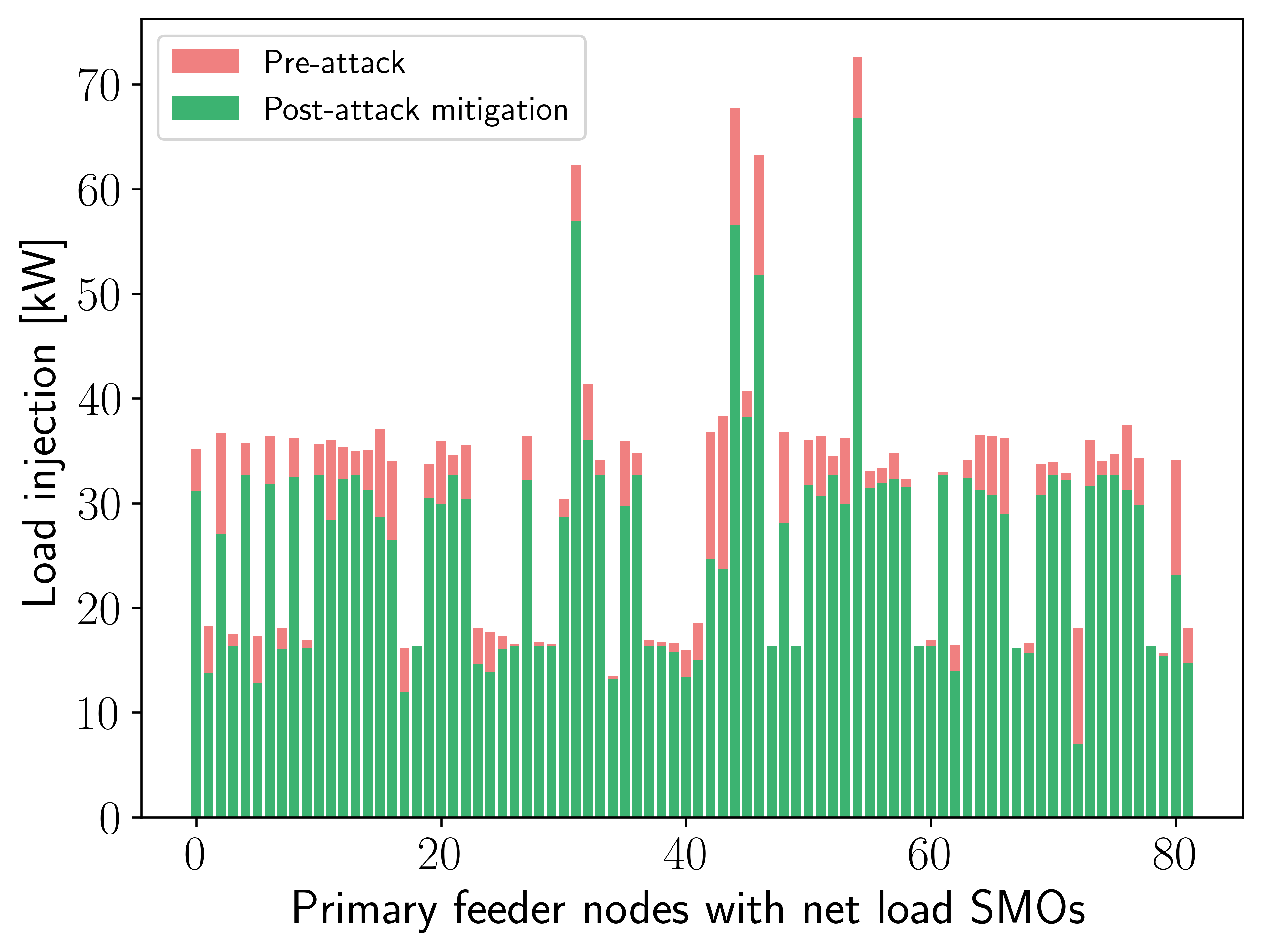}
\caption{Flexible load curtailment for large attack mitigation. \label{fig:attack_large_loads}}
\end{figure}

However, due to the larger scale of the attack, re-dispatching the generator SMOs is no longer sufficient to fully meet the shortfall. Furthermore, as seen in \cref{fig:attack_large}, we are not able to utilize all the upward flexibility of the remaining online SMO 67 since its dispatch is limited by power flow constraints, on nodal voltages and line currents in particular. Thus, we also need to perform some shifting and curtailment of high-wattage flexible loads. These could include EVs and thermostatically controlled loads like HVAC and WHs. In addition, it could also involve some discharging of battery storage systems to reduce the net load. The distribution of net load reductions across the remaining SMOs is shown in \cref{fig:attack_large_loads}, with a total decrease of around $14 \%$ as seen in \cref{tab:metrics}. From \cref{tab:metrics} we also see that the attack would have potentially increased the power import from the transmission grid by over $37 \%$, but the combination of increased local generation and load curtailment helps keep the imported amount almost the same as before. Further details on this attack can be found in SI \cref{sec:attack_2b_details}.

\subsubsection{Effects at the transmission level}
\begin{figure}[htb]
    \centering
    \includegraphics[width=0.9\columnwidth]{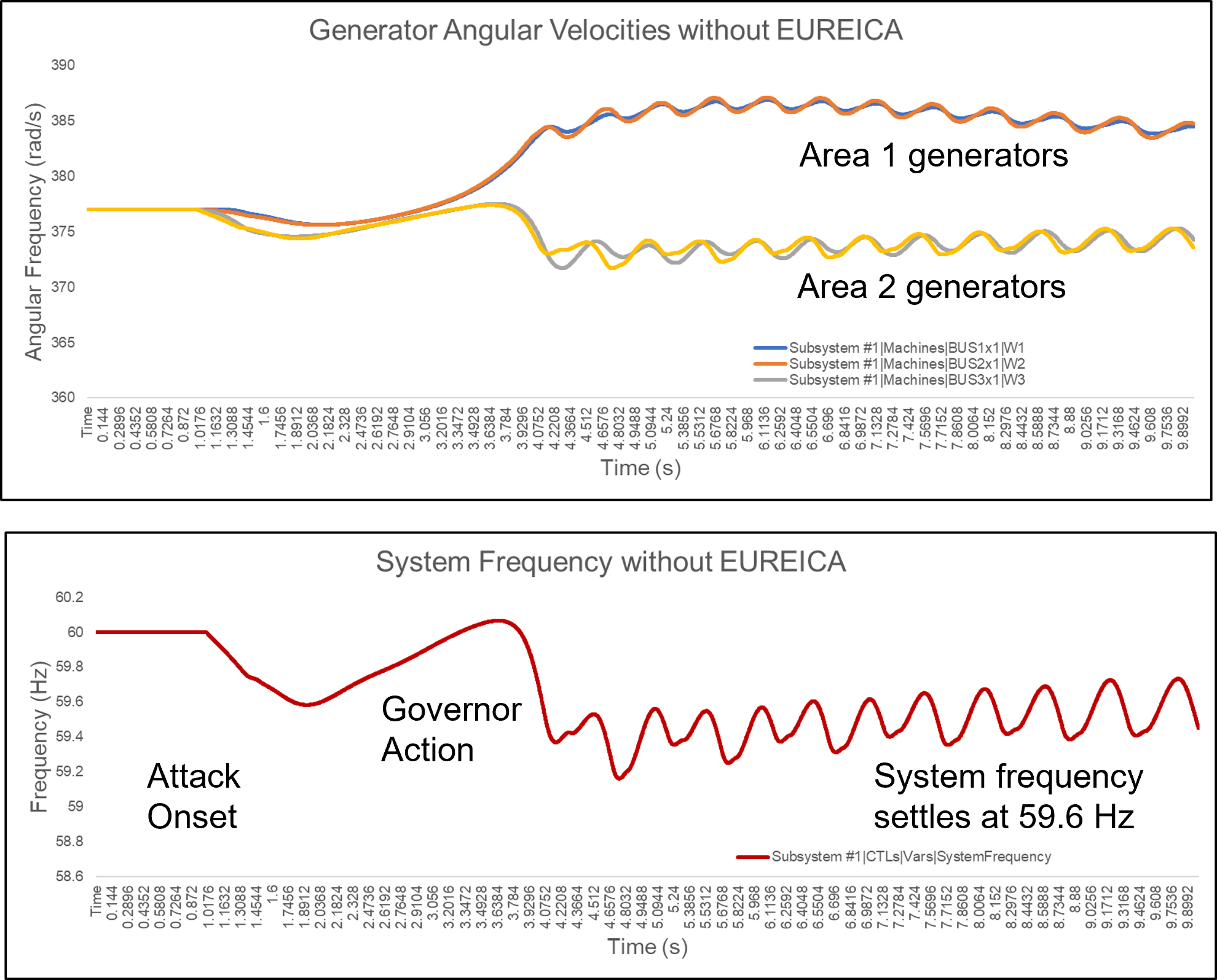}
    \caption{Response without EUREICA; system settles at sub-optimal frequency}
    \label{fig:attack2_wo_eureica}
\end{figure}

\begin{figure}[htb]
    \centering
    \includegraphics[width=0.9\columnwidth]{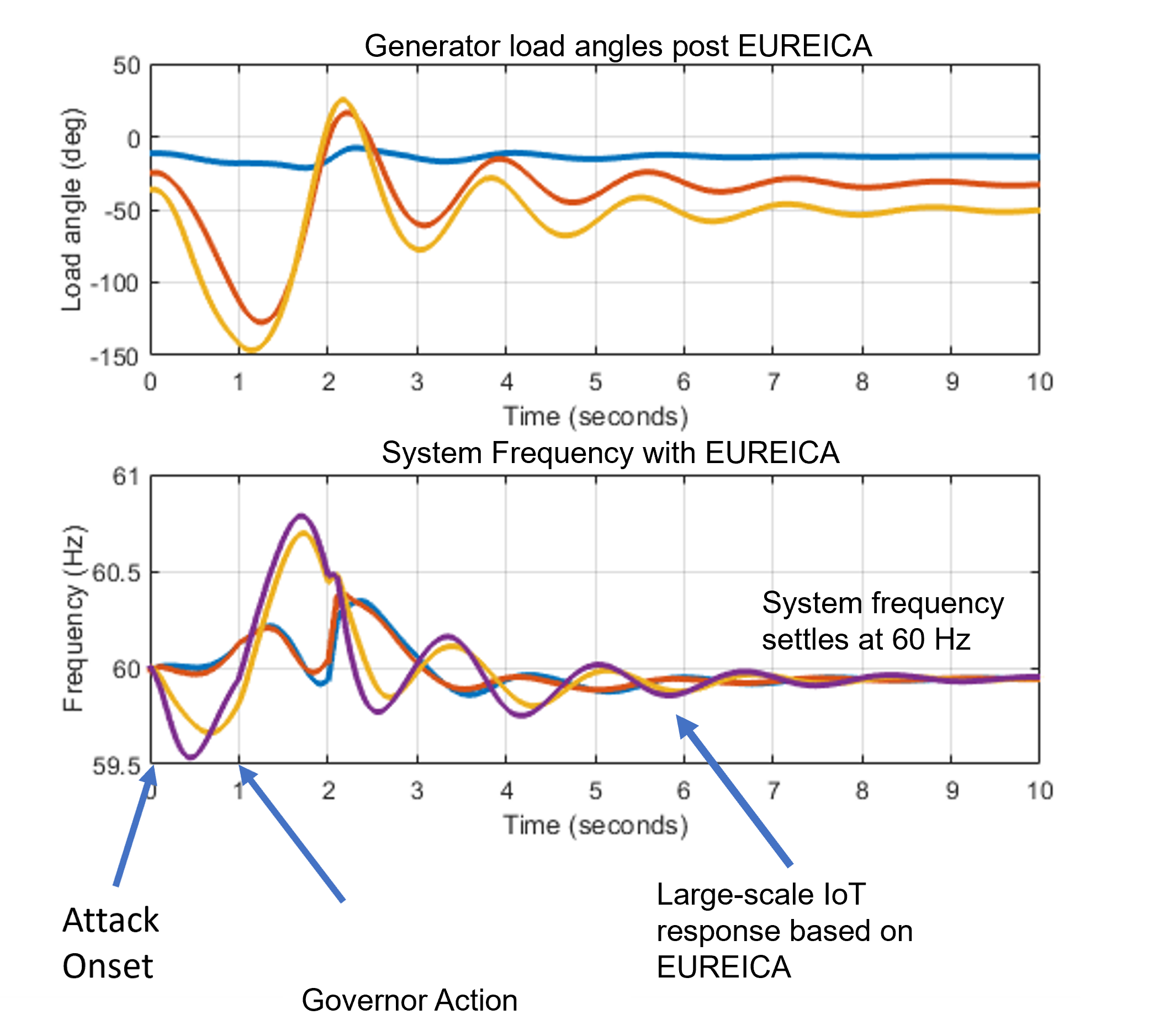}
    \caption{Frequency response with EUREICA; system settles at 60 Hz following demand response and load shedding enabled by the EUREICA framework}
    \label{fig:attack2_with_eureica}
\end{figure}

The overall impact of the generation shortfall and mitigation using EUREICA is simulated in the RTDS using a proxy where the individual feeders are not modeled, but the aggregated effect is studied at the transmission scale. A combined shortfall of 359 MW, corresponding to a simultaneous compromise and outage of 650 kW in all 552 primary feeders in Area 2 triggers a frequency event (see \cref{fig:attack2_wo_eureica}). Left unchecked, this can potentially lead to drastic load shedding or parts of the system being blacked out. To mitigate this situation, the power flow from Area 1 to Area 2 needs to be increased, which was observed in the RTDS, through the action of the governor system, which responds in the timescale of seconds, by increasing generation from the other generators present in the system proportionately based on a droop value. This increases the power flow from the generation-rich Area 1 to Area 2. However, changing the tie-line power flow creates a frequency imbalance, resulting in the system frequency oscillating, and settling at a lower/higher frequency, as shown in \cref{fig:attack2_wo_eureica}. With the EUREICA framework, the frequency mismatch is mitigated by suitably leveraging the flexibility of the remaining generation as well as demand response (DR) mechanisms from flexible loads at both the SMO and SMA levels (see \cref{fig:attack2_with_eureica}). Once the governor response is completed and the system settles at a sub-optimal frequency, a combination of intelligent DR and generation redispatch in Area 2 facilitated by the EUREICA framework allows the system frequency to be restored to normal, ensuring grid resilience, avoiding system stress and increased operational costs.

\subsubsection{Key system metrics, economic, and distributional impacts}
In our simulations, we find that attack mitigation comes at the expense of increased operational costs for the PMO since it needs to dispatch more expensive local resources to a greater extent, rather than importing cheaper power from the main grid (at the LMP rate). The PMOs and SMOs also need to adequately compensate agents for the critical flexibility they provide. As shown in \cref{tab:metrics} for attack 2b, the attack increases the system operating costs by around $7\%$, and the mitigation steps raise the cost by over $31\%$, both relative to the pre-attack case. However, the PMO could recoup this through other revenue streams and cost savings. For example, the transmission system operators may compensate PMOs for locally containing attacks. Being able to leverage local DER flexibility through markets could also reduce the amount of auxiliary backup generation that the PMO needs to maintain, and lower the reserves it may have to otherwise procure from capacity or ancillary service markets. The PMO in turn could also redistribute some of these benefits among the SMOs and SMAs.

\begin{table}
\centering
\caption{Summary of metrics for large-scale attack 2b scenario. \label{tab:metrics}}
\begin{tabular}{@{}p{2.5cm}lll@{}}
\toprule
                                                    & \textbf{Pre-attack} & \textbf{Post-attack} & \textbf{Attack mitigation} \\ \midrule
\textbf{Power import from main grid {[}kW{]}} & 1,325               & 1,821 (+37.4\%)      & 1,328                           \\
\textbf{Total cost {[}\${]}}                        & 10,752              & 11,500 (+7\%)        & 14,156 (+31.7\%)                \\
\textbf{Total load {[}kW{]}}                        & 2,064               & 2,023 (-0.02\%)      & 1,775 (-14\%)                   \\ \bottomrule
\end{tabular}
\end{table}

\subsection{Mitigation of attack 3}

\begin{figure}[htb]
  \centering
  \includegraphics[width=0.75\columnwidth]{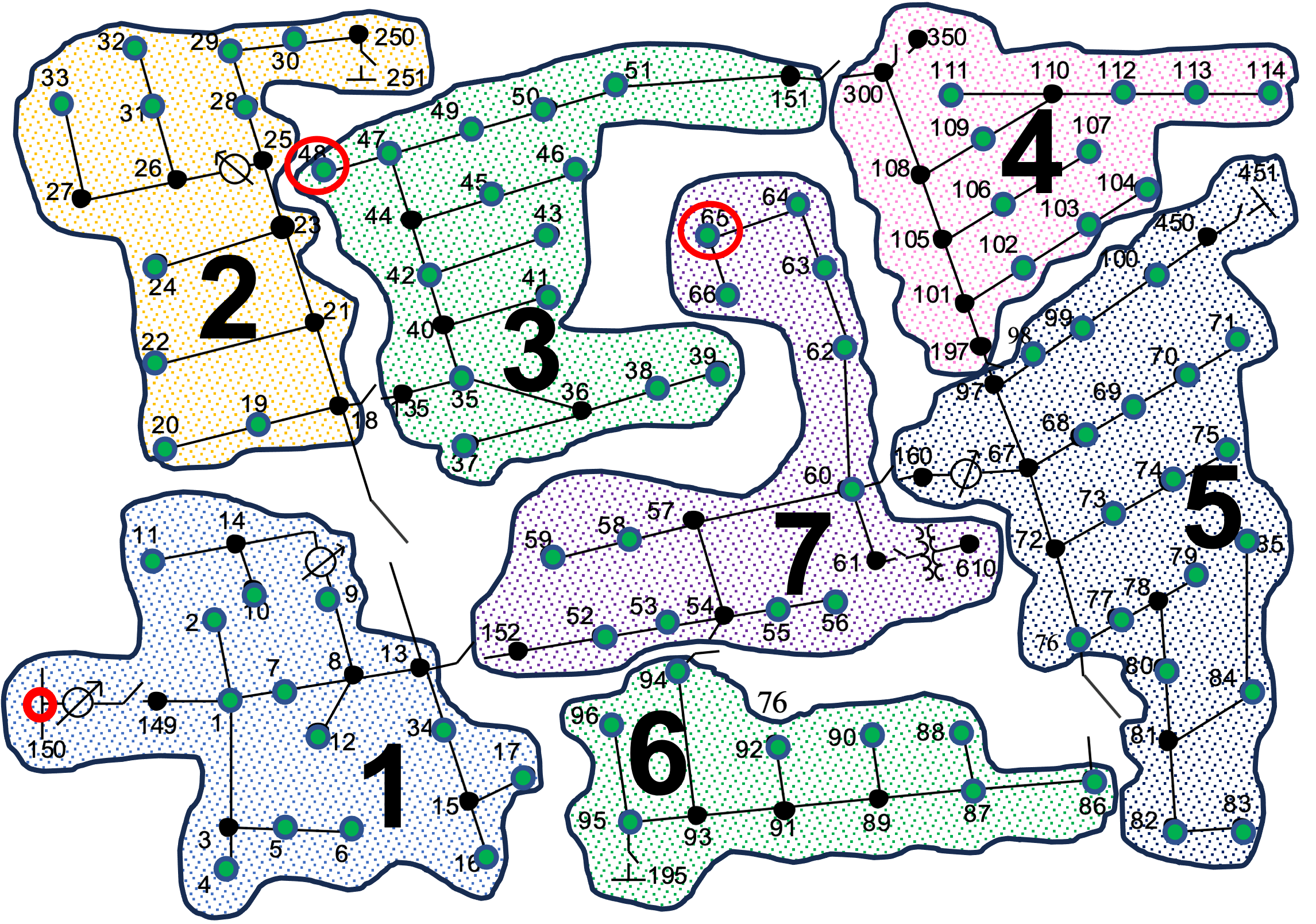}
  \caption{EUREICA IEEE 123-node feeder for reconfiguration module validation.}
  \label{fig:attack4-diagram}
\end{figure}

We now consider the attack scenario where the distribution grid is isolated from the transmission system. In such a case, the distribution grid is fed through an alternate circuit such as from node 350 (see \cref{fig:attack4-diagram} for details). A typical response in such a case is that the distribution grid breaks into several “zones” - creating smaller islands where only a portion of the load is fed through any DERs that may be present. We show below that with the increased awareness provided by the EUREICA framework, a much higher percentage of consumers remain unaffected, by suitably leveraging the DERs at node 48, the microgrid system connected at node 65 (marked by the red circles in \cref{fig:attack4-diagram}, and DR methodologies. In order to ensure feasibility and supply-demand balance with islanding, we also introduce two large diesel generators located at nodes 48 and 65 which may only be called upon when the feeder is islanded. Three cases are presented.

\subsubsection{Critical loads distributed across the feeder}
\begin{figure}[htb]
  \centering
  \includegraphics[width=\columnwidth]{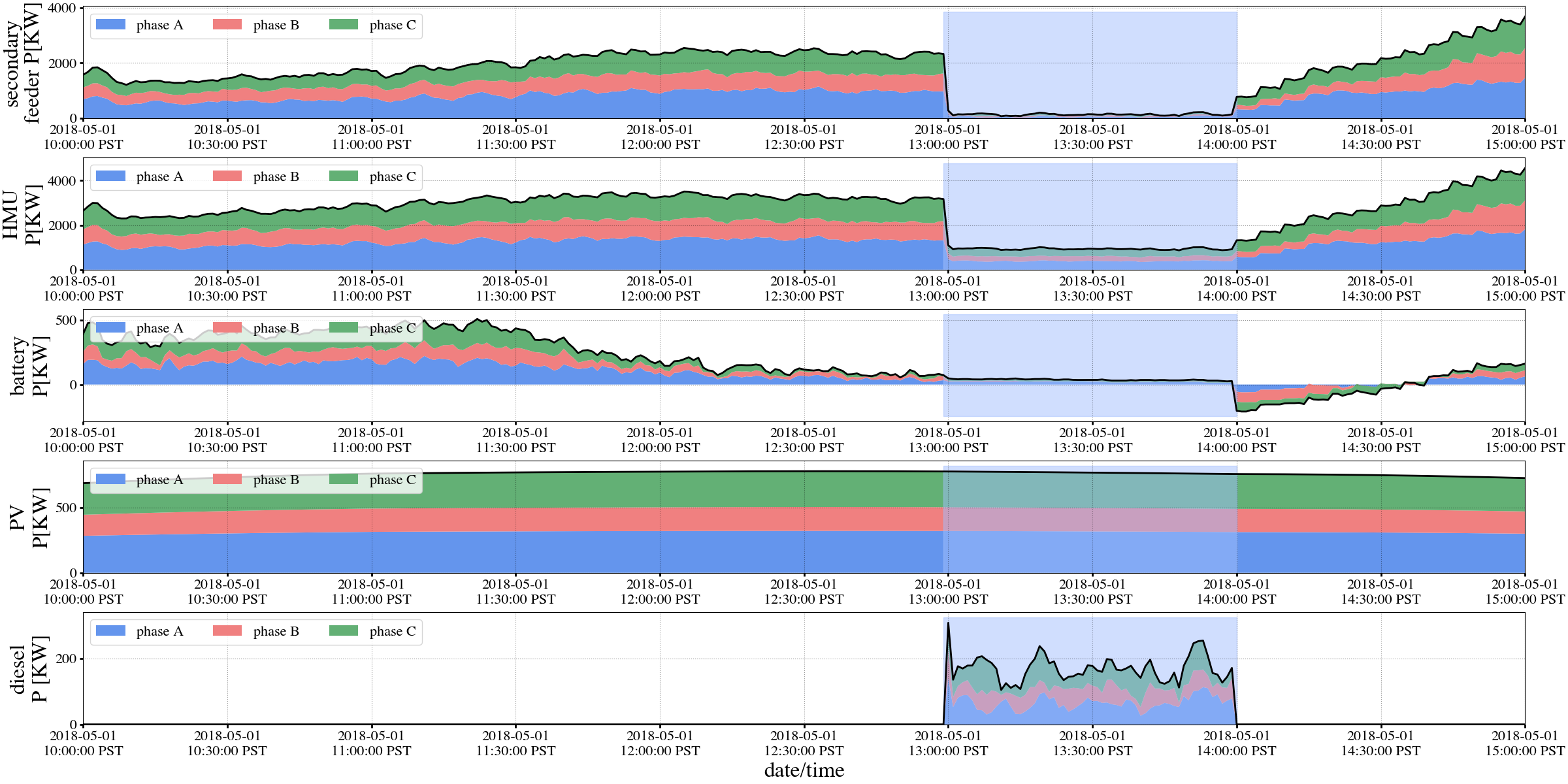}
  \caption{Demand and DG with resilience-based reconfiguration during attack 3.}
  \label{fig:attack4-dgreconf-loads-pnnl}
\end{figure}

In this case, through the proposed resilience-based IoT load restoration with DR optimization strategy (see SI \cref{sec:reconfig_alg}, a feasible reconfiguration path is computed to open or close tie switches and completely or partially shed non-critical grid edge loads using reconfiguration to allow the available generation resources to cover approximately $30\%$ of total load in the system. As seen in \cref{fig:attack4-dgreconf-loads-pnnl}, with almost $70\%$ of the load shed (second graph from the top) between 13:00 and 14:00, and batteries only allowed to discharge, if possible, to supply extra energy (third graph from the top), the burden on the diesel generators is significantly alleviated as they only need to ramp up to about $230$ KW. These results were validated using the HELICS co-simulation platform at PNNL (see SI \cref{sec:val_pnnl} for details). Additional validation results using HELICS and LTDES are included in SI \cref{sec:attack3_pnnl} and SI \cref{sec:attack3_ltdes1}, respectively).

\subsubsection{Critical loads aggregated in a single zone}
\begin{figure}[htb]
  \centering
  \includegraphics[width=\columnwidth]{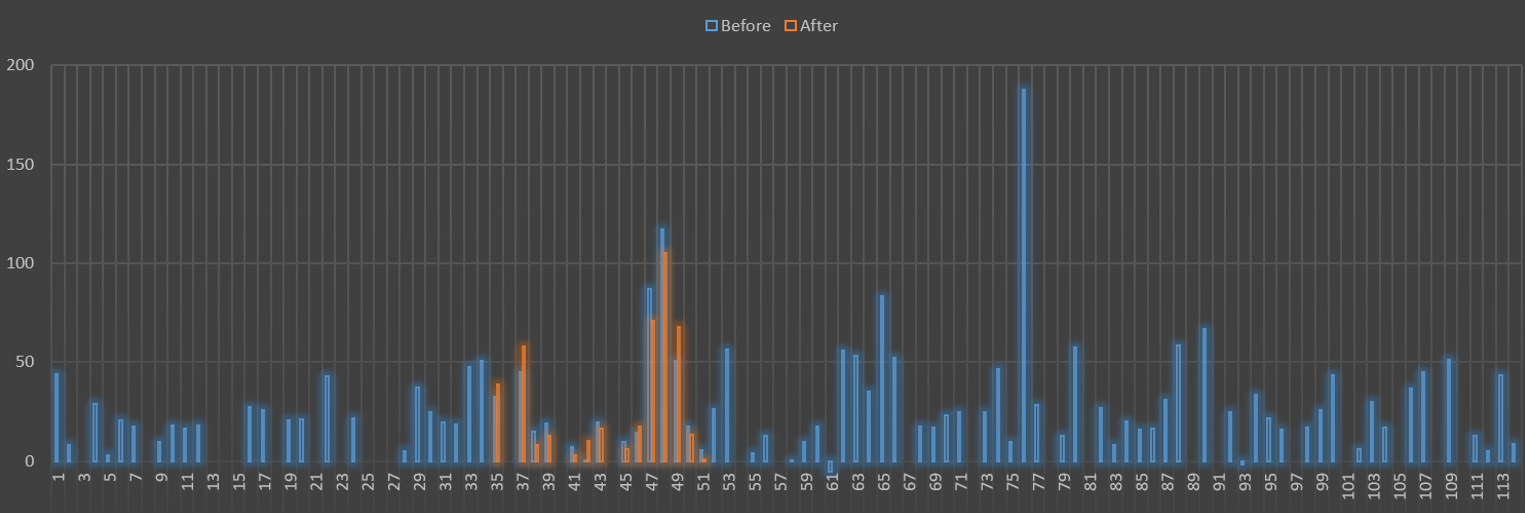}
  \caption{Primary node load change between 12:59 (before) and 13:00 (after attack).}
  \label{fig:attack4-ltdes}
\end{figure}

In this case, the SA from EUREICA helps the reconfiguration algorithm to disconnect or open the switches 18-135 and 151-300 to island zone 3 and pick up only the critical loads in this zone using the DG at node 48, which is a total of 430 kW. The results from this case are shown in \cref{fig:attack4-ltdes}. These results were validated using the DERIM and ADMS-DOTS software at LTDES (see SI \cref{sec:val_ltdes} for details). Additional results are included in SI \cref{sec:attack3_ltdes2}.

\subsubsection{Mitigation with a military microgrid}\label{mm_validation}
\begin{figure}[htb]
     \centering
     \begin{subfigure}{0.95\columnwidth}
         \centering
         \includegraphics[width=\columnwidth]{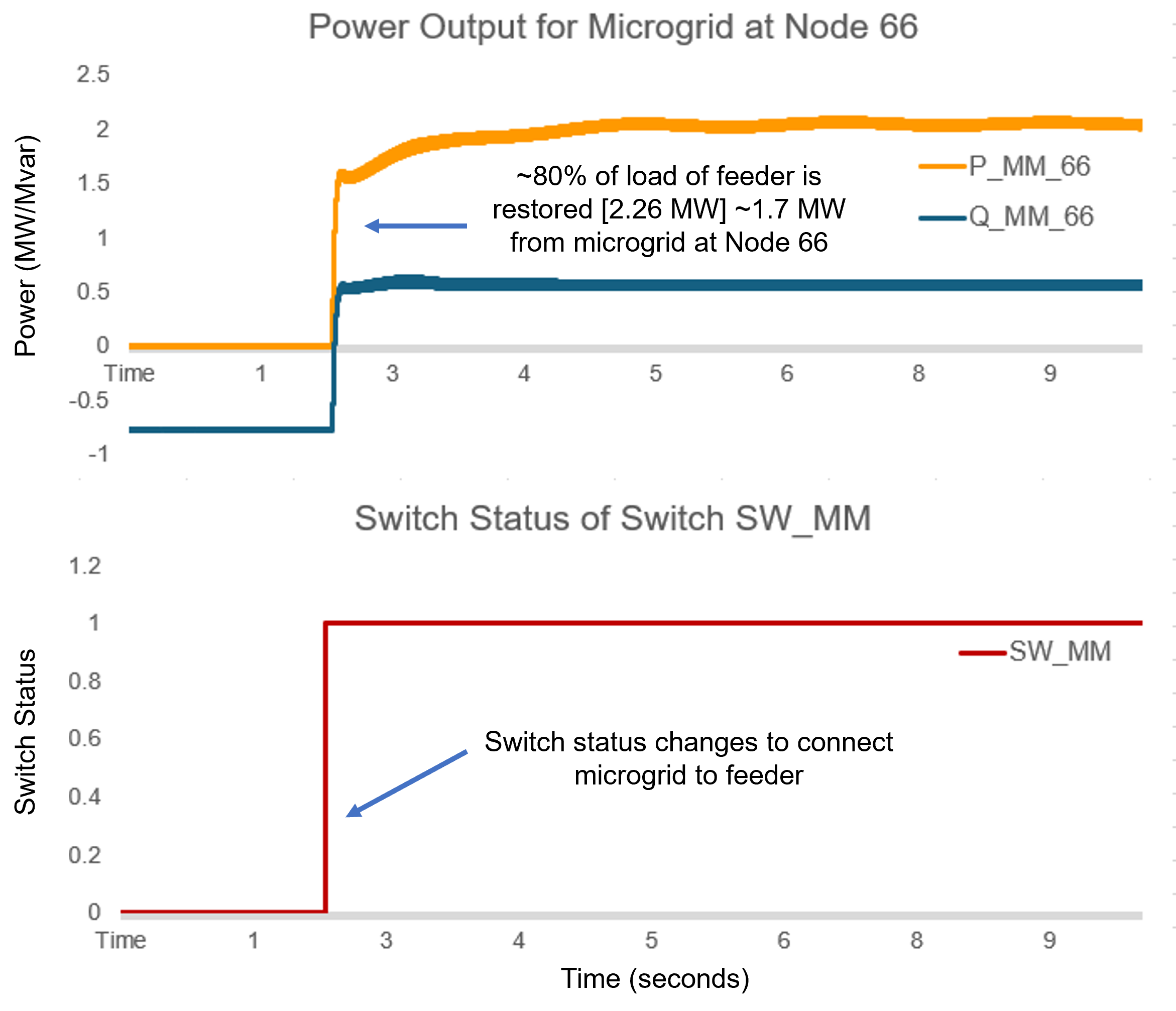}
         \caption{Microgrid response after reconfiguration.}
         \label{attack4_MM}
     \end{subfigure}
     \begin{subfigure}{0.95\columnwidth}
         \centering
         \includegraphics[width=\columnwidth]{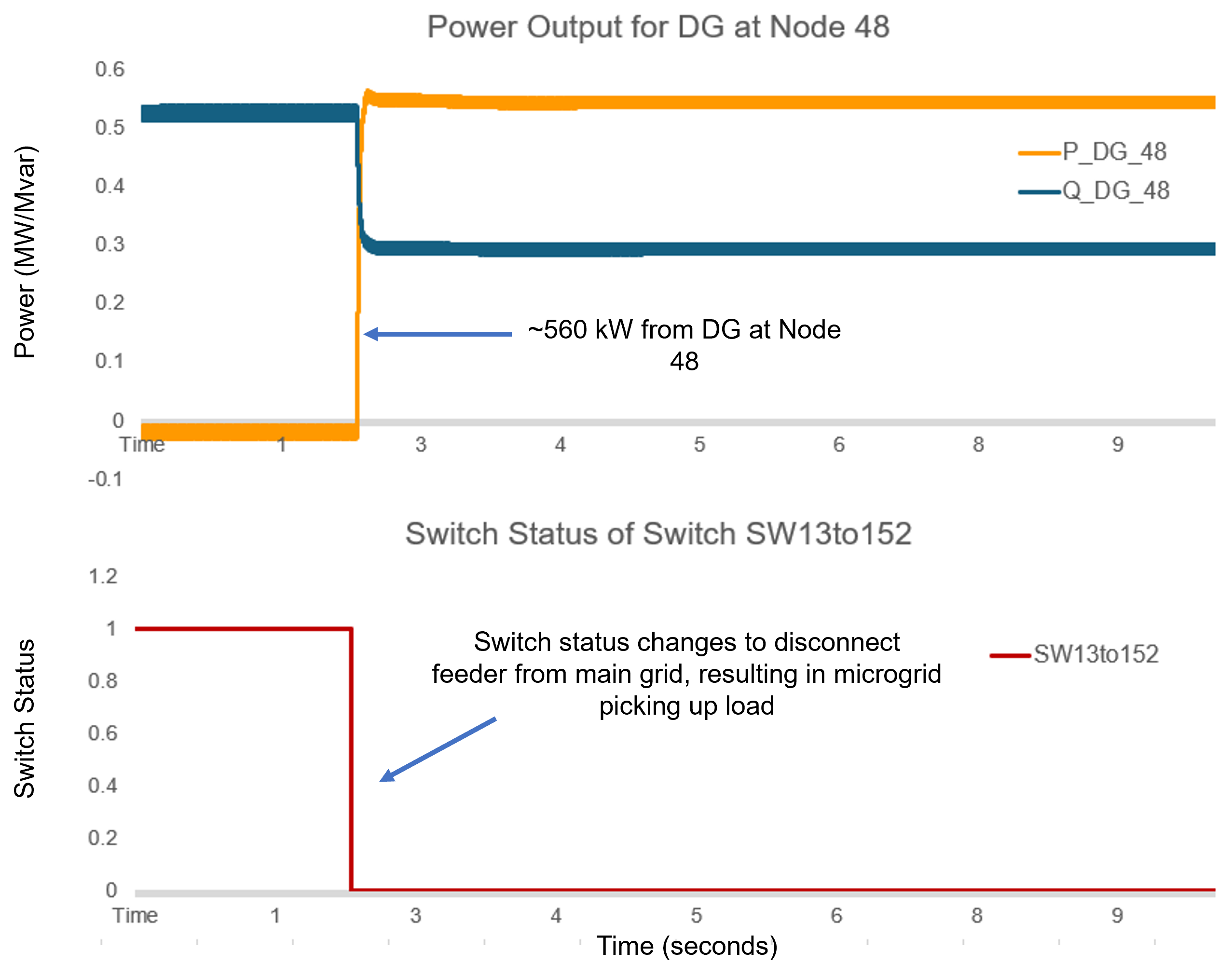}
         \caption{DG response after reconfiguration.}
         \label{fig:attack4_DG}
     \end{subfigure}
        \caption{System response after reconfiguration with microgrid.}
        \label{fig:nrel-attack4}
\end{figure}

We assume that there is a military microgrid at node 66 in the primary circuit, which serves as a backup directly in the distribution system. Under current regulations, defense critical systems have to be disconnected and isolated in the event of contingencies. Since EUREICA has the ability to identify trusted resources, our thesis is that there is confidence in the security of this resource as well as in meeting the power flow requirements, making it feasible to use this additional resource for attack 3 mitigation. First, the fault is isolated using reconfiguration based on the algorithm described in SI \cref{fig:reconfig_alg}. The reconfiguration algorithm returns the most resilient path for implementation, in this case, only one feasible path is present so that is chosen. This islands the feeder by opening the switch between nodes 150 and 149 and connecting the switches to the DG and microgrid at nodes 48 and 66, respectively. Then, a combination of $\approx$ 1.7 MW from the microgrid at node 66, 560 kW from the DG at node 48, and customer-side DR is utilized to pick up approximately 80\% of the total load of the feeder. Some results from this case are shown in \cref{fig:nrel-attack4}, validated using the ARIES platform at NREL (see SI \cref{sec:val_nrel} for details). The complete set of results can be found in SI \cref{sec:attack3_nrel}.

\section{Summary\label{sec:conclusion}}
We have proposed a framework, EUREICA, for achieving grid resilience through the coordination of IoT-Coordinated Assets that are trustable. A local electricity market that has been previously shown to lead to grid reliability and provide services such as voltage support and overall power balance, is leveraged in this framework to ensure grid resilience. The local market accomplishes this through SA to co-located operators. This SA consists of information about DERs and their power injections, as well as their levels of trustability, commitment, and resilience. With this SA, we have shown that a range of cyberattacks can be mitigated using local trustable resources without stressing the bulk grid. The demonstrations have been carried out using a variety of platforms with high fidelity, hardware-in-the-loop, and utility-friendly validation software.

\acknow{This work was supported by the US Department of Energy under Award
DOE-OE0000920 and the MIT Energy Initiative. We gratefully acknowledge several useful discussions with Karan Kalsi at PNNL, and Rob Hovsapian at NREL. V.N. would like to acknowledge the support of a summer internship at NREL.}

\showacknow{} 


\bibliography{refs_manual}

\clearpage
\onecolumn
\appendix
\setcounter{figure}{0}
\setcounter{table}{0} 
\setcounter{subfigure}{0}
\setcounter{section}{0}
\setcounter{subsection}{0}
\setcounter{page}{1}
\makeatletter 
\renewcommand{\thefigure}{S\@arabic\c@figure}
\renewcommand{\thetable}{S\@arabic\c@table}

\renewcommand{\thesection}{\Roman{section}}    
\renewcommand{\thesubsection}{\Roman{section}.\Alph{subsection}}     

\makeatother
\nomenclature{ACOPF}{Alternating Current OPF}
\nomenclature{ADMS}{Advanced Distributed Management System}
\nomenclature{ADP}{Anomalous Data Point}
\nomenclature{AR}{Anomaly Ratio}
\nomenclature{ARIES}{Advanced Research on Integrated Energy Systems}
\nomenclature{CAR}{Cumulative Anomaly Ratio}
\nomenclature{CI}{Current Injection}
\nomenclature{CS}{Commitment Score}
\nomenclature[CVSS]{Common Vulnerability Scoring System}{CVSS}
\nomenclature{CMO}{Consumer Market Operator}
\nomenclature{DCVS}{Device and Communication Vulnerabilities present at the Secondary}
\nomenclature{DER}{Distributed Energy Resources}
\nomenclature{DERIM}{DER Integration Middleware}
\nomenclature{DG}{Distributed Generator}
\nomenclature{DOTS}{Distribution Operations Training Simulator}
\nomenclature{DoS}{Denial of Service}
\nomenclature{DPVs}{Distributed Photovoltaics}
\nomenclature{DRTS}{Digital Real-Time Simulation}
\nomenclature{DSO}{Distribution System Operator}
\nomenclature{DSR}{Distribution System Resiliency}
\nomenclature{EUREICA}{Efficient, Ultra-REsilient, IoT-Coordinated Assets}
\nomenclature{EV}{Electric Vehicle}
\nomenclature{FL}{Federated Learning}
\nomenclature{AHP}{Analytic Hierarchy Process}
\nomenclature{GE}{General Electric}
\nomenclature{HELICS}{Hierarchical Engine for Large-scale Infrastructure Co-simulation}
\nomenclature{HIL}{Hardware-in-the-Loop}
\nomenclature{HVAC}{Heating, Ventilation, and Air Conditioning}
\nomenclature{ICAs}{IoT-Coordinated Assets}
\nomenclature{ICS}{Industrial Control Systems}
\nomenclature{IEEE}{Institute of Electrical and Electronics Engineers}
\nomenclature{IoT}{Internet of Things}
\nomenclature{LEM}{Local Electricity Market}
\nomenclature{LMP}{Locational Marginal Price}
\nomenclature{MadIoT}{Manipulation of Demand via IoT}
\nomenclature{MCDM}{Multiple-Criteria Decision-Making}
\nomenclature{NREL}{National Renewable Energy Lab}
\nomenclature{NVD}{National Vulnerability Database}
\nomenclature{OPF}{Optimal Power Flow}
\nomenclature{OT}{Operational Technology}
\nomenclature{PAC}{Proximal Atomic Coordination}
\nomenclature{PCC}{Point of Common Coupling}
\nomenclature{PM}{Primary Market}
\nomenclature{PMA}{Primary Market Agent}
\nomenclature{PMO}{Primary Market Operator}
\nomenclature{PNNL}{Pacific Northwest National Lab}
\nomenclature{PNR}{Primary Node Resiliency}
\nomenclature{PRM}{Primary Resilience Manager}
\nomenclature{RS}{Resilience Score}
\nomenclature{RTDS}{Real-Time Digital Simulator}
\nomenclature{SA}{Situational Awareness}
\nomenclature{SM}{Secondary Market}
\nomenclature{SMA}{Secondary Market Agent}
\nomenclature{SMO}{Secondary Market Operator}
\nomenclature{SOCP}{Second-Order Conic Program}
\nomenclature{SRM}{Secondary Resilience Manager}
\nomenclature{STNR}{Secondary Transformer Node Resiliency}
\nomenclature{TS}{Trustability Score}
\nomenclature{WH}{Water Heater}

\printnomenclature

\section{Overall EUREICA framework\label{sec:eureica_framework_high_level}}

\begin{figure}[htb]
\centering
\includegraphics[width=\linewidth]{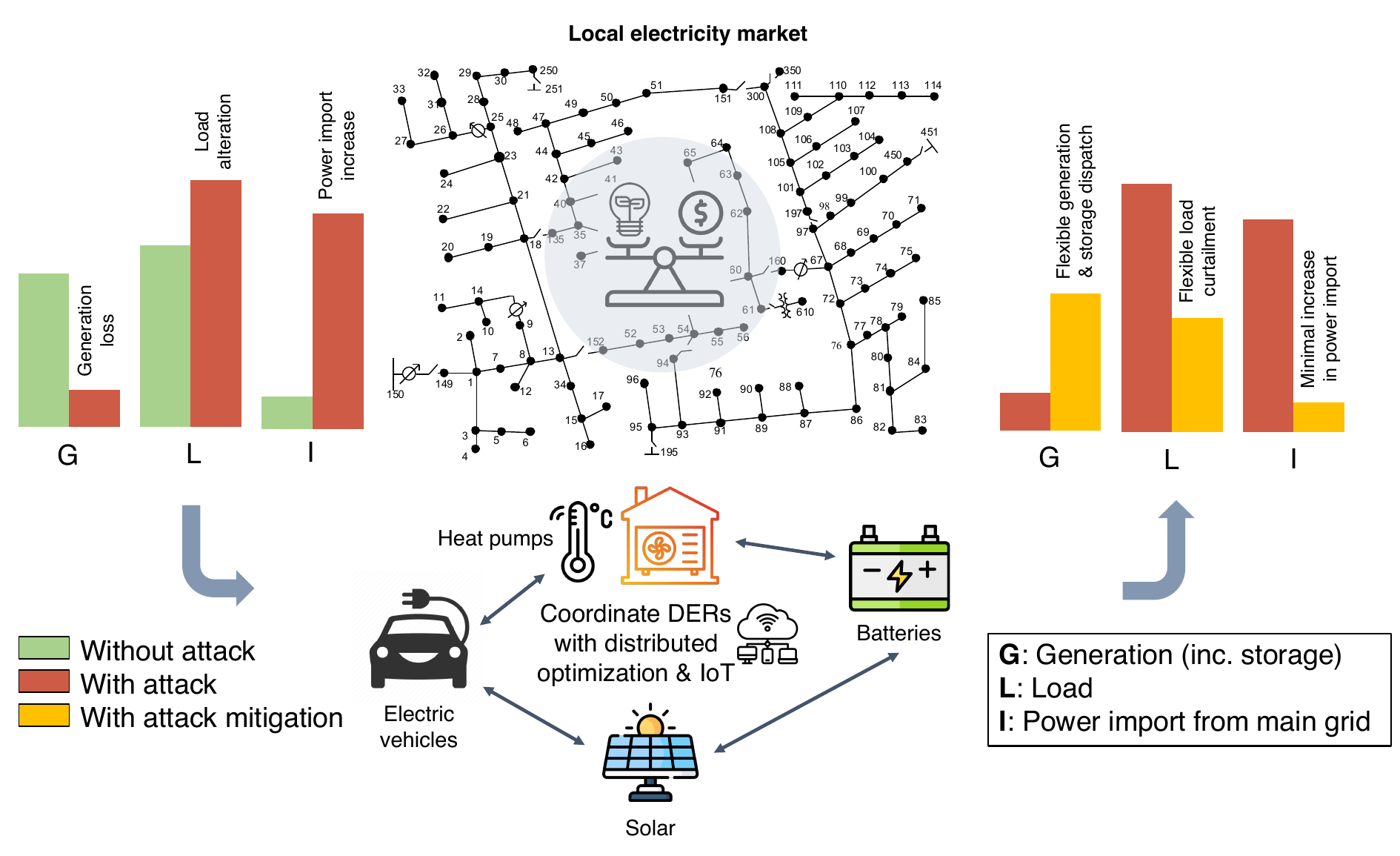}
\caption{Summary of EUREICA framework.}
\label{fig:eureica_framework}
\end{figure}

\cref{fig:eureica_framework} shows a simplified high-level overview of the EUREICA framework.

\section{IEEE 123-NODE DISTRIBUTION NETWORK\label{sec:ieee123}}

\begin{figure}[htb]
\centering
\includegraphics[width=\linewidth]{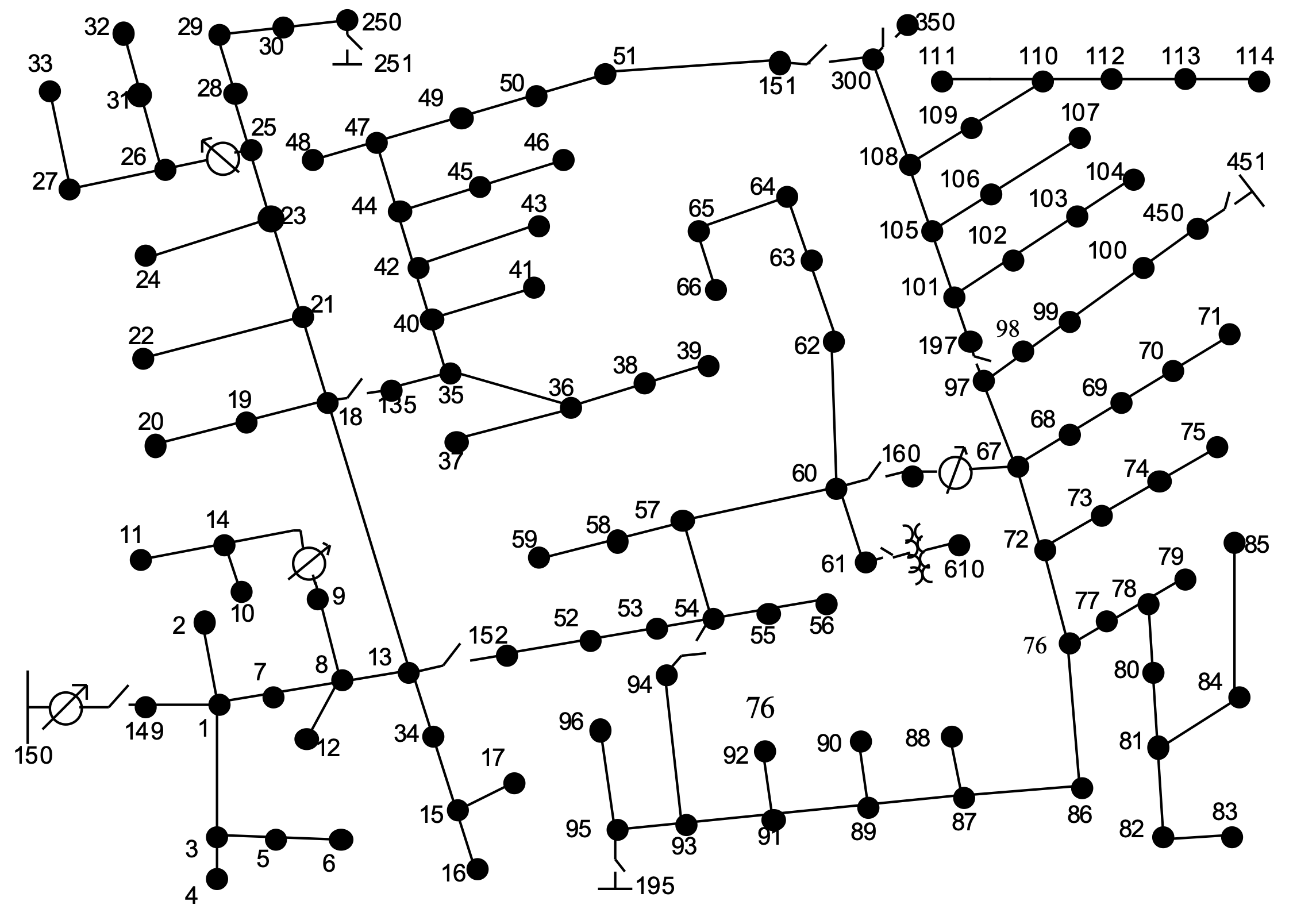}
\caption{IEEE 123-node test feeder network.}
\label{fig:network}
\end{figure}

\cref{fig:network} shows the IEEE 123-node test feeder \citeSI{kersting2001radial}, which is the distribution network model used to simulate and validate all our attack scenarios.

\subsection{MODIFIED IEEE 123-NODE MODEL WITH HIGH DER PENETRATION\label{sec:network_pnnl_high_der}}

\begin{figure}[htb]
\centering
\includegraphics[width=\linewidth]{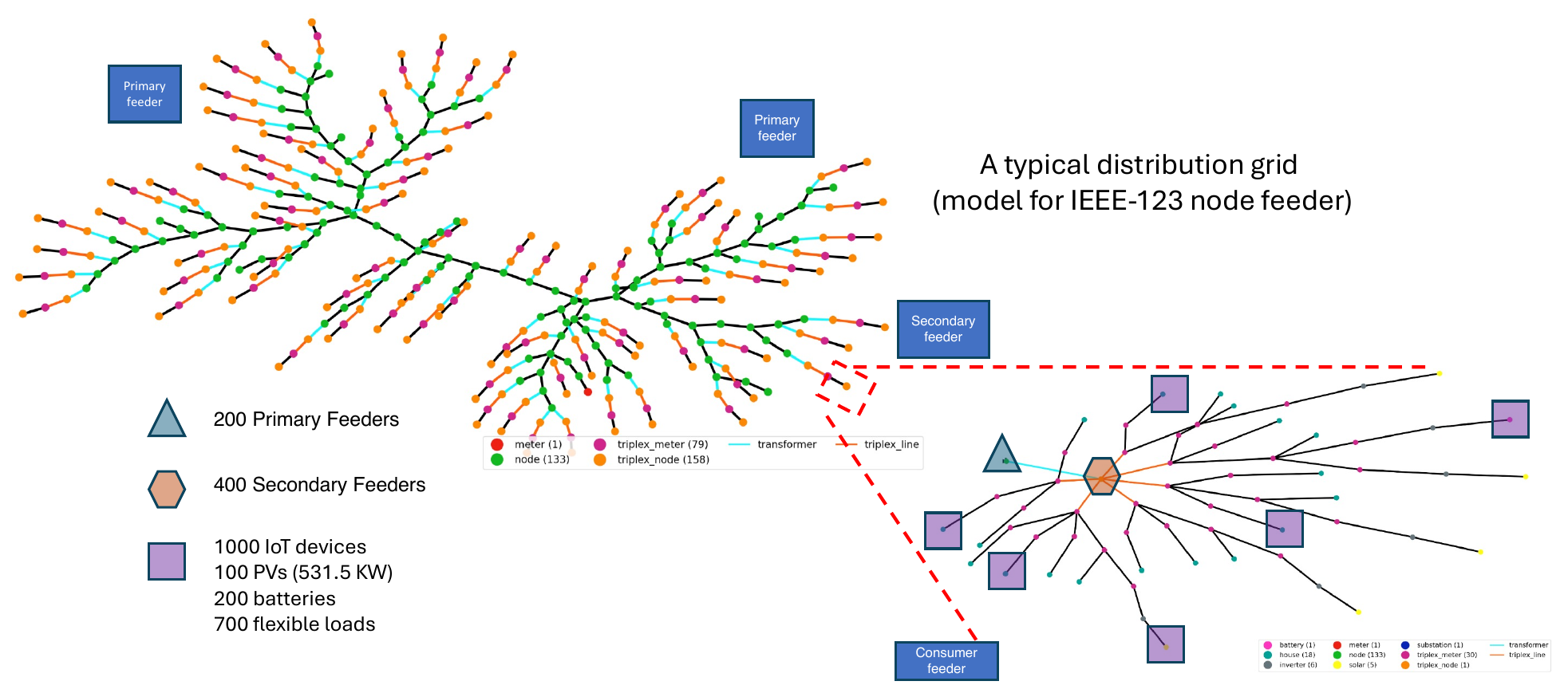}
\caption{Modified IEEE 123-node GridLAB-D with added DERs.}
\label{fig:network_pnnl_high_der}
\end{figure}

\section{EXAMPLE INSTANCE OF LEM}\label{sec:lem_example}

\begin{figure}[htb]
\centering
\includegraphics[width=\linewidth]{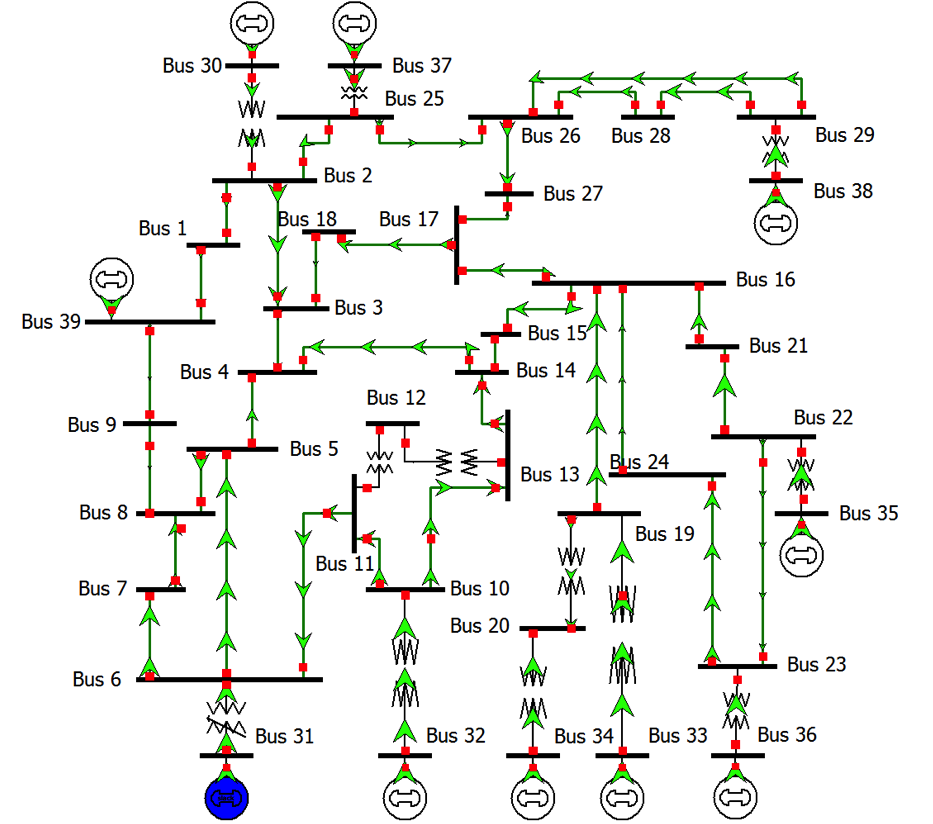}
\caption{IEEE 39-bus transmission system.}
\label{fig:ieee39}
\end{figure}

Here, we sketch out a possible (hypothetical) instantiation of our LEM for the city of Boston, MA, which is located in the New England (NE) region in the US. \cref{fig:ieee39} shows the IEEE 39-bus transmission system \citeSI{athay1979practical} which is a synthetic representation of the entire NE region, with a peak load of 6254 MW and a total of around 7 million homes. This corresponds to $\approx$ 162 MW and 180,000 homes per bus. Given that the IEEE 123-node distribution feeder has a peak load of roughly 3.6 MW, we can estimate that there will be 44 such primary feeders per transmission bus and 4100 homes per feeder. Thus, the city of Boston with a total of 300,000 homes \citeSI{bostonhouses}, will be served by 73 primary feeders across 2 transmission buses.

\begin{figure}[htb]
\centering
\includegraphics[width=\linewidth]{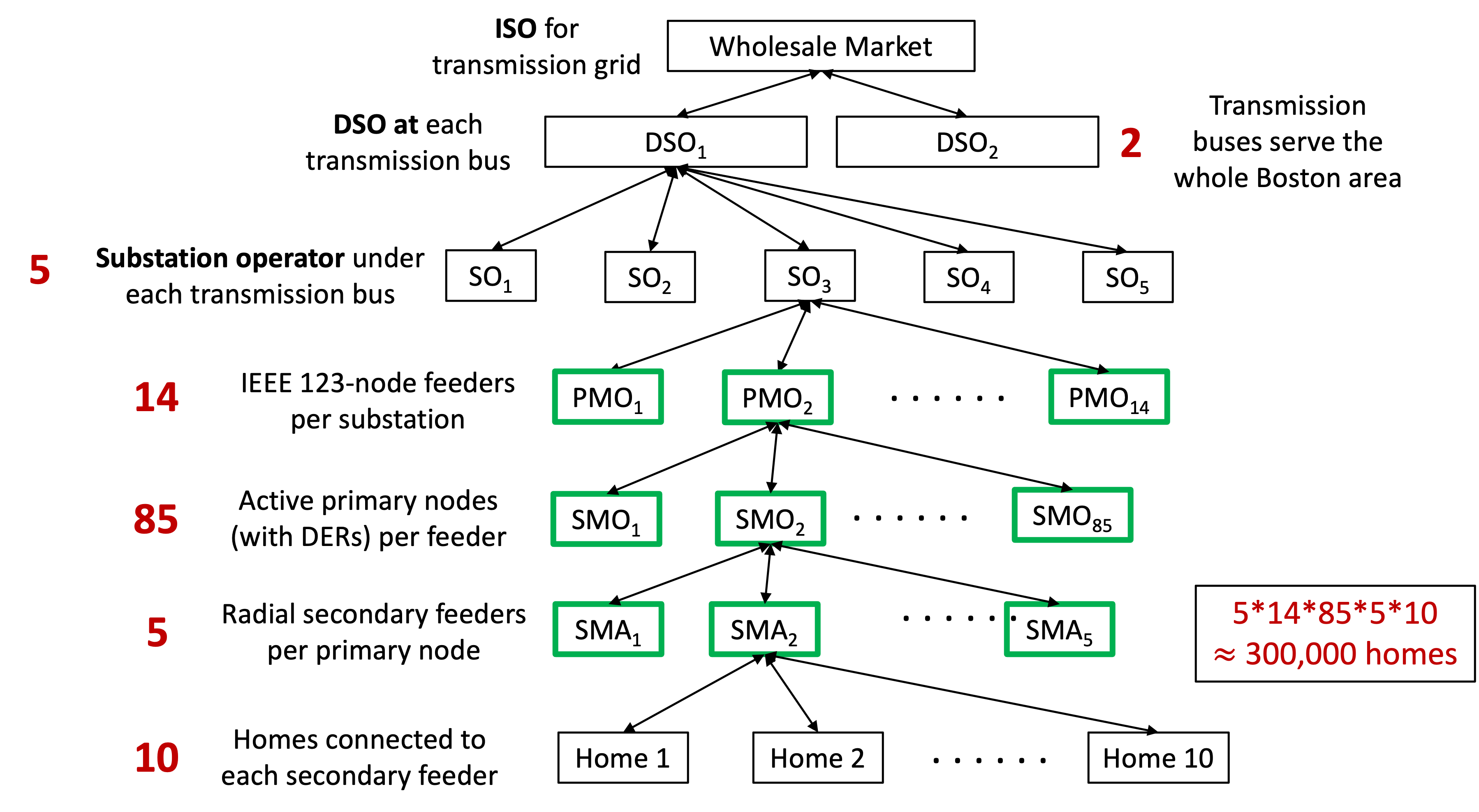}
\caption{Example of hypothetical LEM for the city of Boston, MA.}
\label{fig:lem_boston}
\end{figure}

\cref{fig:lem_boston} shows a breakdown of different entities to form a hierarchical LEM for Boston. Note that the main market operators and agents that are relevant for this work are marked in green.

\section{DETAILS OF THE LOCAL ELECTRICITY MARKET\label{sec:sm_details}}

\begin{figure}[htb]
\centering
\includegraphics[width=\columnwidth]{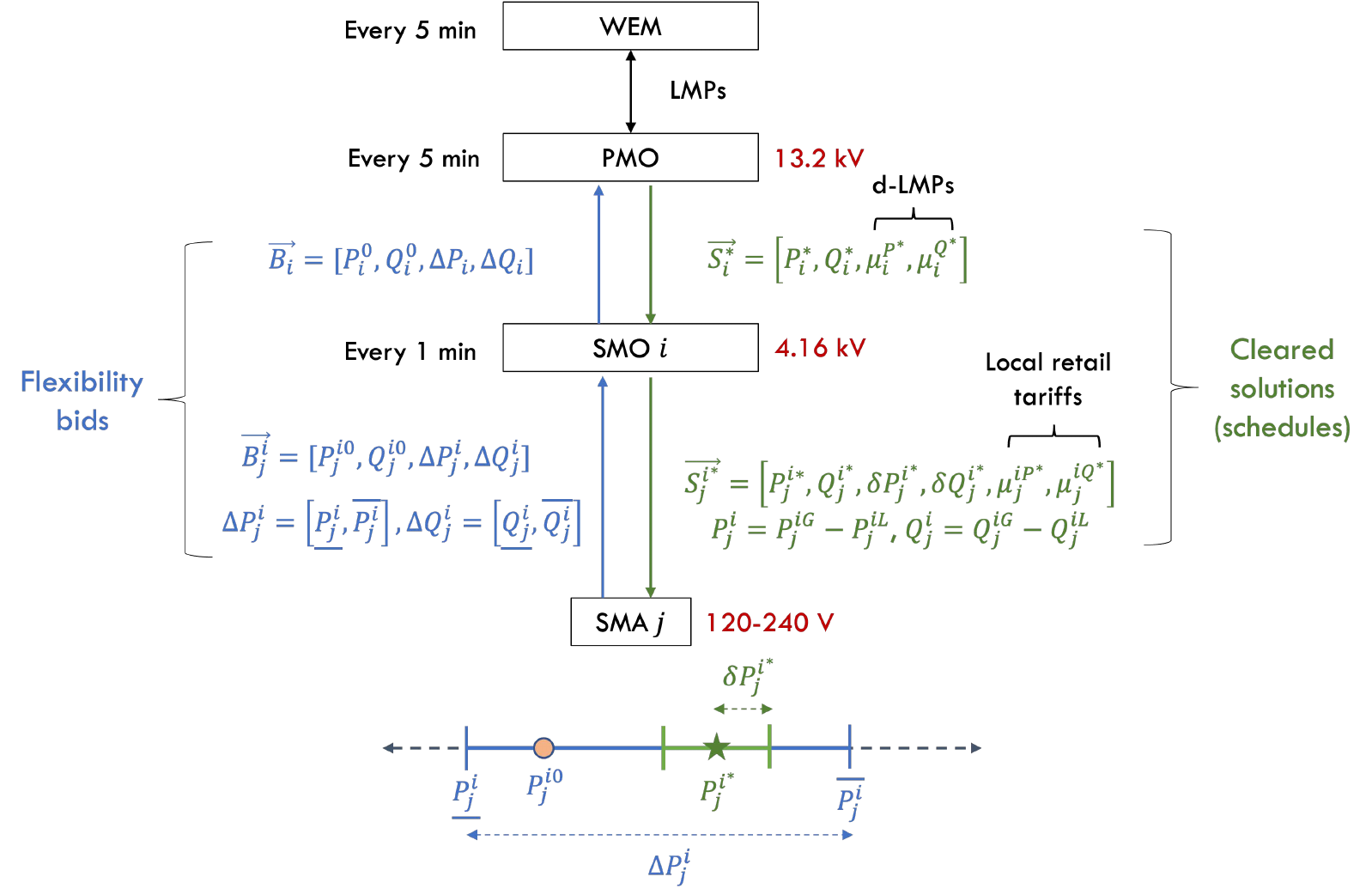}
\caption{Overall inputs and outputs in the LEM.}
\label{fig:SMbids}
\end{figure}

\cref{fig:SMbids} shows the inputs and outputs for different levels of the hierarchical LEM. For both the SM and the PM, the inputs consist of the baseline power injections and flexibility bids, while the outputs are the market schedules (setpoints for power injections) and their associated flexibility ranges, along with the corresponding electricity prices of tariffs.

\subsection{Objective functions for optimization in the Secondary Market\label{sec:sm_obj}}
The four objective functions considered in the SM clearing are defined as:
\begin{enumerate}
    \item[O1.] Maximization of aggregate resilience, $f_i^1$, given by the following, where $RS_j^i$ denotes the resilience score of SMA $j$ under SMO $i$
    $$f_i^1 = - \sum_{j=1}^n RS^i_j ((P_j^i - P_j^{i0})^2 +  (Q_j^i - Q_j^{i0})^2)$$
    \item[O2.] Minimization of net cost, $f_i^2$ to the SMO for running the SM 
    $$ f_i^2  = \sum_{j=1}^n \mu_j^{iP} P_j^i + \mu_j^{iQ} Q_j^i$$ 
    \item[O3.] Maximization of total flexibility, $f_i^3$ that the SMO can extract from all its SMAs 
    $$f_i^3 = - \sum_{j=1}^n (\delta P_j^i + \delta Q_j^i)$$ 
    \item[O4.] Minimization of disutility of the SMAs, $f_i^4$ arising from flexibility provision $$f_i^4 = \sum_{j=1}^n \beta_j^{iP}(P_j^i - P_j^{i0})^2 + \beta_j^{iQ}(Q_j^i - Q_j^{i0})^2.$$
\end{enumerate}

\subsection{Three-phase SM optimization problem\label{sec:sm_ci}}

\begin{subequations}
\label{eq:SMopt_ci}
\begin{align}
& \min \sum_{j \in \mathcal{N}_{J,i}} \{f_{j,1}^i,f_{j,2}^i,f_{j,3}^i,f_{j,4}^i\} \label{eq:cost} \\
& f_{1,j}^i \succ f_{2,j}^i \succ f_{3,j}^i \succ f_{4,j}^i \label{eq:ranking}, \; \; \Phi = \{a, b, c\} \\ 
& f_{j,1}^i = -C^i_j \left(\sum_{\phi \in \Phi}(P_j^{i,\phi} - P_j^{i0,\phi})^2 + (Q_j^{i,\phi} - Q_j^{i0,\phi})^2 \right)\nonumber \\
& f_{j,2}^i  = \sum_{\phi \in \Phi} \mu_j^{iP,\phi} P_j^{i,\phi} + \mu_j^{iQ,\phi} Q_j^{i,\phi} \nonumber \\
& f_{j,3}^i = -\sum_{\phi \in \Phi}(\delta P_j^{i,\phi} + \delta Q_j^{i,\phi}) \nonumber\\
& f_{j,4}^i = \beta_j^{iP}\sum_{\phi \in \Phi}\left(P_j^{i} - P_j^{i0}\right)^2 + \beta_j^{iQ}\sum_{\phi \in \Phi}\left(Q_j^{i} - Q_j^{i0}\right)^2 \nonumber \\
& \text{subject to:} \nonumber\\
& P_j^{i,\phi} - \delta P_j^{i,\phi} \geq \underline{P}_j^{i,\phi} \; Q_j^{i,\phi} - \delta Q_j^{i,\phi} \geq \underline{Q}_j^{i,\phi}  \label{eq:PQ_lowerlim} \\
& P_j^{i,\phi} + \delta P_j^{i,\phi} \leq \overline{P}_j^{i,\phi}, \; Q_j^{i,\phi} + \delta Q_j^{i,\phi} \leq \overline{Q}_j^{i,\phi}  \label{eq:PQ_upperlim} \\ 
& \delta P_j^{i,\phi}, \; \delta Q_j^{i,\phi} \geq 0, \; 0 \leq \mu_j^{iP} \leq \overline{\mu}^{iP}, 0 \leq \mu_j^{iP} \leq \overline{\mu}^{iQ}  \; \label{eq:flex_fairness} \\ 
& \sum_{t_s}^{t_s + \Delta t_p} \sum_{j \in \mathcal{N}_{J,i}} \sum_{\phi \in \Phi} \left(\mu_j^{iP,\phi}(t) P_j^{i,\phi}(t) + \mu_j^{iQ,\phi}(t) Q_j^{i,\phi}(t)\right) \Delta t_s  \nonumber \\
& \leq \sum_{\phi \in \Phi} \left(\mu_i^{P^*,\phi}(\hat{t}_p) P_i^{\phi^*}(\hat{t}_p) + \mu_i^{Q^*,\phi}(\hat{t}_p) Q_i^{\phi^*}(\hat{t}_p)\right) \Delta t_p \label{eq:budgetPQ} \\
& \sum_{j \in \mathcal{N}_{J,i}} P_j^{i,\phi}(t_s) = P_i^{\phi^*}(\hat{t}_p), \quad \sum_{j \in \mathcal{N}_{J,i}} Q_j^{i,\phi}(t_s) = Q_i^{\phi^*}(\hat{t}_p) \label{eq:PQbalance1}
\end{align}
\end{subequations}

\subsection{Bidding in the Primary Market\label{sec:pm_bidding}}
The bid submitted by each SMO $j \in \mathcal{N}$ into the PM $\mathcal{B}_i^{P}$ defined as:

\begin{align}
   \mathcal{B}_i^{S} =\{P_i^{0},Q_i^{0},\underline{P}_i,\underline{Q}_i,\overline{P}_i,\overline{Q}_i,\alpha_i^P, \alpha_i^Q, \beta_i^P,\beta_i^Q\}. \label{eq:pm_bid1}
\end{align}\begin{align}
    & P_i^0(t_p) = \sum_{j \in \mathcal{N}_i}  P_j^{i*}(t_p),\; Q_i^0(t_p) = \sum_{j \in \mathcal{N}_i}  Q_j^{i*}(t_p) \label{eq:aggregation}  \\
    & \underline{P}_i = \sum_{j\in \mathcal{N}_i} P^i_j - \delta P_j^{i*}, \overline{P}_i =\sum_{j\in \mathcal{N}_i} P_j^{i*} + \delta P_j^{i*} \nonumber \\
    & \underline{Q}_i = \sum_{j\in \mathcal{N}_i} Q_j^{i*} - \delta Q_j^{i*}, \overline{Q}_i = \sum_{j\in \mathcal{N}_i} Q_j^{i*} + \delta Q_j^{i*}\nonumber
\end{align}

In the above, (i) $P_i^0, Q_i^0$ denote the baseline active and reactive power injection bids of the SMOs, (ii) ($\underline{P}_i, \underline{Q}_i$) and ($\overline{P}_i, \overline{Q}_i$) denote the downward and upward flexibilities (around their nominal values) in active and reactive power, respectively, (iii) $\alpha_i^P, \alpha_i^Q$ are the local net generation costs, and (iv) $\beta_i^P, \beta_i^Q$ are the flexibility disutility parameters of SMO $i$ for P and Q, respectively. The SMO computes (iii) and (iv) as weighted averages of all their SMA retail tariffs, and SMA disutility parameters, respectively, as follows:
$$
\alpha_i^P = \frac{\sum_{j \in \mathcal{N}_i \mu_j^{iP*} |P_j^{i*}|}}{\sum_{j \in \mathcal{N}_i |P_j^{i*}|}}, \beta_i^P = \frac{\sum_{j \in \mathcal{N}_i \beta_j^{iP*} |P_j^{i*}|}}{\sum_{j \in \mathcal{N}_i |P_j^{i*}|}}
$$

\subsection{Objective functions for optimization in the Primary Market\label{sec:pm_obj}}
We define the following functions
\begin{align}
    f^P(\textbf{y}^P) &=\sum_{i \in \mathcal{N}} f^P_i(\textbf{y}^P_i) = \sum_{i \in \mathcal{N}} \Big[f_{i}^{\text{Load-Disutil}}(\textbf{y}^P_i) \nonumber \\
    & + f_{i}^{\text{Gen-Cost}}(\textbf{y}^P_i)\Big] + \xi\Big[\sum_{(ki) \in \mathcal{E}} f_{ki}^{\text{Loss}}(\textbf{y}^P_i)\Big] \label{eq:cost:1} \\
    f_{i}^{\text{Load-Disutil}}(\textbf{y}^P_i)&= \beta_{i}^{P}(P_{i}^{L}-P_{i}^{L0})^2+\beta_{i}^{Q}(Q_{i}^L-Q_{i}^{L0})^2 \label{eq:cost:2} \\
	f_{i}^{\text{Gen-Cost}}(\textbf{y}^P_i)&= \begin{cases}\alpha_{i}^{P} (P_{i}^{G})^2 +\alpha_{i}^{Q} (Q_{i}^{G})^2, & \\
	\lambda_{i}^{P} P_{i}^{G}+\lambda_{i}^{Q} Q_{i}^{G},\text{if } i & \text{ is PCC}
	\end{cases} \label{eq:cost:3}\\
	f_{ki}^{\text{Loss}}(\textbf{y}^P_i)&=R_{ki}|I_{ki}|^2 \label{eq:cost:4}
\end{align}
The objective function used in \cref{eq:cost:1} used is a weighted linear combination of (i) maximizing social welfare in \cref{eq:cost:2}, (ii) minimizing total generation costs in \cref{eq:cost:3} and (iii) minimizing electrical line losses in \cref{eq:cost:4}. The total cost includes paying the locational marginal price (LMP) $\lambda$ for importing power from the transmission grid at the point of common coupling (PCC), as well as the payments to local generator PMAs that provide net positive injections into the PM. We divide by suitable base values to convert all quantities to per unit (between 0 and 1 p.u.). Thus, it is reasonable to combine all the terms into a single objective function using a simple weighted sum. The hyperparameter $\xi$ controls the tradeoff between penalizing line losses versus optimizing for other objectives. The coefficients $\alpha_i, \beta_i$, are communicated by each PMA $i$ as part of their bids, while $\xi$ is a global hyperparameter common to all PMAs, and determined by the PMO. Here, $R$ denotes the network resistance matrix and $\mathcal{E}$ denotes the set of all edges in the network.

\subsection{Computation of commitment scores\label{sec:commit_app}}
We describe here the details of computing the commitment reliability score, mentioned in \cref{sec:commit}. From the SM clearing, the SMAs $j$ are directed by their SMO $i$ to keep their net injections within the intervals $[P_j^{i*} - \delta P^{i*}_j, P_j^{i*} + \delta P_j^{i*}]$. We first compute the deviations (if any) in their actual responses $\hat{P}^i_j$ from this range, where $\llbracket \cdot \rrbracket$ denotes the indicator function: 
\begin{align}
    e^{iP}_j(t_s) & = \llbracket\hat{P}^i_j > \overline{P}_j^{i*}\rrbracket (\hat{P}_j^i - \overline{P}_j^{i*}) + \llbracket\hat{P}^i_j < \underline{P}_j^{i*}\rrbracket(\underline{P}_j^{i*} - \hat{P}^i_j) \nonumber \\ & + \llbracket\underline{P}_j^{i*} \leq \hat{P}^i_j \leq \overline{P}_j^{i*}\rrbracket\max (\hat{P}^i_j - \overline{P}_j^{i*},\underline{P}_j^{i*} - \hat{P}^i_j) \label{eq:commit1}
\end{align}
We then obtain relative deviations by comparing these with the magnitudes of their corresponding baseline setpoints:
\begin{equation}
    \overline{e^{iP}_j} (t_s) = \frac{e^{iP}_j(t_s)}{|P_j^{i*}(t_s)|}, \quad \overline{e^{iQ}_j} (t_s) = \frac{e^{iQ}_j(t_s)}{|Q_j^{i*}(t_s)|} \label{eq:commit2}
\end{equation}
These are then normalized to unit vectors to compare the deviations among all SMAs overseen by the SMO. This allows the SMO to assess their relative performance across all its SMAs.
\begin{equation}
        \widetilde{\mathbf{e^{iP}}}(t_s) = \frac{\mathbf{\overline{e^{iP}}}(t_s)}{{\|\mathbf{\overline{e^{iP}}}}(t_s)\|}, \; \widetilde{\mathbf{e^{iQ}}}(t_s) = \frac{\mathbf{\overline{e^{iQ}}}(t_s)}{\|\mathbf{\overline{e^{iQ}}}(t_s)\|} \label{eq:commit3}
\end{equation}
The scores are then updated, with the score being increased when the SMAs follow their contracts and decreased otherwise:
\begin{align}
C^i_j(t_s) & = \begin{cases} 1 &\mbox{if } t_s = 0 \\
C^i_j(t_s - 1) - \frac{\widetilde{e^{iP}_j} (t_s) + \widetilde{e^{iQ}_j} (t_s)}{2} & \mbox{if } t_s > 0 \end{cases} \label{eq:commit4}
\end{align}
Finally, we perform min-max normalization across all the SMAs' scores to ensure that $0 \leq C_j \leq 1 \; \forall \; \text{SMAs } j$.
$$
    \overline{C}^i_j = \frac{C_j^i - \max_j C^i_j}{\max_j C^i_j - \min_j C^i_j}
$$

\section{TRUSTABILITY SCORES AND RESILIENCY METRICS\label{sec:ts_resmetric}}

\subsection{Computation of IoT trustability scores\label{sec:iot-ts}}
The IoT trustability score (TS) is computed utilizing the federated self-learning concept \citeSI{Sarker2023ResiliencyIoTs}. Anomalies in IoT data are the key factor in the formulation of the IoT TS. Another contributing factor is the IoT device's market commitment history. More details of the features are shown in \cref{tab:feat}.

\begin{table}[htb]
\centering
\caption{Features considered for each type of data.}
\label{tab:feat}
\begin{tabular}{|c|c|}
\hline
\textbf{Data Source} & \textbf{Features} \\ \hline
\begin{tabular}[c]{@{}c@{}}IoTs network\\  packet\end{tabular}  & \begin{tabular}[c]{@{}c@{}}Source/Destination IP,  Source/Destination port, \\ Packet length, Protocols, Intra-packet arrival time\end{tabular} \\ \hline
HVAC & \begin{tabular}[c]{@{}c@{}}Timestamp, Load, Indoor temperature, \\ outdoor temperature, Temperature setpoint,\\ Indoor area, Building thermal insulation\end{tabular} \\ \hline
PV & \begin{tabular}[c]{@{}c@{}}Timestamp, Power generation, \\ Rating, Solar irradiance\end{tabular} \\ \hline
Battery & \begin{tabular}[c]{@{}c@{}}Timestamp, Charging/Discharging rate,\\ SoC, KW capacity\end{tabular} \\ \hline
EV & \begin{tabular}[c]{@{}c@{}}Timestamp, Charging rate, \\ SoC\end{tabular} \\ \hline
\end{tabular}
\end{table}

To detect anomalies, we learn the IoT data's expected behavior and prediction for the short time steps. Predicted data is compared with measured data for anomaly detection. For prediction, we are using an autoencoder neural network for federated unsupervised learning. There will be one autoencoder model for each IoT device to train on its physical data and one more autoencoder model to train only on IoT network packet data. For each type of data, there is a tolerance value $T_{err}$ for the relative error (RE). If any data point ($DP$) crosses $T_{err}$, then that is flagged as an anomalous data point ($ADP$). So, for any reporting time period $\Delta t$, non-anomaly ratio ($NAR$) is calculated using,
 \begin{equation}
     NAR=1- \frac{\text{Total ADP number over }{\Delta t}^{}}{\text{Total DP number over }{\Delta t}}
 \end{equation}
Next, the cumulative non-anomaly ratio ($CNAR$) is computed, where $T$ is the fixed total time period and is always divisible by $\Delta t$.
\begin{equation}\label{cnar}
    CNAR_t=\sum_{j=1}^{\frac{T}{\Delta t}}\frac{T}{j\Delta t} NAR_{t-j\Delta t}
\end{equation}
IoT Trustability Score ($TS$) for time $t$ and building/house $i$ is calculated:
\begin{equation}\label{its}
    TS_{t,i}=w_t\times NAR_t+ w_{t-}\times\frac{CNAR_t}{CNAR_{max}}
\end{equation}
where
\begin{equation} \label{wtits}
    (I) \; w_t \geq w_{t-} \quad \quad
    (II) \; w_t + w_{t-}=1
\end{equation}
Here, $CNAR_{max}$ is calculated using \eqref{cnar} with the maximum $NAR$ being $NAR=1$ for the whole time period $T$. Finally, to get the overall $TS_t$ of any observation node with IoTs at time $t$, we average the $TS_{t,i}$ of all the clients $i$ of that observation node to calculate $TS_t$:
\begin{equation}
    TS= \frac{\sum_{i=1}^{M} TS_{t,i}}{M}
\end{equation}
where $M$ is the total number of clients or buildings/houses at that observation node.

\subsection{Secondary Transformer and Primary Node Resiliency Metric (STNR and PNR)\label{sec:pnr}}

Secondary transformer node resiliency (STNR) is computed using multiple resiliency factors and TS.
\begin{equation}
STNR_j = \prod_{i=1}^{n_c} F_i^{W_i}
\end{equation}
where $n_c$ is the total number of factors for the category of the secondary level node, $F_i$ is the value for each factor, and $W_i$ is the normalized weight for each factor. These factors $f$ influencing resiliency are determined and assigned weights to aggregate into the PNR score. Factors that can be determined directly from the secondary level configuration are described in \cref{fig:res_score}. All the device and communication vulnerabilities present at the secondary (DCVS) level of a primary node are identified using the national vulnerability database (NVD) \citeSI{booth2013national}. Then DCVS factor is calculated as,
\begin{equation}
    DCVS=\frac{1}{\sum_{i=1}^{N_s} CVSS_i}
\end{equation}
where $N_s$ is the number of total vulnerabilities present at the secondary level. Here, the common vulnerability scoring system (CVSS) is one of several methods to measure the impact of vulnerabilities in devices known as Common Vulnerabilities and Exposures (CVE). It is an open set of standards used to assess the vulnerability of software and assign severity along a scale of 0-10. The National Institute of Standards and Technology (NIST) analyzes all identified vulnerabilities and enlists these in the NVD. In case of the absence of any vulnerability, DCVS will be equal to $1$. 

Weight assignment and aggregation are managed by fuzzy multiple-criteria decision-making (MCDM), specifically the fuzzy analytic hierarchy process (Fuzzy AHP). A weighted average of the STNR results in the primary node resiliency (PNR):
\begin{equation}\label{eq:pnr}
PNR_k =  \frac{\sum_{j=1}^n (STNR_j \times W_j )}{\sum_{j=1}^n W_j }
\end{equation}
where $W_i$ is the weighted coefficient for the $i^{th}$ secondary feeder node.

\subsection{Distribution System Resiliency (DSR)}
Let $F=(f_{ij})\in R_+^{m \times n}$ be the factors value matrix, where $f_{ij}$ is value of factor $i$ of primary node $j$. The higher the value of $f_{ij}$, the more the node will contribute to the resiliency metric in regard to that factor. Now, following the data envelopment analysis (DEA) method, each node $p$ can choose a set of weights $w^p=(w^p_1,...w^p_m)$, where, $ \sum_{i=1}^{m}w^p_i=1$. Now the relative contribution (RC) of the node $p$ to the total contribution of all the nodes towards DSR, as measured by node $p$’s weight selection can be evaluated as,
\begin{equation}\label{eq:ratio}
    RC^p=\frac{\sum_{i=1}^{m}w^p_i f_{ip}}{\sum_{i=1}^{m}w^p_i \sum_{j=1}^{n}\left (f_{ij}\right)}
\end{equation}
Now, each node wants to maximize this ratio in \cref{eq:ratio} to have the best set of weights so that they can contribute to the maximum possible value in DSR. Using the weight vector for each node, a combination of multiplicative and additive methods are used to get the DSR.
\begin{equation} \label{eq:dsr}
    DSR= \sum_{j=1}^{n} \left (\prod_{i=1}^{m} \left (f_{ij}  \right )^{w^j_i} \right )
\end{equation}

\begin{figure}[htb]
\centering
\includegraphics[width=\columnwidth]{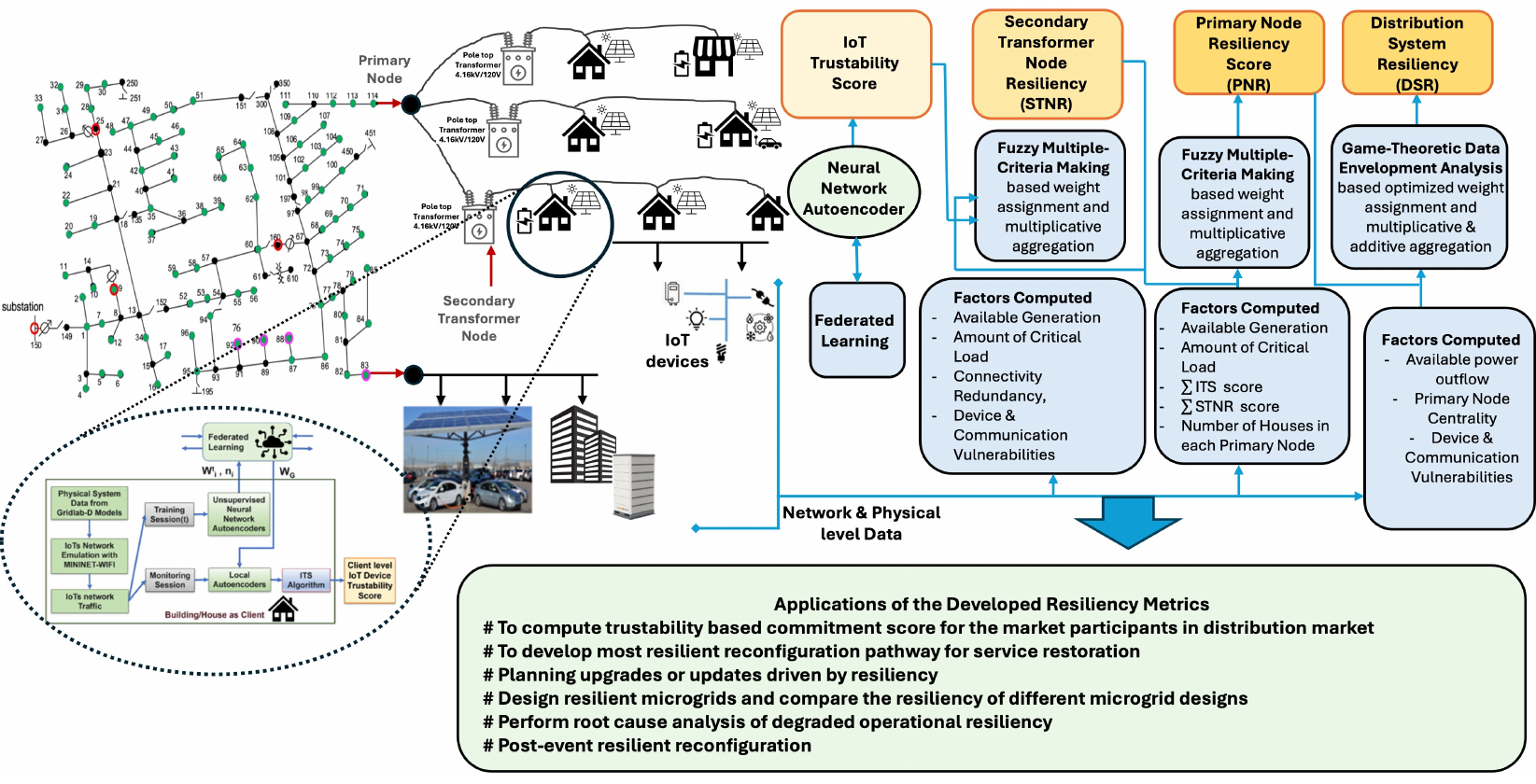}
\caption{Overview of the developed resilience score for the distribution system with IoTs}
\label{fig:res_score}
\end{figure}

Details related to the computation of DSR are shown in \cref{fig:res_score}. 

\section{POWER SYSTEM MODELS}
\subsection{Branch flow model\label{sec:bf}}
The branch flow OPF problem is formally stated as follows, where $R$ and $X$ denote the network resistance and reactance matrices respectively, $v$ and $I$ denote the nodal voltage magnitudes and branch currents respectively, and $\mathcal{E}$ denotes the set of all edges in the network. The primal decision variables here for each PMA $i$ are $\textbf{y}^{P,BF}_i = [P_i^G, Q_i^G, P_i^L, Q_i^L, v_i, I_{ki}] \; \forall i \in \mathbf{N}, \; (ik) \in \mathcal{E}$, where $\mathcal{E}$ is the set of all network edges or branches.
\begin{align}
    & \underset{\textbf{y}^P}{\min} \;\; f^{S-W}(\textbf{y}^P) \label{eq:pmo} \\
    & \text{subject to:} \nonumber\\
	& v_{i} - v_{k} = \left(R_{ki}^{2} + X_{ki}^{2}\right) |I_{ki}|^2 - 2 \left(R_{ki} P_{ki} + X_{ki} Q_{ki}\right)\nonumber\\
	& P_{i}^{G} - P_{i}^{L} = -P_{ki} + R_{ki} |I_{ki}|^2 + \sum_{k:\left(i,k\right) \in \mathcal{E}} P_{ik}\nonumber\\
	& Q_{i}^{G} - Q_{i}^{L} = -Q_{ki} + X_{ki} |I_{ki}|^2 + \sum_{k:\left(ik\right) \in \mathcal{E}} Q_{ik}\nonumber\\ 
  	& P_{ki}^{2} + Q_{ki}^{2} \leq \overline{S}_{ki}^{2}, \; P_{ki}^{2} + Q_{ki}^{2} \leq v_{i} |I_{ki}|^2, \;\underline{v}_i \leq v_i \leq \overline{v}_i  \nonumber\\
	& \underline{P}_i^{G} \leq P_i^{G} \leq \overline{P}_i^{G}, \; \underline{P}_i^{L} \leq P_i^{L} \leq \overline{P}_i^{L} \nonumber\\
	& \underline{Q}_i^{G} \leq Q_i^{G} \leq \overline{Q}_i^{G}, \; \underline{Q}_i^{L} \leq Q_i^{L} \leq \overline{Q}_i^{L} \nonumber\\
\end{align} 

\subsection{Current injection model\label{sec:ci}}
The decision vector for the CI model is given by $\textbf{y}^{P,CI}_i =[P_i^{\phi},Q_i^{\phi},V_{i}^{\phi,R},V_{i}^{\phi,I},I_{i}^{\phi,R},I_{i}^{\phi,I},I_{ik}^{\phi,R},I_{ik}^{\phi,I}]$, where $\phi \in \{a,b,c\}$ are the phases and $\mathcal{E}$ is the set of all network edges or branches. The primal decision variables for each SMO $i$ obtained by solving the optimization problem consist of (i) active ($P_i^{\phi*}$) and reactive ($Q_i^{\phi*}$) power setpoints (ii) real and imaginary components of nodal voltages ($V_{i}^{\phi,R*}, V_{i}^{\phi,I*}$) and current injections ($I_{i}^{\phi,R*}, I_{i}^{\phi,I*}$). Note that these are solved for each non-zero phase $\phi \in \mathcal{P} = \{a,b,c\}$. The CI-OPF problem formulation is given by:
\begin{subequations}
\label{eq:ciopf}
\begin{align}
&\min _x f^{obj}(x)\\
&I^R=\operatorname{Re}(\mathrm{Y} V), \; I^I=\operatorname{Im}(\mathrm{YV}) \label{eq:ohmslaw} \\
&P_i^\phi=V_i^{\phi, R} I_i^{\phi, R}+V_i^{\phi, I} I_i^{\phi, I} \quad \forall i \in \mathcal{N}, \phi \in \mathcal{P} \label{eq:bilinearP} \\
&Q_i^\phi=-V_i^{\phi, R} I_i^{\phi, I}+V_i^{\phi, I} I_i^{\phi, R} \quad \forall i \in \mathcal{N}, \phi \in \mathcal{P} \label{eq:bilinearQ}\\
& (I_{ik}^{\phi,R})^2+(I_{ik}^{\phi,I})^2\le \overline{I_{ik}^{\phi}}^2 \quad \forall i\in \mathcal{N}, \phi\in \mathcal{P}, (ik) \in \mathcal{E} \label{eq:Ibranchlims} \\
& \underline{{V_{i}^\phi}}^2\le({V_{i}^{\phi,R}})^2 + ({V_{i}^{\phi,I}})^2 \le \overline{V_{i}^{\phi}}^2 \quad \forall i\in \mathcal{N}, \phi\in \mathcal{P} \label{eq:Vmaglims} \\
& \underline{P_i^\phi} \leq P_i^\phi \leq \overline{P_i^\phi}, \;  \underline{Q_i^\phi} \leq Q_i^\phi \leq \overline{Q_i^\phi}
\end{align}
\end{subequations}
where $Y$ is the 3-phase bus admittance matrix for the network, and $V$ and $I$ are matrices of nodal voltages and currents respectively. Problem \cref{eq:ciopf} is nonconvex due to bilinear constraints \cref{eq:bilinearP,eq:bilinearQ}, and the ring constraint \cref{eq:Vmaglims} on voltage magnitudes. We obtain a convex relaxation by using McCormick envelopes (MCE), which represent the convex hull of a bilinear product $w=xy$ by using upper and lower limits on $x, \; y$. Thus, we replace the bilinear equality with a series of linear inequalities, denoted as $\text{MCE}(w) = \{w=xy: x\in [ \underline{x}, \overline{x}],  y\in [ \underline{y}, \overline{y}]\}$:
\begin{equation} \label{eq:mce}
  MCE(w,\underline{x},\overline{x},\underline{y},\overline{y}) = 
    \begin{cases}
      w\ge \underline{x}y+x\underline{y}-\underline{x}\underline{y} \\
      w\ge \overline{x}y+x\overline{y}-\overline{x}\overline{y} \\
      w\le \underline{x}y+x\overline{y}-\overline{x}\underline{y} \\
      w\le \overline{x}y+x\underline{y}-\underline{x}\overline{y}
    \end{cases}       
\end{equation}
We introduce auxiliary variables for each of the four bilinear terms $\{a_i^{\phi},b_i^{\phi},c_i^{\phi},d_i^{\phi}\} = \{V_i^{\phi, R} I_i^{\phi, R},V_i^{\phi, I} I_i^{\phi, I},V_i^{\phi, R} I_i^{\phi, I},V_i^{\phi, I} I_i^{\phi, R}\}$ allowing us to convert constraints \cref{eq:bilinearP,eq:bilinearQ} to linear constraints with MCE constraints on each of the auxiliary variables. We also need additional constraints on the nodal current injections and nodal voltages in order to define the MCE constraints. These voltage and current bounds can be determined by applying a suitable preprocessing method using the nodal $P$ and $Q$ limits from the SMO bids \citeSI{Ferro2020ANetworks}. The resulting bounds will also implicitly satisfy constraints \cref{eq:Ibranchlims} and \cref{eq:Vmaglims}. Thus, we can replace constraints \cref{eq:bilinearP,eq:bilinearQ,eq:Ibranchlims,eq:Vmaglims} with the following set of constraints in order to obtain the relaxed CI-OPF problem, which reduces to a linear program that can be solved easily. However, we do incur the overhead of computing the tightest possible $V$ and $I$ bounds to obtain a good convex relaxation, which in turn ensures that the relaxed solutions are feasible for the original problem.
\begin{subequations}
\label{eq:linear_mce}
\begin{align}
& P_i^\phi=a_i^{\phi} + b_i^{\phi}, \; \; Q_i^\phi=-c_i^{\phi} + d_i^{\phi} \quad \forall i \in \mathcal{N}, \phi \in \mathcal{P} \label{eq:linearPQ}\\
& \underline{I_{i}^{\phi,R}} \le I_{i}^{\phi,R} \le \overline{I_{i}^{\phi,R}}, \; \underline{I_{i}^{\phi,I}} \le I_{i}^{\phi,I} \le \overline{_{i}^{\phi,I}}  \label{eq:Ilims} \\
& \underline{V_{i}^{\phi,R}} \le V_{i}^{\phi,R}\le \overline{V_{i}^{\phi,R}}, \; \underline{V_{i}^{\phi,I}} \le V_{i}^{\phi,I}\le \overline{V_{i}^{\phi,I}} \label{eq:Vlims} \\
& a_i^{\phi} \in MCE(V_{i}^{\phi,R}I_{i}^{\phi,R},\underline{V_{i}^{\phi,R}},\overline{V_{i}^{\phi,R}},\underline{I_{i}^{\phi,R}},\overline{I_{i}^{\phi,R}}) \\
& b_{i}^{\phi} \in MCE(V_{i}^{\phi,I}I_{i}^{\phi,I},\underline{V_{i}^{\phi,I}},\overline{V_{i}^{\phi,I}},\underline{I_{i}^{\phi,I}},\overline{I_{i}^{\phi,I}}) \\
& c_{i}^{\phi} \in MCE(V_{i}^{\phi,R}I_{i}^{\phi,I},\underline{V_{i}^{\phi,R}},\overline{V_{i}^{\phi,R}},\underline{I_{i}^{\phi,I}},\overline{I_{i}^{\phi,I}}) \\
& d_{i}^{\phi} \in MCE(V_{i}^{\phi,I}I_{i}^{\phi,R},\underline{V_{i}^{\phi,I}},\overline{V_{i}^{\phi,I}},\underline{I_{i}^{\phi,R}},\overline{I_{i}^{\phi,R}}) \label{eq:mce_ineqs}
\end{align}
\end{subequations}

\section{DISTRIBUTED OPTIMIZATION\label{sec:dist_opt}}

For a given global optimization (primal) problem with equality and inequality constraints for $K$ number of nodes (or agents):
\begin{equation}
    \min _x \sum_{i=1}^K f_i(x) \text { s.t. } G x=b, \quad H x \leq d 
\end{equation} 
We can decompose this into $\mathcal{S} = \{S_1, S_2,\dots S_K\}$ coupled optimization problems, known as atoms (representing each SMO $i$). We separate the vector of all decision variables $x$ into two sets: $\mathcal{L} = \{L_i, \forall i \in [K]\}$ and $\mathcal{O} = \{O_i, \forall i \in [K]\}$ which is a partition of decision variables into those that are owned and copied by atom $i$, respectively. We can similarly also decompose the constraints into sets owned by each atom $\mathcal{C} = \{C_i, \forall i \in [K]\}$. These variable copies across multiple atoms can then be used to satisfy coupled constraints and global objectives. Note that for a number $K$, $[K] =\{1, 2, \dots K\}.$

The decomposed (or atomized) optimization problem is shown in \cref{eq:opt_atom}, where $a_j$ and $f_j(a_j)$ are the primal decision variables (both owned and copies) and individual objective functions corresponding to each SMO atom, respectively. $G_j$ and $H_j$ are the atomic constraint submatrices of $G$ and $H$, while $b_j$ and $d_j$ are subvectors of $b$ and $d$ of the right hand side constraint vectors $b$ and $d$, respectively. $B$ is the directed graph incidence matrix defining the owned and copied atomic variables. This incidence matrix allows us to fully parallelize the distributed optimization by defining coordination or consensus constraints, which enforce that all the copied variables for each atom $j$ must equal the values of their corresponding owned values in every other atom $i \neq j$. $B_j$ and $B^j$ denote the incoming and outgoing edges for atom $j$ respectively. Here 
\begin{align}
& \min _{a_j} \sum_{j \in K} f_j\left(a_j\right) \label{eq:opt_atom} \\
& \text { s.t. } G_j a_j=b_j, \; H_j a_j \leq d_j, \; B_j a=0 \; \forall j \in [K]  \nonumber\\
& B_{im} \triangleq \begin{cases}
-1, & \text { if } i \text { is 'owned" and } m \text { a related "copy" } \\
1, & \text { if } m \text { is "owned" and } i \text { a related "copy" } \\
0, & \text { otherwise } \end{cases} \nonumber
\end{align}

The augmented Lagrangian is first atomized or decomposed for each node or SMO, introducing dual variables $\mu$ and $\nu$ corresponding to primal equality and coordination constraints respectively. Note that the inequality constraints are handled directly during the primal minimization step by appropriately defining the feasible set.
\begin{align}
\mathcal{L}(a, \mu, \nu) & =\sum_{j \in K}\left[f_j\left(a_j\right)+\mu_j^T\left(G_j a_j-b_j\right)+\nu_j^T B_j a\right] \nonumber\\
& =\sum_{j \in K}\left[f_j\left(a_j\right)+\mu_j^T\left(G_j a_j-b_j\right)+\nu^T B^j a_j\right] \nonumber\\
& \triangleq \sum_{j \in K} \mathcal{L}_j\left(a_j, \mu_j, \nu\right) \label{eq:aug_lang}
\end{align}

\subsection{PAC algorithm\label{sec:pac}}

We can then apply the prox-linear approach of \citeSI{aljanaby2005survey} to \cref{eq:aug_lang} and obtain the proximal atomic coordination (PAC) algorithm \citeSI{Romvary2022AOptimization,Haider2020TowardGrids}:

$$
\begin{aligned}
& a_j[\tau+1]=\underset{a_j \in \mathbb{R}^{\left|T_j\right|}}{\operatorname{argmin}}\left\{\begin{array}{c}
\mathcal{L}_j\left(a_j, \bar{\mu}_j[\tau], \bar{\nu}[\tau]\right) \\
+\frac{1}{2 \rho}\left\|a_j-a_j[\tau]\right\|_2^2
\end{array}\right\} \\
& \mu_j[\tau+1]=\mu_j[\tau]+\rho \gamma_j \tilde{G}_j a_j[\tau+1] \\
& \bar{\mu}_j[\tau+1]=\mu_j[\tau+1]+\rho \hat{\gamma}_j[\tau+1] \tilde{G}_j a_j[\tau+1]
\end{aligned}
$$

Communicate $a_j$ for all $j \in[K]$ with neighbors
$$
\begin{aligned}
& \nu_j[\tau+1]=\nu_j[\tau]+\rho \gamma_j[B]^{O_j} a[\tau+1] \\
& \bar{\nu}_j[\tau+1]=\nu_j[\tau+1]+\rho \hat{\gamma}_j[\tau+1][B]^{O_j} a[\tau+1]
\end{aligned}
$$
Communicate $\bar{\nu}_j$ for all $j \in[K]$ with neighbors.

The primal and dual variables are initialized as follows, $\forall j \in [K]$:
$$
\begin{aligned}
& a_j[0] \in \mathbb{R}^{|| T_j \mid} \\
& \mu_j[0]=\rho \gamma_j \tilde{G}_j a_j[0] \\
& \bar{\mu}_j[0]=\mu_j[0]+\rho \hat{\gamma}_j[0] \tilde{G}_j a_j[0] \\
& \nu_j[0]=\rho \gamma_j[B]^{O_j} a[0] \\
& \bar{\nu}_j[0]=\nu_j[0]+\rho \hat{\gamma}_j[0][B]^{O_j} a[0]
\end{aligned}
$$

\subsection{NST-PAC algorithm\label{sec:nst-pac}}

This work employs an enhanced, accelerated version called NST-PAC developed in \citeSI{Ferro2022AHubs}. It is a primal-dual method incorporating both $L2$ and proximal regularization terms. The convergence speed is increased by using time-varying gains and Nesterov-accelerated gradient updates for both the primal and dual variables. The iterative NST-PAC algorithm consists of the following steps at each iteration $\tau$:
\begin{align}
a_j[\tau+1] & = \underset{a_j}{\operatorname{argmin}}\big\{\mathcal{L}_j\left(a_j, \hat{\mu}_j[\tau], \hat{\nu}[\tau]\right)\\
& +\frac{\rho_j \gamma_j}{2}\left\|G_j a_j-b_j\right\|_2^2 + \frac{\rho_j \gamma_j}{2}\left\|B_j a_j\right\|_2^2 \nonumber\\
& +\frac{1}{2 \rho_j}\left\|a_j-a_j[\tau]\right\|_2^2\big\} \nonumber\\
 \hat{a}_j[\tau+1] &= a_j[\tau+1]+\alpha_j[\tau+1]\left(a_j[\tau+1]-a_j[\tau]\right) \nonumber\\
 \mu_j[\tau+1]& = \hat{\mu}_j[\tau]+\rho_j \gamma_j\left(G_j \hat{a}_j[\tau+1]-b_j\right) \nonumber\\
 \hat{\mu}_j[\tau+1]&= \mu_j[\tau+1]+\phi_j[\tau+1]\left(\mu_j[\tau+1]-\mu_j[\tau]\right) \nonumber\\
 \text {Communicate } & \hat{a}_j \text { for all } j \in[K] \text { with neighbors } \nonumber\\
 \nu_j[\tau+1]&= \hat{\nu}_j[\tau]+\rho_j \gamma_j B_j \hat{a}_j[\tau+1] \nonumber\\
 \hat{\nu}_j[\tau+1]&= \nu_j[\tau+1]+\theta_j[\tau+1]\left(\nu_j[\tau+1]-\nu_j[\tau]\right)\nonumber \\
 \text {Communicate } & \hat{\nu}_j \text { for all } j \in[K] \text { with neighbors } \nonumber
\end{align}
The algorithm further protects privacy by masking both the primal and dual variables. Masking is implemented by using iteration-varying and atom-specific parameters $\alpha_j[\tau]$, $\phi_j[\tau]$ and $\theta_j[\tau]$. Masking the dual variables (or shadow prices), in particular, is desirable since these may reveal sensitive data related to costs, operating constraints, or other preferences of SMOs. Instead, masked variables $\hat{a}$ and $\hat{\nu}$ are exchanged between atoms. By iteratively solving the local, decomposed optimization problems across all SMOs, NST-PAC (and PAC) provably converge to the globally optimal ACOPF (relaxed) solutions for the whole primary feeder \citeSI{Romvary2022AOptimization,Ferro2022AHubs}.

\section{{FEDERATED LEARNING}\label{sec:fl}}
\begin{figure}[htb]
    \centering
    \includegraphics[width=0.6\columnwidth]{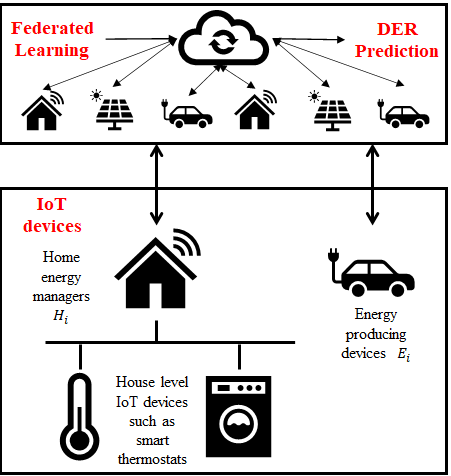}
    \caption{An overview of the DER prediction process using federated learning}
    \label{intro}
\end{figure}

When it comes to DER forecasting, the challenges of privacy as well as the requirement of large training data sets, can be met using a distributed machine learning paradigm, federated learning (FL)~\citeSI{yang2019federated,li2020federated,lim2020federated,buildFL2020}. FL is a machine learning framework where each device participates in training a central model without sending actual data, but only exchanges gradient information in the training phase and sends prediction estimates during deployment. A general overview of the proposed DER prediction process is shown in \cref{intro}. In the figure, various IoT devices including those at a house level such as smart thermostats and smart washers, and energy-producing devices such as PV and EVs are considered. The future smart grid will include a wider range of devices that will be capable of computation and communication. The house-level devices are grouped under a home energy manager $H_i$ to enable aggregation at a house level while energy-producing devices such as EV and PV (grouped under $E_i$) are assumed to directly participate in a transactive environment. Both $H_i$ and $E_i$ can be considered typical IoT-based DERs connected to a power grid, whose energy consumption needs to be predicted. Using a communication infrastructure, our goal is to exchange information between the DERs in the bottom local layer and the global decision makers at the top layer in a private and secure manner so as to lead to an accurate prediction of the DER consumption/generation. The DER prediction is then utilized to formulate a grid service, to mitigate the load swings and \mbox{\lq\lq peaks"} that occur in the distribution grid due to a lack of situational awareness. The FL-based DER prediction is used to anticipate the load swings and mitigate them by proactively controlling the DERs \citeSI{venkataramanan2022forecast}.

\begin{figure}[htb]
    \centering
    \includegraphics[scale=0.45]{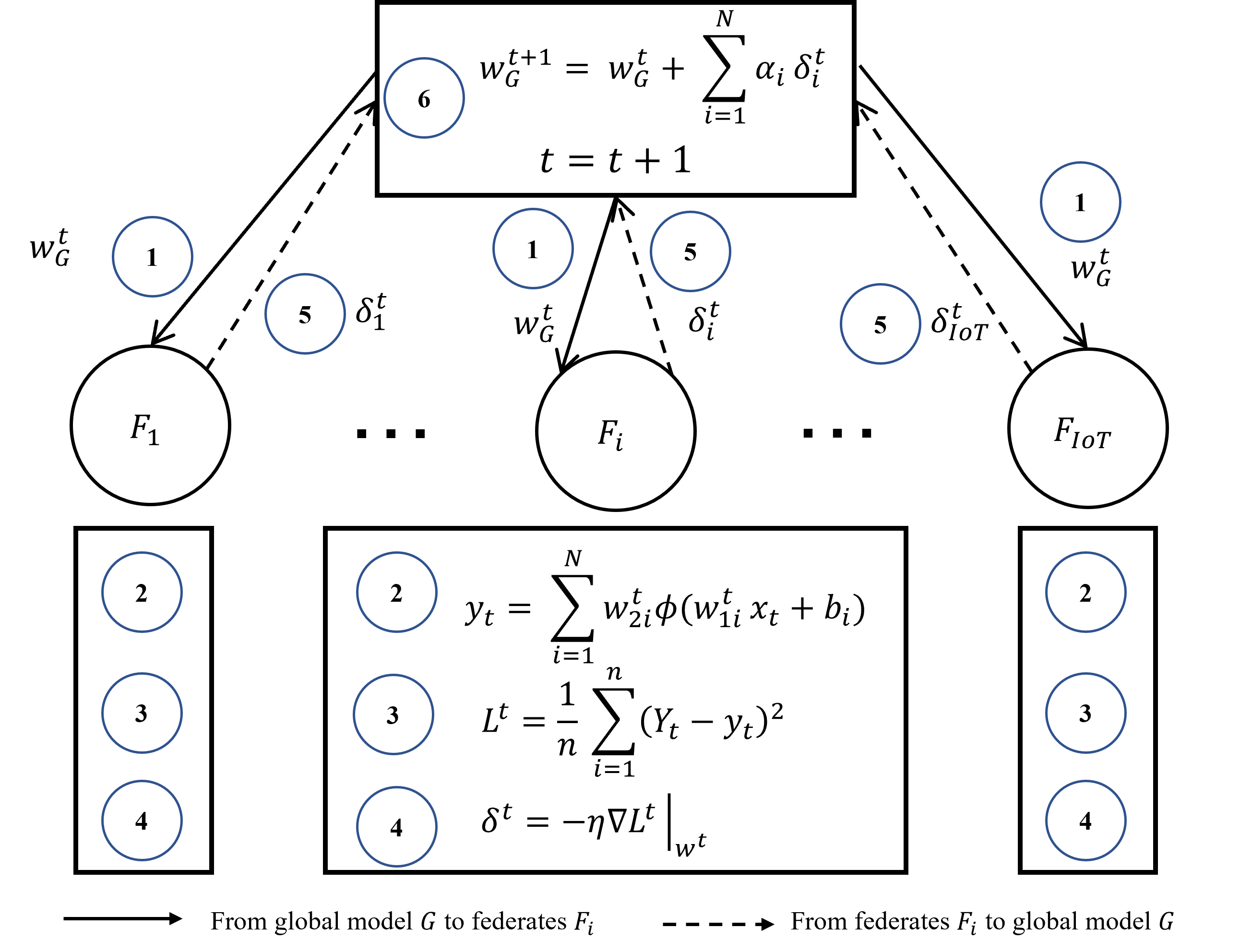}
    \caption{The schematic of neural network training using federated learning is shown here. The steps (1)-(6) are repeated until $L^t \leq \epsilon$ }
    \label{block}
\end{figure}

A typical process of DER-forecast can occur in the following manner. Collect the input-output pair $[x_t, \hat{P}^t(T)]$ for a federate $F_i$ for several samples $n$. The features used in this work are \mbox{$x_t=[P^t (T-15), P^t (T-30), P^t (T-60), P^t (T-120)]$}, where \mbox{$P^t(T-m)$} denotes the actual power consumption, and $m$ denotes the minutes prior to time $T$. The number of samples $n$=2880, was obtained by collecting data every 15 minutes over a period of 30 days. The overall training schematic of the FL-based neural network is shown in \cref{block}. 

\section{DETAILED COMMUNICATION STEPS AND TIMELINE\label{sec:comm_details}} 

\begin{figure}[htb]
\centering
\includegraphics[width=\columnwidth]{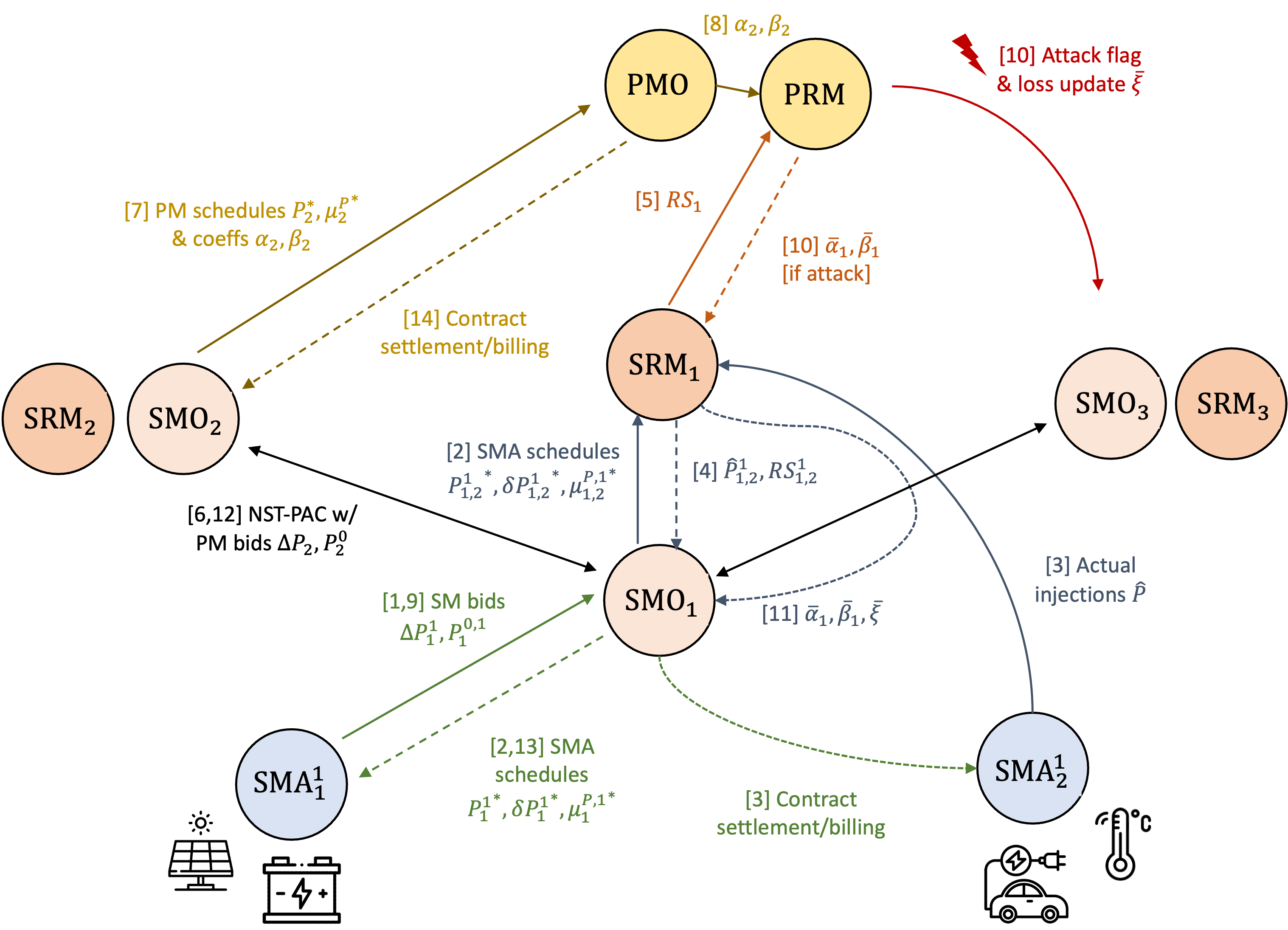}
\caption{Diagram showing a more detailed communication scheme and steps for information exchange between the various market operators, agents, and resilience managers at the secondary and primary levels.\label{fig:detailed_comm}}
\end{figure}

See \cref{fig:detailed_comm} for more details on the communication among different market agents, operators, and resilience managers.

\section{MITIGATION USING THE MARKET OPERATORS AND RESILIENCE MANAGERS\label{sec:attack_mitig}}

Our market framework consisting of the SM and PM provides situational awareness (SA) in the form of available power injections at various nodes at the primary and secondary levels. Once the market is cleared, during execution, the actual injections from the SMA and PMA are monitored by the SRM and PRM, respectively (see \cref{fig:prmsrm}). These injections are then utilized by these managers to compute commitment scores, trustability scores, and resilience scores (RS), as shown in \cref{sec:commit_app} and \cref{sec:iot-ts}. In what follows, we will show how the SA from the market operators and the RS from the resilience managers can be utilized to mitigate all the attacks described in \cref{sec:usecases}. 

As a result of continuous monitoring, any unexpected deviation from the agents' nominal performance in the form of change in the net injection at the PCC, raises a flag. Any such flag makes the operators shift from the nominal operating mode to the resilience mode. Minimal visibility regarding actual injections from all PMA is assumed to be available. Rather, we assume that each SRM only locally observes the actual injection from the corresponding SMA, and each SRM communicates that information to the PRM. More importantly, the attack scenarios considered also assume that this important communication to the PRM from all SRMs is completely sabotaged (as was the case in the Ukraine 2015-16 attacks). Despite this loss of communication, the PRM is able to step in and mitigate the attack as the flag raised is independent of this communication loss and is due to a physical impact of the agents' deviation from nominal performance. Subsequently, the PMO redispatches trustworthy PMAs so as to bring the power import from the bulk grid down to pre-attack levels. The new setpoints for the PMAs/SMOs are in turn suitably disaggregated to compute new setpoints for the SMAs through a re-dispatch by the SM. Before proceeding to the results, we propose a specific mitigation strategy that leverages the SA provided by our approach.
\begin{figure}[htb]
\centering
\includegraphics[width=\columnwidth]{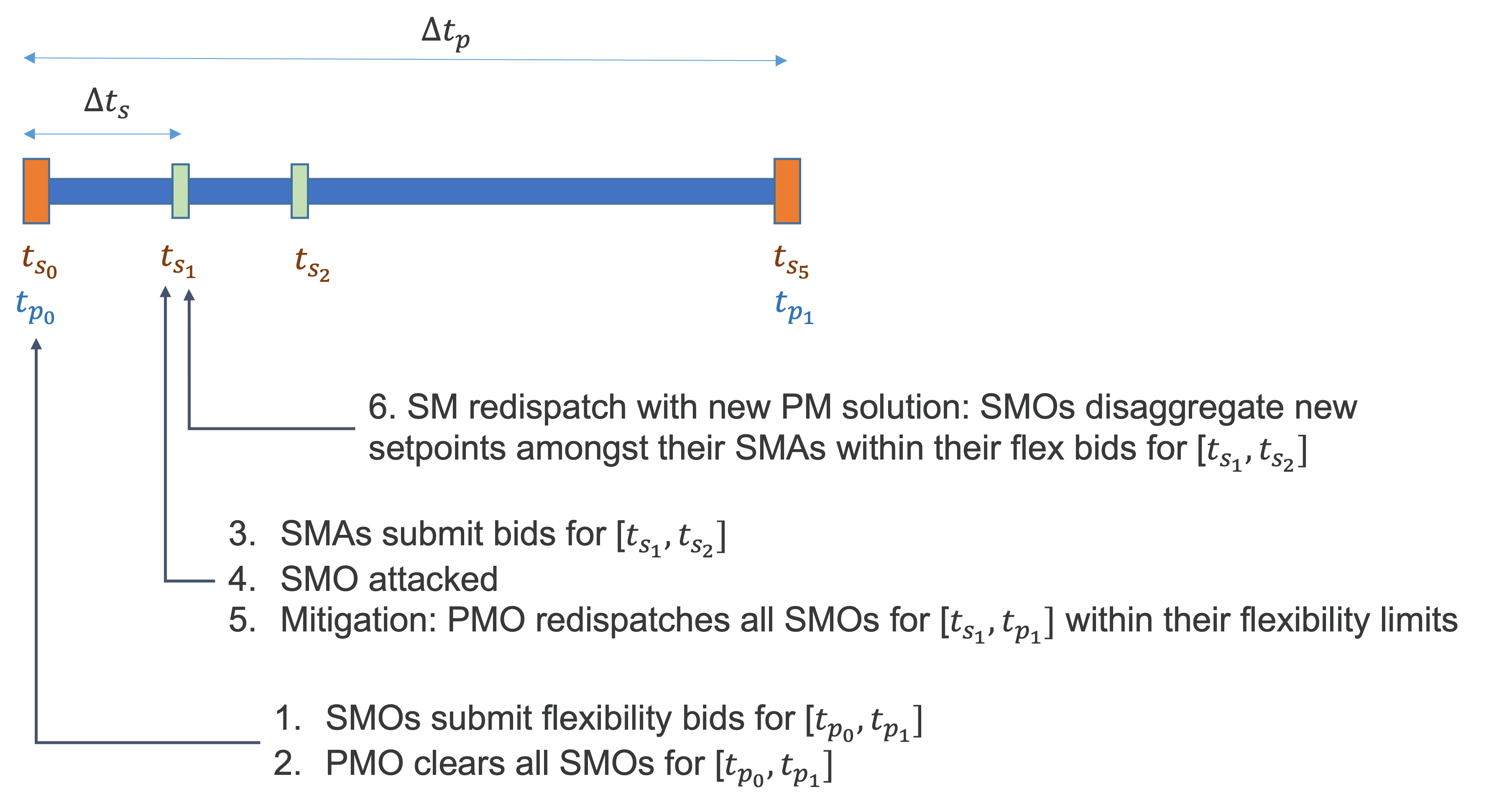}
\caption{Timeline of attack detection and mitigation. \label{fig:attack_timeline}}
\end{figure}

\subsection{Algorithm (A) for redispatch by the PMO in a balanced network \label{sec:costcoeff1}} 

We first consider a balanced, equivalent single-phase network using the BF model. The starting point for the overall mitigation sequence is the awareness that an attack has occurred. This is realized by the PRM in the form of a change in the net load from $P_{PCC}$ to $\overline{P}_{PCC}$, which denotes the net load from the entire primary feeder at the substation before and after the attack, respectively. This can be detected by the PRM at the substation or point of common coupling (PCC) since this is the power it imports from the main transmission grid. We propose a redispatch algorithm that the corresponding SMOs can carry out based on the ratio between these two values. To describe this redispatch algorithm, we begin with the cost function in \cref{eq:cost:1} for the PM ACOPF problem. For ease of exposition, this can be rewritten in a simplified manner as: 
\begin{gather}
    \sum_{i=1}^n \left(\frac{1}{2} \; \alpha_i P_i^{G^2} + \beta_i \left(P_i^L - P^{L0}_i\right)^2\right) + \xi \cdot losses \label{eq:update_obj}\\
    \overline{\alpha}_i = \Delta_{\alpha} \alpha_i, \; \overline{\beta}_i = \Delta_{\beta} \beta_i, \; \overline{\xi} = \Delta_{\xi} \xi; \; \alpha, \beta, \xi, \Delta > 0 \label{eq:update1}\\
    \Delta_{\alpha} = \Delta_{\beta} = \frac{|P_{PCC}|}{|\overline{P}_{PCC}|}, \; \Delta_{\xi} = \frac{|\overline{P}_{PCC}|}{|P_{PCC}|} \label{eq:update2}
\end{gather} 
We note that a change in the power import from the main grid causes $\Delta_\alpha, \Delta_\beta, \Delta_\xi$ to deviate from unity. Suppose that several distributed local generator SMOs are attacked, as in Attack 1(a). This would increase net feeder load, i.e.  $|\overline{P}_{PCC}| > |P_{PCC}|$ (note that both $P_{PCC}, \overline{P}_{PCC} < 0$ since net loads are negative injections), thus causing $\Delta_\alpha<1$. Applying this cost coefficient update would lower the cost coefficients from $\alpha_i$ to  $\alpha_i^\prime$. This results in dispatching more local generation from remaining online SMOs instead of importing power from the bulk grid. As the SMOs also have information about the flexibility in their corresponding SMAs in the form of $\delta P^*, \delta Q^*$ (see \cref{sec:sm_clear}), the overall hierarchical PM-SM market structure automatically provides the solutions of the new dispatch. Similarly, a value of $\Delta_{\beta} < 1$  reduces the disutility coefficients to encourage more demand response via load shifting and/or curtailment, by utilizing the downward flexibility provided by the SMOs bidding into the PM, and subsequently also by the SMAs bidding into the SM. In contrast to these two values, when the net import from the main grid increases, then $\Delta_{\xi} > 1$ penalizes electrical line losses more heavily in the objective function. As a result, the redispatch discourages imports from the transmission grid in favor of dispatching more local DERs. This is because distribution grids are more lossy (have higher resistance to reactance ratios), and hence prioritizing the loss minimization makes it more efficient to utilize local generation closer to the loads being served.

After deriving the multiplicative coefficient update factors $\Delta_{\alpha}, \Delta_{\beta}, \Delta_{\xi}$, the PRM can broadcast these common values to all the SRMs simultaneously, who in turn send them to their corresponding SMOs. The SMOs update each of their objective function coefficients using these factors and then perform distributed optimization to redispatch the PM, resulting in new $P$ and $Q$ setpoints for SMOs, along with new nodal distribution LMPs (d-LMPs). This is followed by each SMO also re-dispatching their SM, in order to disaggregate the new setpoints among their SMAs. A timeline of the key events is shown in \cref{fig:attack_timeline}.

\subsection{Algorithm (B) for redispatch in an unbalanced, 3-phase network\label{sec:costcoeff2}}

For the unbalanced 3-phase case, we use a modified algorithm for the coefficient update. The update rule here is more sophisticated since in this case, the variables are now 3-phase vectors rather than scalars. 
\begin{gather}
\Delta=\mathbf{P}_{PCC}-\overline{\mathbf{P}}_{PCC} \\
 Z_i\left(\delta_i\right)=1+\frac{RS_i \Delta^{\top} \delta_i}{\mu \sum_i RS_i} \Longrightarrow \gamma_{i \delta}=\frac{1}{Z_i\left(\delta_i\right)} \\
\overline{\boldsymbol{\alpha}}_i=\gamma_{i \alpha} \boldsymbol{\alpha}_i, \quad \overline{\boldsymbol{\beta}}_i=\gamma_{i \beta} \boldsymbol{\beta}_i, \quad \overline{\boldsymbol{\xi}}=\left(\frac{\sum_i \gamma_{i \alpha}+\gamma_{i \beta}}{2n}\right)^{-1} \boldsymbol{\xi} \label{eq:update_ci_model}
\end{gather}
Note here that $\textbf{P}_{PCC}, \overline{\textbf{P}}_{PCC}$ are the 3-phase power imports from the tie line before and after the attack. $\boldsymbol{\alpha}_i, \boldsymbol{\beta}_i$ are $3\times 1$ vectors representing cost and disutility coefficient for each phase at SMO node $i$, and $\boldsymbol{\xi}$ is a 3-phase hyperparameter that penalizes line losses in the objective function. A DG attack that increases net load would result in $\gamma_{i\alpha}, \gamma_{i\beta} < 1$ and $\overline{\boldsymbol{\xi}} < \boldsymbol{\xi}$. Thus, these coefficient updates work using a similar intuition to algorithm (A) in that it favors local DER generation and load flexibility over transmission imports. A key difference here is that the PRM also takes into account the RS of each SMO during the redispatch so that it relies more heavily on resilient SMOs for attack mitigation. The PRM updates the coefficients $\alpha_i, \beta_i$ and $\xi$ to $\alpha_i^\prime, \beta_i^\prime$ and $\xi^\prime$, and sends the new coefficient values to all SRMs. The SRMs send these new objective functions to the corresponding SMOs, and the rest of the mitigation procedure follows in the same manner as in the previous section.

\section{RESILIENCE-DRIVEN RECONFIGURATION ALGORITHM\label{sec:reconfig_alg}}

\begin{figure}[htb]
    \centering
    \includegraphics[width=\columnwidth]{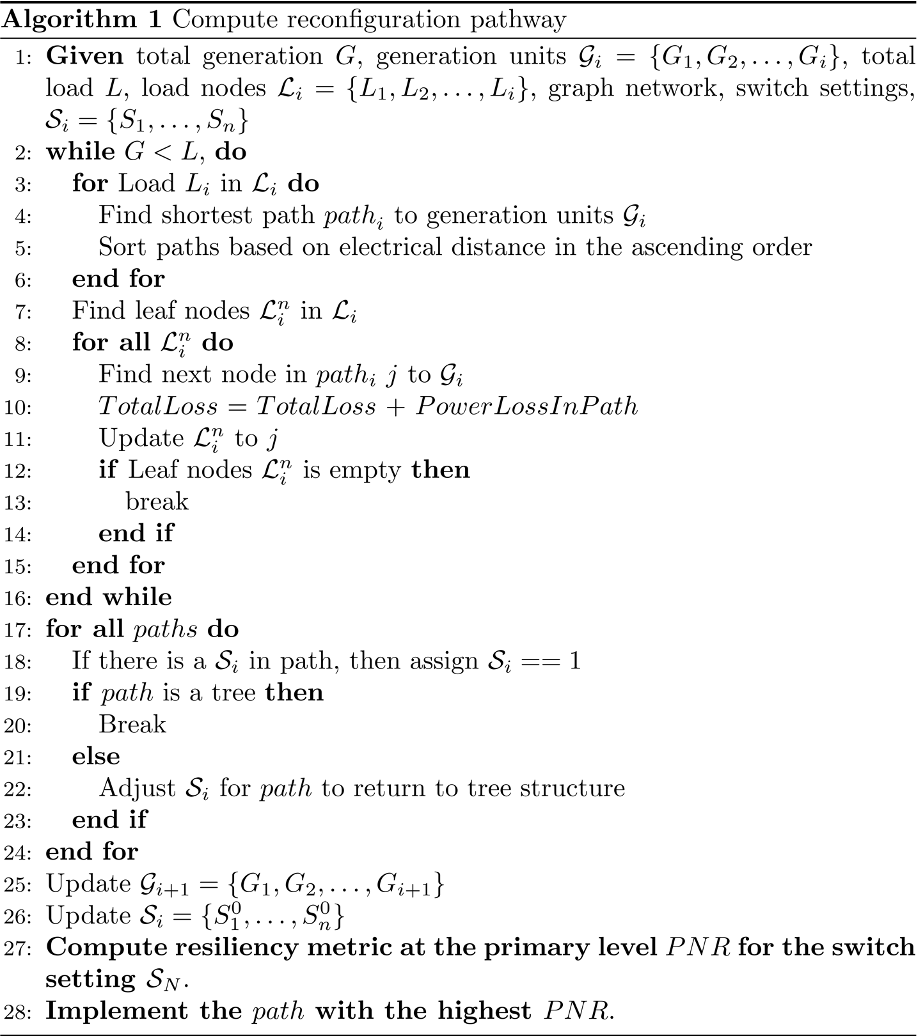}
    \caption{Resilience-based reconfiguration algorithm.}
    \label{fig:reconfig_alg}
\end{figure}

All possible shortest paths are computed between each generation source and critical load pairs present within the system using the graph network. If the generation is not enough to supply the total critical load, then the algorithm searches for the next available generation. This will continue until the critical load demand is met. As the generation sources are assigned to critical loads, if any source's capacity is more than the assigned load, the source's partial remaining capacity will be utilized for other loads. Once we have all the feasible paths for reconfiguration, we will compute the resiliency metric for each path (see \cref{sec:pnr} for how to compute the resiliency metrics and $PNR$), which will support the operator in finding the most resilient path to restore. The reconfiguration paths will be determined based on the stress levels of the grid and the corresponding degree of the failed SMA node, the tolerance bands and flexibility of the ICA, and the security levels and privacy needs of the SMA. This process is outlined in \cref{fig:reconfig_alg}.

\section{VALIDATION PLATFORMS\label{sec:validation}}
\subsection{PNNL\label{sec:val_pnnl}}
Since the proposed framework is intended to be deployed in a large-scale distribution system with a high penetration of renewables, we choose our first validation platform to be driven by HELICS (Hierarchical Engine for Large-scale Infrastructure Co-simulation), an open-source cyber-physical energy co-simulation framework for energy systems \citeSI{trevor2024Helics}. As the core engine of the platform, HELICS provides time management and data exchanges between the simulators, also known as federates. Moreover, through standard procedures and application programming interfaces (APIs), data exchange between federates is performed either as values or messages. Within this platform, GridLAB-D is the distribution system simulator used to simulate all of the use cases we discussed above. GridLAB-D is versatile, with the capability of simulating large 3-phase unbalanced distribution systems, with agent-based and information-based modeling tools and extensive data collection tools for end-use technologies and interface APIs for co-simulation \citeSI{Chassin_4517260}. Thus, it allows us to modify the standard IEEE 123-node system to include (1) residential (either single or multi-family) and commercial buildings with or without heating, ventilation, and air conditioning (HVAC) systems, (2) edge devices with IoT connectivity including HVAC type appliances, small electronics and lighting, (3) distributed generators (DGs) such as photovoltaic (PV) panels, battery energy storage systems (BESSs), and diesel generators. The distribution model is also assumed to have all its loads connected via smart meters with load-shedding capability to regulate energy demand according to a distribution system operator’s command. Specifically, the IoT-enhanced version of the IEEE 123-node system consists of the assets detailed in \cref{tab:distr-feeder-conf}.

\begin{table}[htb]
    \caption{GridLAB-D IEEE 123-node test feeder features with IoT-enhanced model.}
    \label{tab:distr-feeder-conf}
    \centering
    \begin{tabular}{p{0.2\columnwidth}|p{0.2\columnwidth}|p{0.1\columnwidth}|p{0.2\columnwidth}}
        \hline
        & & \Centering{Number} & \Centering{Capacity} \\ \hline
        \RaggedRight{Standard IEEE 123-node test feeder} & \RaggedRight{Spot loads} & \Centering{$85$} & \RaggedRight{$3,985.7$ kVA} \\ \hline
        \RaggedRight{EUREICA IEEE 123-node test feeder} & \RaggedRight{Houses - Demand response (HVACs in all, WHs in 348)} & \Centering{$1,008$} & \RaggedRight{variable ($4$ KW avg/house) ($20\%$ to $30\%$ critical)} \\ \cline{2-4} 
         & \RaggedRight{Distributed generators (DGs)} & \Centering{$380$} & \RaggedRight{$1,745.8$ kVA} ($\approx44\%$ system penetration) \\ \cline{3-4}
         & \RaggedLeft{PVs} & \Centering{$207$} & \RaggedRight{$880.84$ kVA} \\ \cline{3-4}
         & \RaggedLeft{BESSs} & \Centering{$173$} & \RaggedRight{$865$ kVA}  \\ \hline
         & \RaggedRight{Community PVs} & \Centering{$12$} & \RaggedRight{$96$ kVA}  \\ \hline
    \end{tabular}
\end{table}

\subsection{NREL\label{sec:val_nrel}}

\begin{figure}[htb]
    \centering
    \includegraphics[scale=0.18]{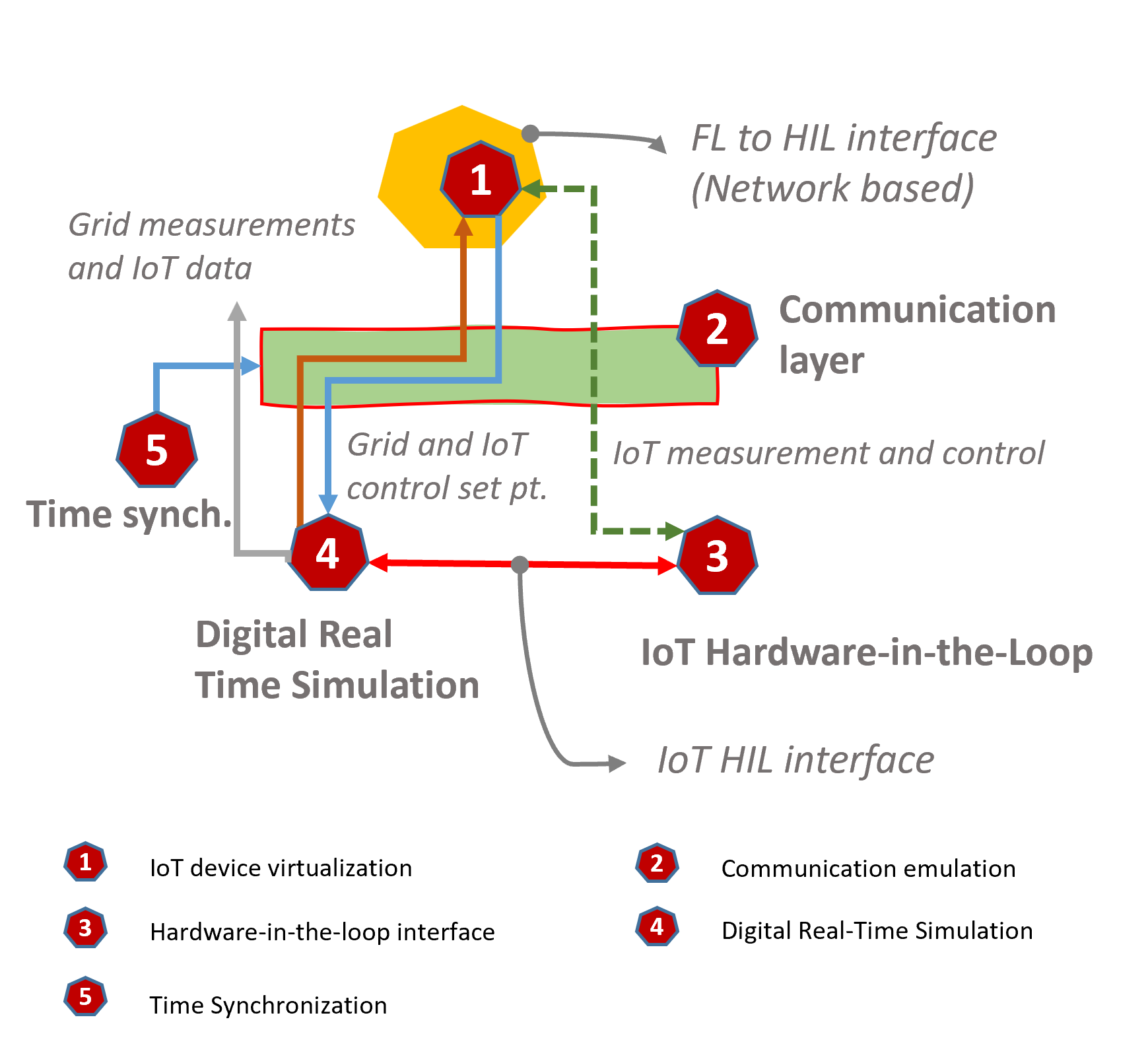}
    \caption{ARIES-DRTS Validation Platform at NREL}
    \label{fig:aries_drts}
\end{figure}

The objective of validating the EUREICA framework at NREL is to evaluate the feasibility of implementing the framework in \textit{real-time}. Since electrons flow in the grid in real-time, it is critical that the operations proposed should also function in real-time, and be in compliance with operational requirements. The Advanced Research on Integrated Energy Systems (ARIES) at NREL is a cutting-edge virtual emulation platform that encompasses actual Distributed Energy Resource (DER) hardware systems, such as wind turbines, photovoltaic (PV) arrays with controllers, batteries, and storage systems \citeSI{kurtz2021aries}. The digital real-time simulation (DRTS) cluster of ARIES is used for validating the performance of the EUREICA modules. The overall validation platform is shown in \cref{fig:aries_drts}. It constitutes 5 components - (i) IoT device virtualization (using Raspberry Pis and Typhoon HIL to characterize IoT devices), (ii) Communication emulation (using analog and network connections to emulate communication at the speed of actual communication in the field) (iii) hardware-in-the-loop interface (to provide increased fidelity for components under study) (iv) digital real-time simulation (using RTDS and Typhoon HIL to emulate the power grid in real-time) (v) Time synchronization (to bring together the hardware components using a time server to ensure accuracy of simulation). This validation platform is used to determine the performance of the EUREICA framework for all the modules. 

\subsection{LTDES\label{sec:val_ltdes}}
\begin{figure}[htb]
\centering
\includegraphics[width=\columnwidth]{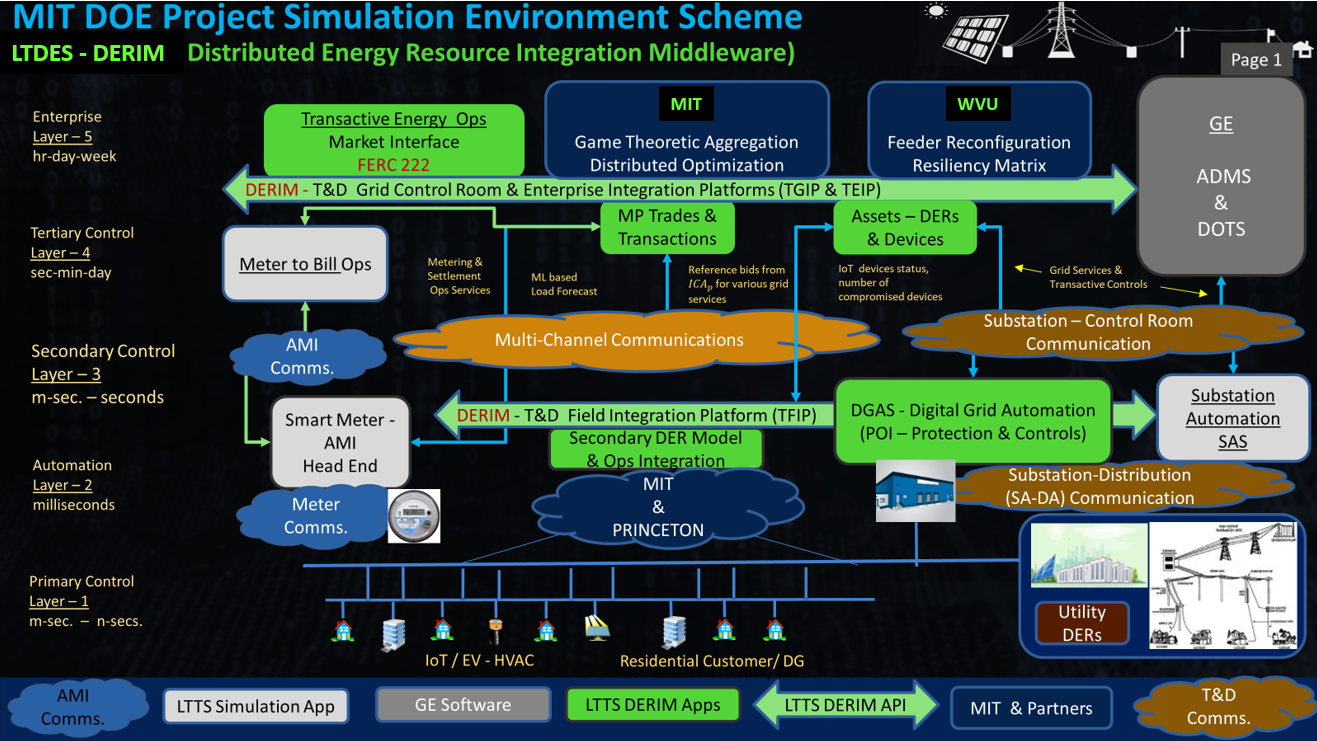}
\caption{DERIM interface with ADMS-DOTS in the LTDES validation platform. \label{figure4-ltdes}}
\end{figure}

\begin{figure}[htb]
\centering
\includegraphics[width=\columnwidth]{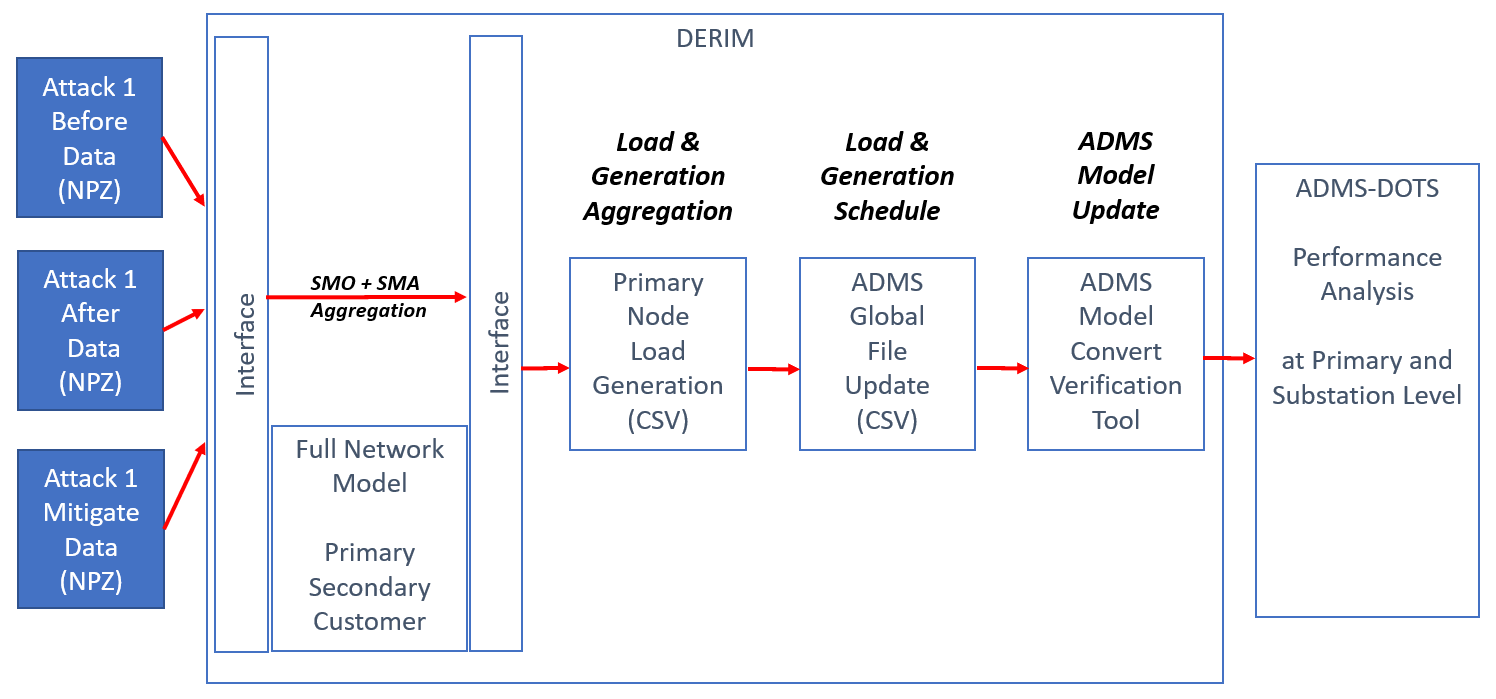}
\caption{Attack 1a validation process workflow in the LTDES validation platform with DERIM and ADMS-DOTS.\label{fig11-ltdes}}
\end{figure}

A training simulator-based platform was also used to validate the EUREICA framework. In particular, the General Electric (GE) ADMS DOTS (Advanced Distribution Management System-Distribution Operations Training Simulator), used for training operators and dispatchers, was used as the validation platform. Rather than users waiting to experience challenging events on the job, dispatchers are able to familiarize themselves with advanced application functionality and gain an understanding of how they interact with other subsystems of the ADMS. Into this ADMS-DOTS we introduce DERIM, a Distributed Energy Resource Integration Middleware, an interface that allows integration of various DERs with the ability to communicate as dictated by the EUREICA framework (see \cref{figure4-ltdes}). This integrated system uses the same software components, programmatic and user interfaces as the real-time ADMS, and creates an effective training and testing environment to operate with the actual network model, data, and functions in a controlled and safe environment. \cref{fig11-ltdes} shows an example of what the validation process looks like for attack 1a.


\section{Kundur 2-area system\label{sec:kundur}}

\begin{figure}[htb]
\centering
\includegraphics[width=\columnwidth]{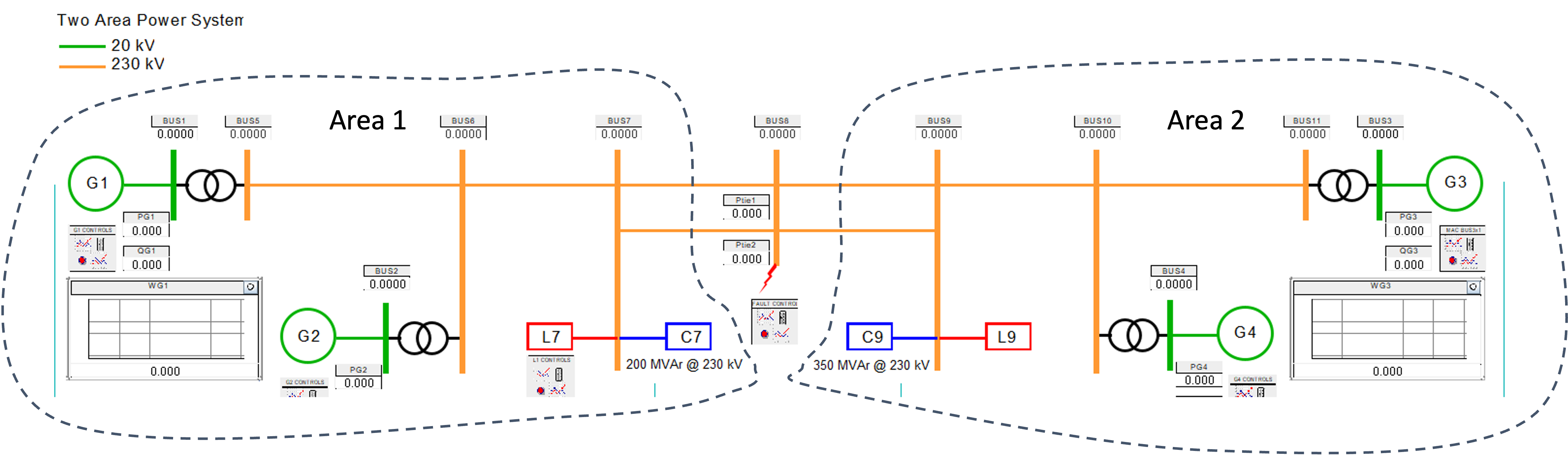}
\caption{Schematic of Kundur 2-area power system \label{fig:kundur}}
\end{figure}

\cref{fig:kundur} is a diagram of the Kundur 2-area transmission system commonly used as a test case to study dynamic stability, power interchange, oscillation damping, etc. The system contains 11 buses, four generators, and two areas. The two areas are connected with weak tie lines \citeSI{kundur2007power}.

\section{Numerical simulation setup for markets\label{sec:sim_setup}}
All use cases considered are based on an IEEE 123 test feeder (see \cref{fig:network_schematic}), which is radial, unbalanced, and multi-phase. The feeder was modeled in the GridLAB-D environment (see SI \cref{sec:val_pnnl} for more details) and augmented to have a high penetration of DERs. For all attacks except attack 3, we assumed that the switch settings are in their nominal positions such that we have one primary feeder, with 85 active nodes with SMOs/PMAs (out of the 123 in total). A PMO was assumed to be at the slack bus (substation), at either 115 or 69kV, with the SMOs at 4.16kV, and each SMA at 120-240V. The flexibility bids for the SMAs and the SMOs were randomly generated, allowing each to offer flexibilities of up to $\pm 30\%$ around their baseline power injections \citeSI{Olsen2014Demand2020_2}. We used 5-minute real-time market LMPs from the California ISO and assumed the Q-LMP to be 10\% of the P-LMP. Note that for all attack scenarios except attack 3, we used the CI model to represent the feeder as is. 

For attack 2, however, we considered a modified version of the feeder and deployed the BF model instead. Here, we modified the original IEEE-123 feeder to consider a case where we have a few large distributed generators (PV, batteries, diesel generators) concentrated at just five primary feeder (SMO) nodes numbered 25, 40, 67, 81, and 94. This is in contrast to the other attacks where there were instead a larger number of smaller DERs distributed throughout the network. Another distinguishing factor of this scenario is that the originally unbalanced feeder was converted to an equivalent balanced 3-phase model by (i) assuming all switches to be at their normal positions, (ii) converting single-phase spot loads to 3-phase, (iii) assuming cables to be 3-phase transposed, (iv) converting configurations 1 thru 12 to symmetric matrices and (v) modeling shunt capacitors as 3-phase reactive power generators \citeSI{Haider2020TowardGrids}. Each SMO was assumed to have between 3-5 SMAs with the number chosen uniformly at random. Since the injection data in the original IEEE-123 model was only available up to the primary feeder node level, we artificially randomly disaggregated the injections at each SMO amongst its SMAs, which could be net loads or generators.

We then performed a co-simulation of both the PM and SM for all attack scenarios. We refer the reader to \citeSI{nair2022hierarchical,nair2023local} for the behavior of this market structure for a nominal scenario when there is no attack. In what follows, we only consider the three attack scenarios described above.  We also note that our flexibility bids were synthetically created, so the resulting flexible ranges in our simulations may be quite large at times and not realistic in some cases. However, our proposed framework can be generally applied to cases where there is less DER flexibility as well.

\section{MARKET SIMULATIONS OF OTHER ATTACKS\label{sec:mkt_other_attacks}}
\subsection{Mitigation of Attack 1b\label{sec:attack1b}}

We note that in Attack 1b, there is a loss in net generation, and therefore the power imported from the bulk grid increases. It is also assumed that the communication from all  SRM to the PRM is disrupted, while the communication from PRM to SRM remains intact. That is, the PRM loses observability but is still able to communicate the redispatch of the new coefficients to the SRM. We do not consider the case when such observability is not lost, a discussion of which is beyond the scope of this paper. With the redispatch, the PM-SM framework identifies all of the new trustable PMAs through the SA computations described in Section 3 with the overall power balance met at all points in the distribution grid. 

\begin{figure}[htb]
\centering
\begin{subfigure}[t]{0.49\columnwidth}
    \includegraphics[width=\columnwidth]{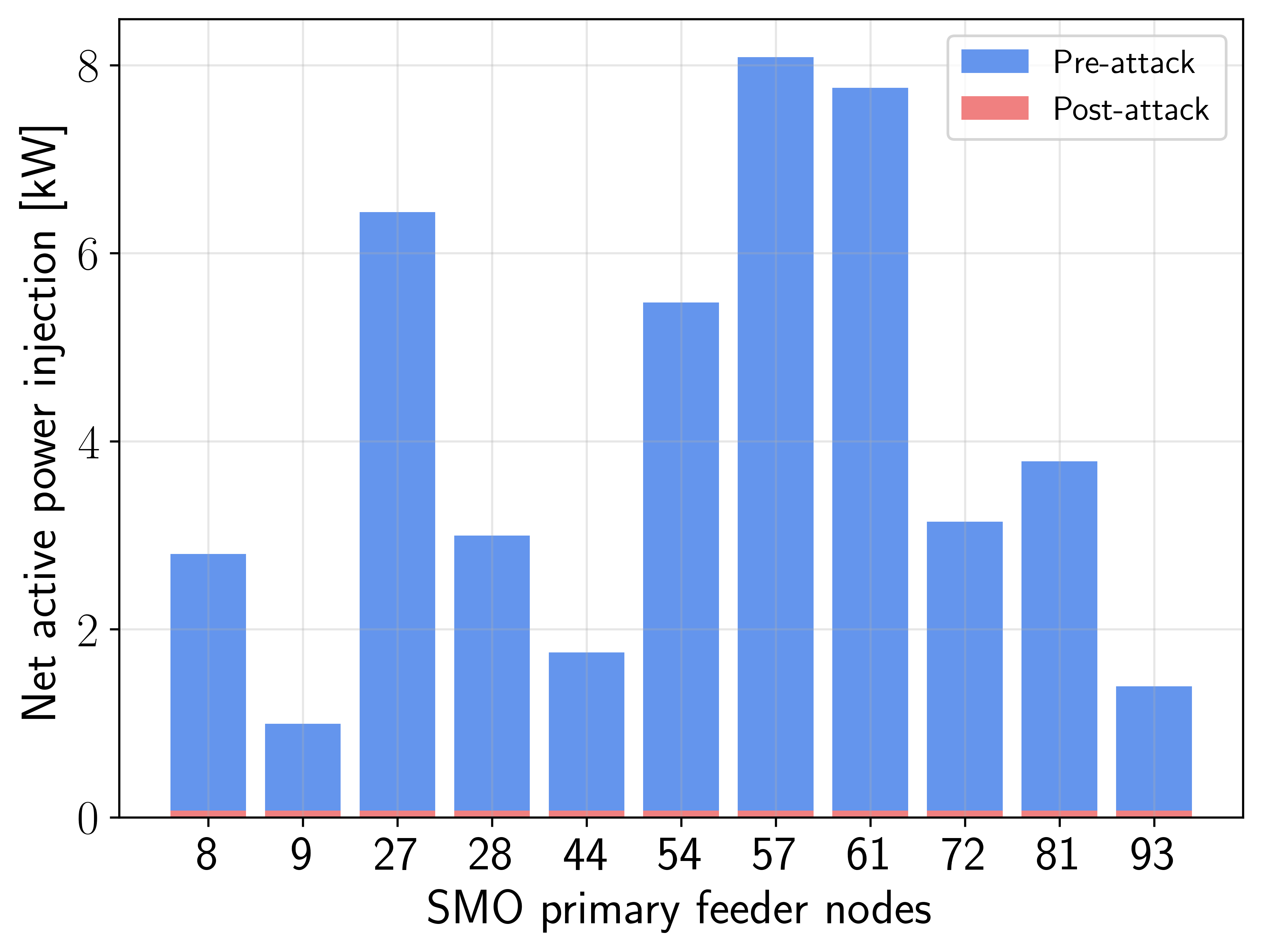}
\caption{Generation with and without attack 1b. \label{fig:attack1b_gen}}
\end{subfigure}
\begin{subfigure}[t]{0.49\columnwidth}
    \includegraphics[width=\columnwidth]{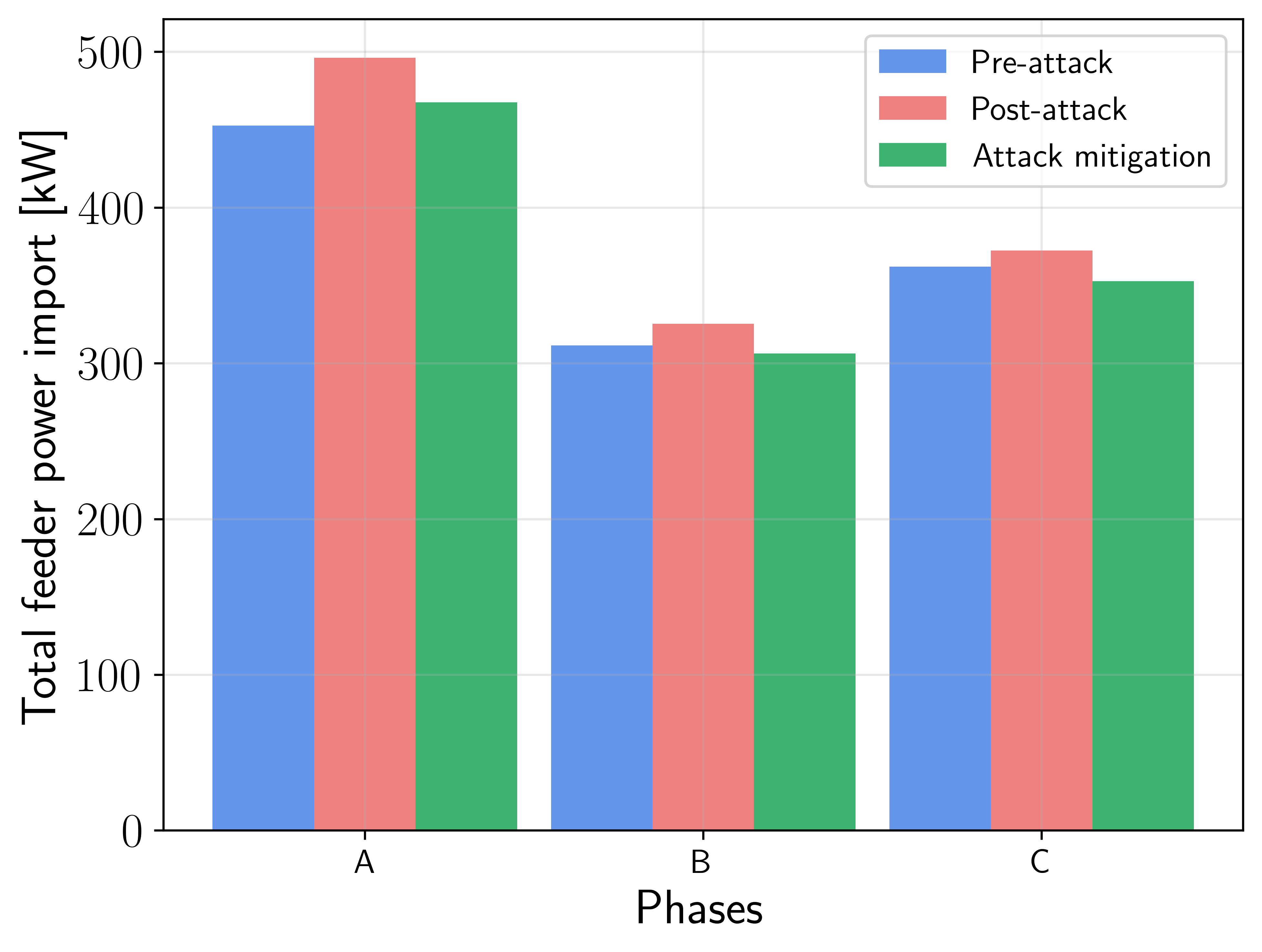}
\caption{Feeder power import. \label{fig:attack1b_import}}
\end{subfigure}
\caption{Effects of attack 1b on SMO net generation and power import.}
\end{figure}

\begin{figure}[htb]
\centering
\includegraphics[width=\columnwidth]{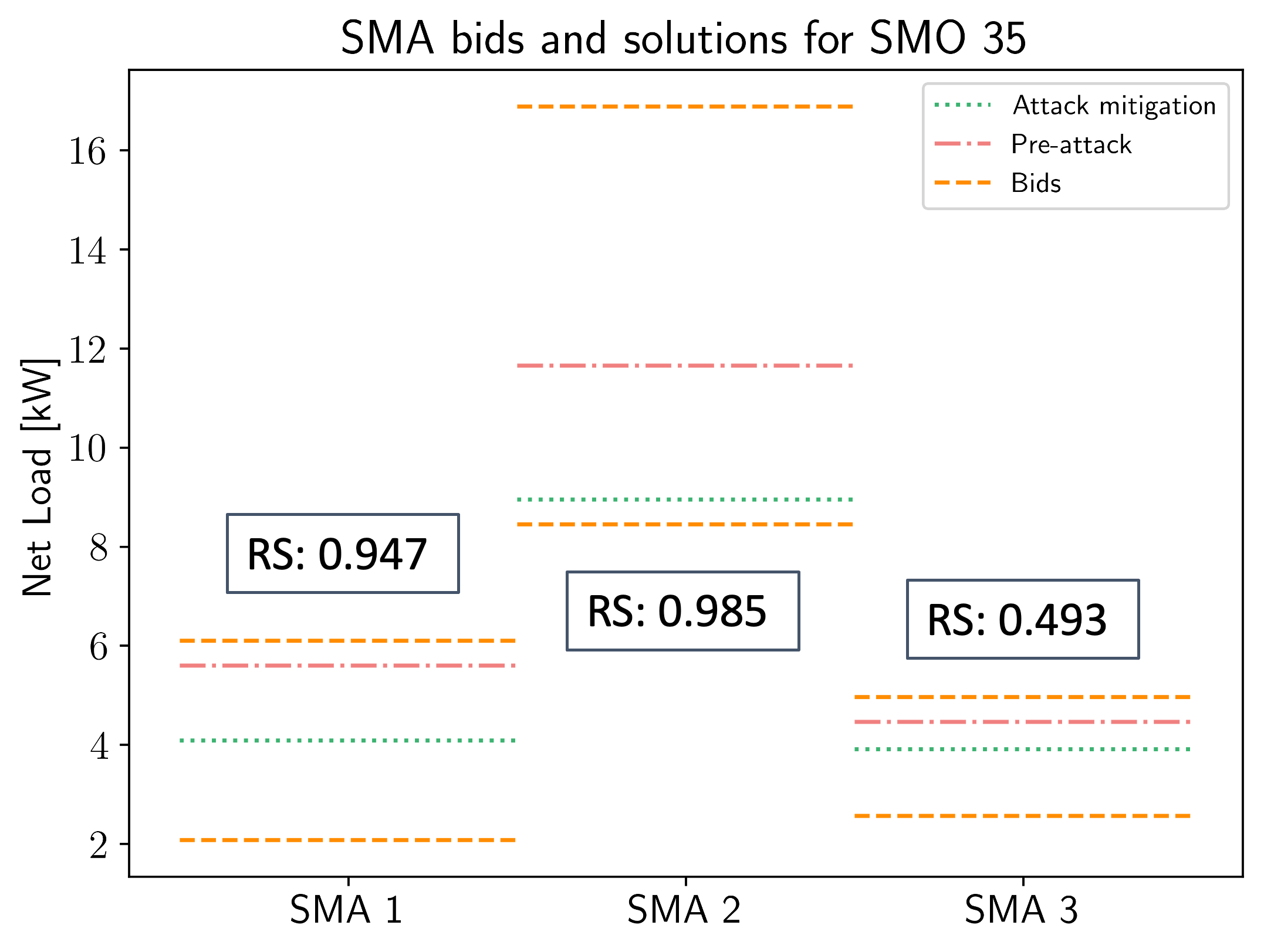}
\caption{Dis-aggregation of changes in the setpoints for SMO (from the PM) at node 35 across its 3 SMAs (in the SM), resulting from attack 1b mitigation, along with each SMA's RS. All 3 SMAs are on phase A. \label{fig:attack1b_icas}}
\end{figure}

\begin{figure}[htb]
\centering
\includegraphics[width=\columnwidth]{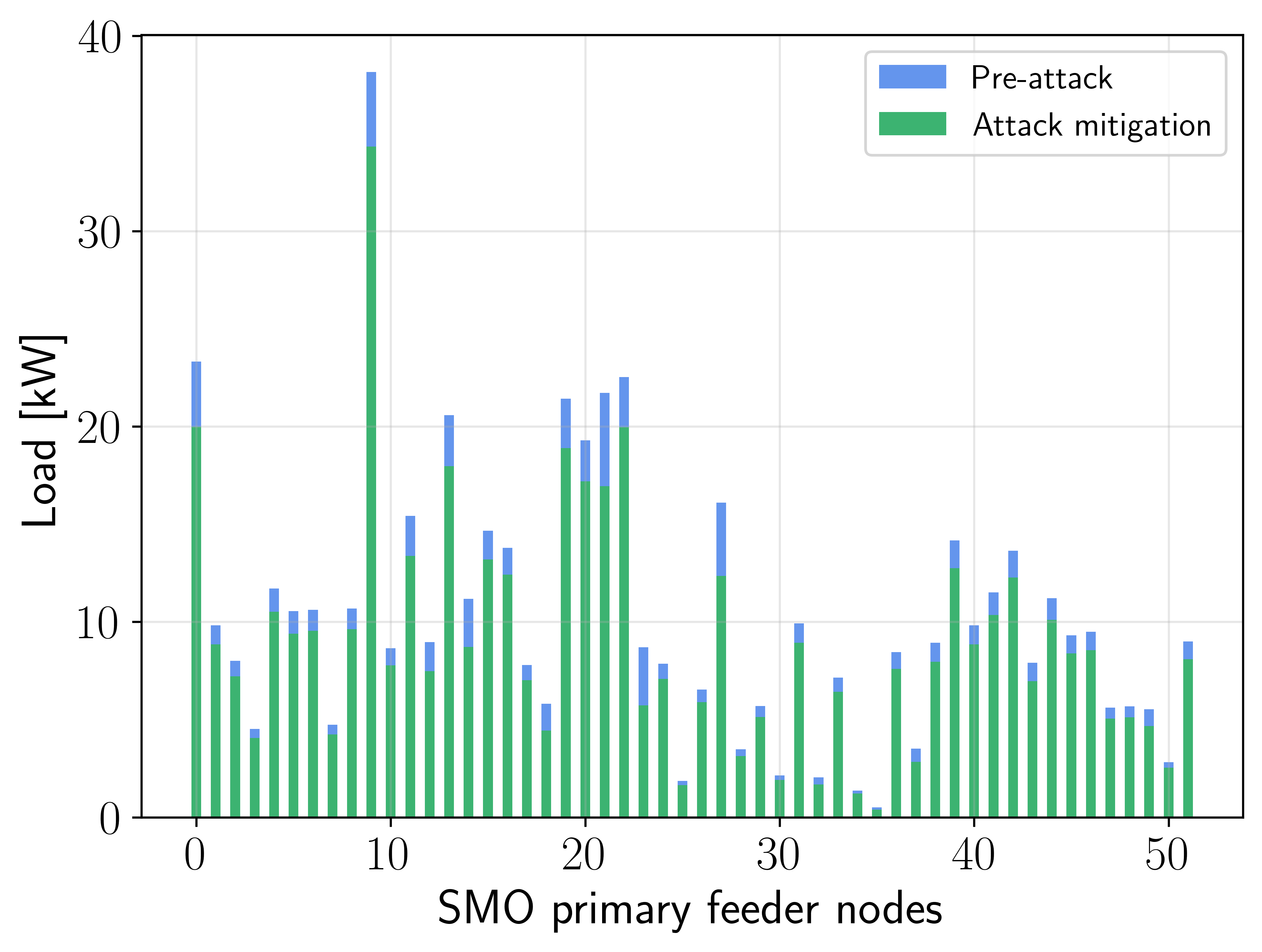}
\caption{Curtailment of flexible loads for attack 1b mitigation. \label{fig:attack1b_curtail}}
\end{figure}

\begin{figure}[htb]
\centering
\includegraphics[width=\columnwidth]{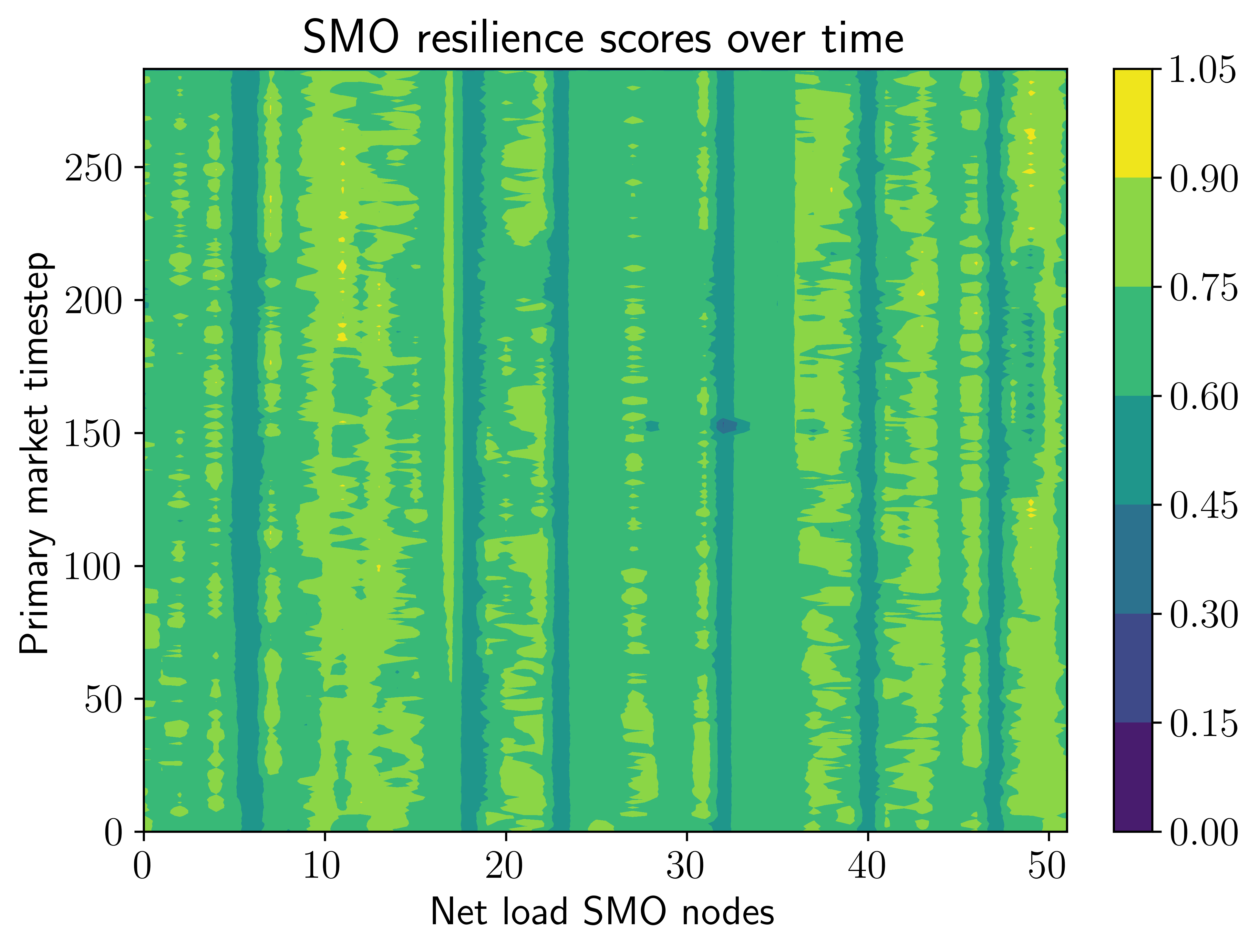}
\caption{Locational-temporal trends of RS across all flexible SMO nodes and over the whole simulation period of 24h. \label{fig:attack1b_rs_trends}}
\end{figure}

The steps in mitigating this attack are as follows. Due to the attack, 45 kW of net-generation compromised as shown in \cref{fig:attack1b_gen}. The PMO alerts other trustable PMAs/SMOs to redispatch their generation assets in the PM. Trustable PMAs/SMOs will curtail flexible loads to respond and mitigate the attack as in \cref{fig:attack1b_curtail}. The redispatch is also influenced by the resilience scores of different SMOs over time shown in \cref{fig:attack1b_rs_trends}. SMOs redispatch the SM which provides correct setpoints to all their SMAs. An example of this in \cref{fig:attack1b_icas} shows how the SMO at node 35 disaggregates its new setpoint amongst its 3 SMAs.
As a result of mitigation, the total import from the main grid stays at the same level as shown in \cref{fig:attack1b_import}. 

\subsection{Mitigation of Attack 1c\label{sec:attack1c}}
\begin{figure}[htb]
\centering
\begin{subfigure}[t]{\columnwidth}
    \includegraphics[width=\columnwidth]{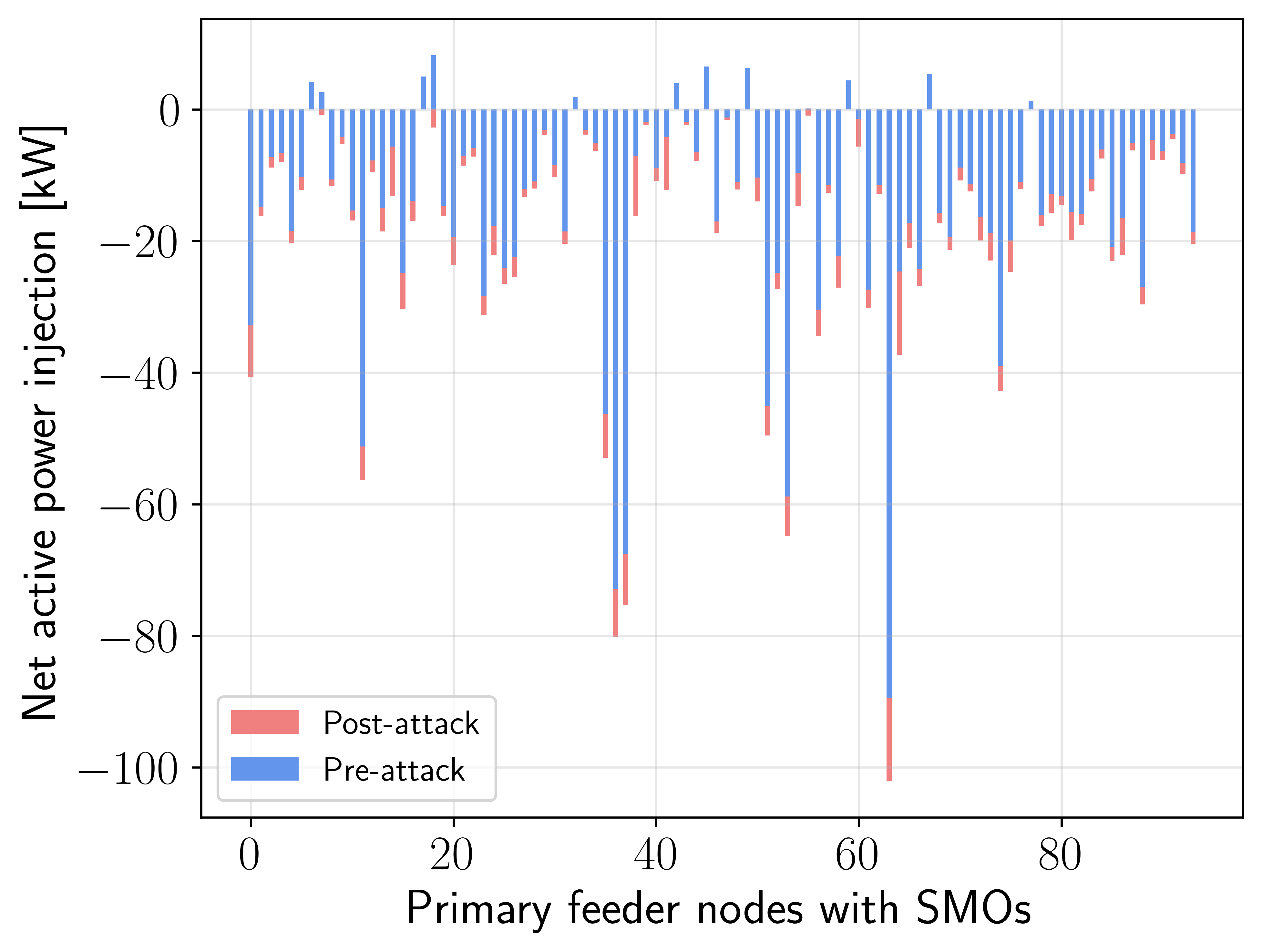}
    \caption{SMO injections.  \label{fig:attack1c_smo_b4_a4}}
\end{subfigure}
\begin{subfigure}[t]{0.6\columnwidth}
    \includegraphics[width=\columnwidth]{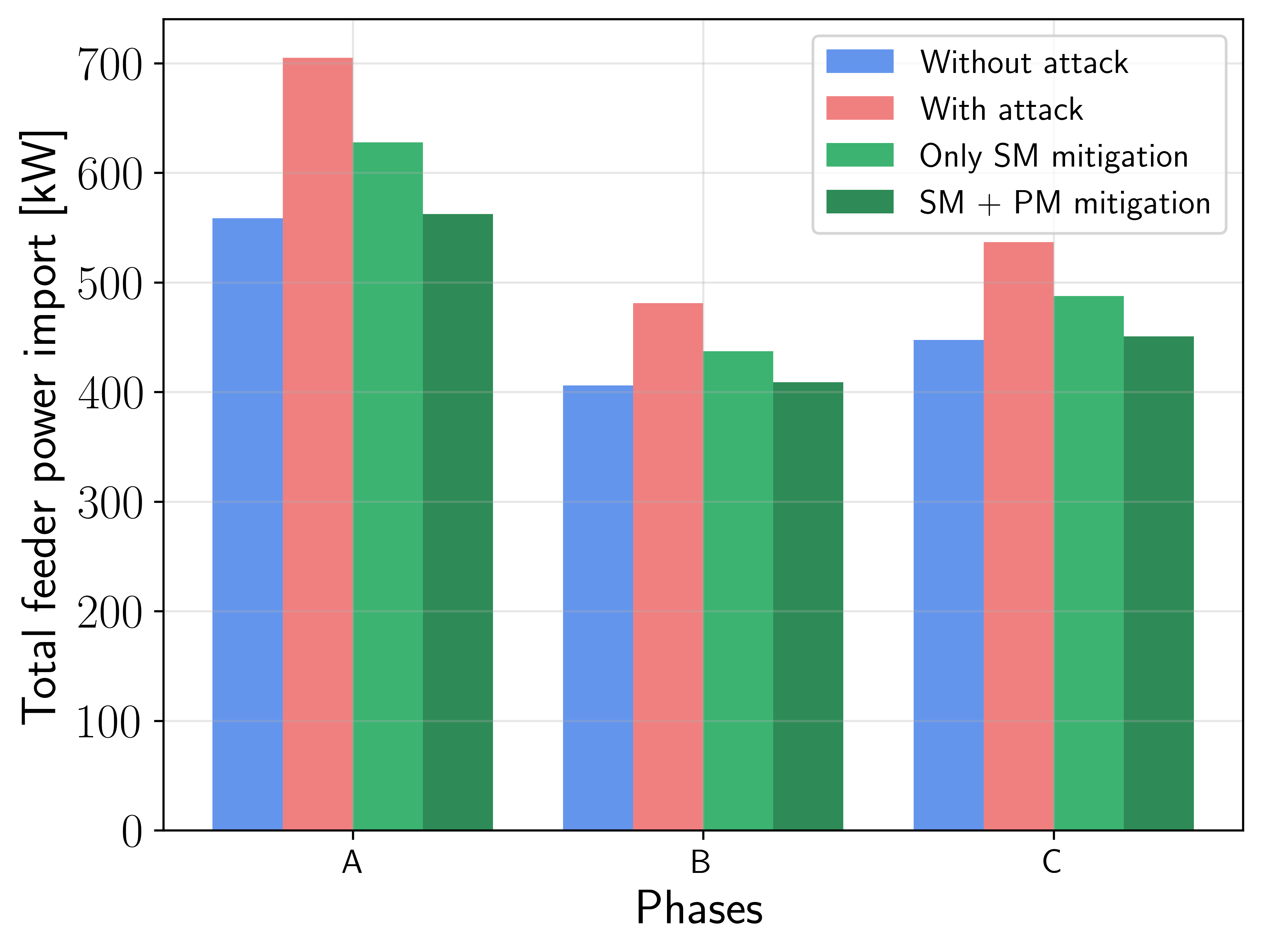}
    \caption{3-phase power imports from the main grid.  \label{fig:attack1c_import}}
\end{subfigure}
\caption{Effects of attack 1c on SMO injections and power import.}
\end{figure}

Attack 1c is a more distributed attack where individual SMAs representing secondary feeders are attacked directly. We considered a case where a large number of DGs including solar PV and batteries are attacked. A total of 53 SMA nodes with DGs were compromised and taken offline, resulting in a total loss in generation capacity of 157 kW. This leads to a decrease in the net injections across all the SMOs as seen in \cref{fig:attack1c_smo_b4_a4} - there are no longer any SMOs with net generation after the attack and the loss of local generation also increases the net load at the SMOs. This leads to an increase in power import from the main transmission grid as in \cref{fig:attack1c_import}. 

In the case of all other attacks, the mitigation strategy involves the PM redispatch occurring 1st followed by the SM redispatch. There, only the PM is directly involved in attack mitigation while the SM is only used to disaggregate the new SMO setpoints amongst their SMAs. However, in the case of attack 1c, the SM redispatch occurs first at the secondary feeder level and is then followed by the PM redispatch at the primary feeder level. Thus, both the SM and PM are actively involved in attack mitigation here. We see in \cref{fig:attack1c_import} that we can partially mitigate the attack by leveraging the flexibility of SMAs in the SM. However, SM mitigation alone is not sufficient. We need to also utilize the inter-SMO flexibility in the PM to fully mitigate and restore the feeder import back down to the pre-attack level. A summary of the attack metrics is shown in \cref{tab:attack_1c}.

\begin{table}[htb]
\centering
\caption{Attack 1c summary.\label{tab:attack_1c}}
\begin{tabular}{@{}ccc@{}}
\toprule
& \textbf{Power import {[}kW{]}} & \textbf{Total net load {[}kW{]}} \\ 
\midrule
\textbf{Pre-attack} & 1412    &    1457                      \\       
\textbf{Post-attack}    & 1722   & 1716                \\ 
\textbf{SM mitigation only}    & 1553   & 1547                \\
\textbf{SM + PM mitigation}    & 1422   & 1417                \\ 
\bottomrule
\end{tabular}
\end{table}

\begin{figure}[htb]
\includegraphics[width=\columnwidth]{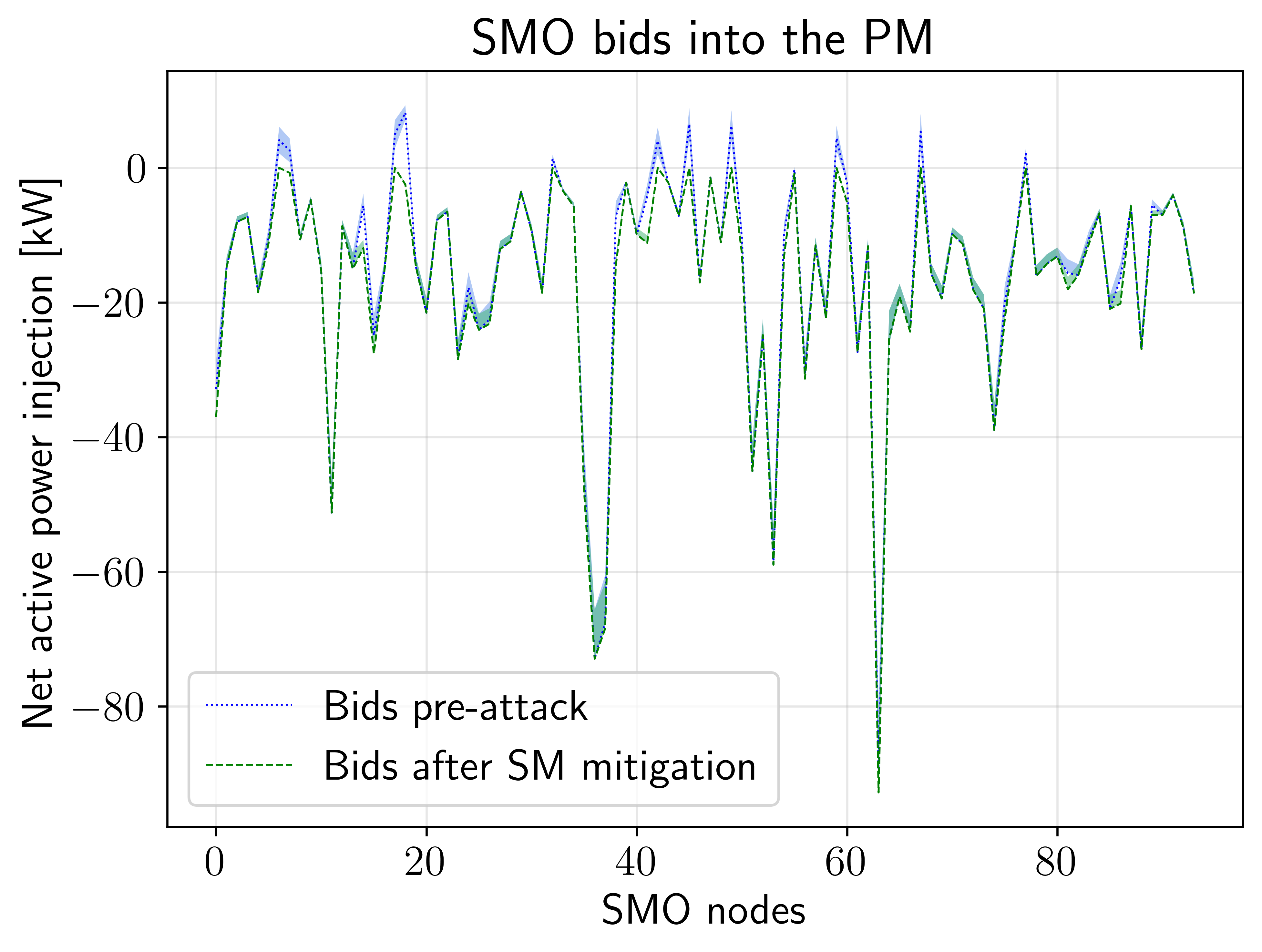}
\caption{Comparison of SMO flexibility bids into the PM before and after the attack. The dashed and dotted lines indicate the baseline values while the shaded regions are the flexibility bids around the baseline. \label{fig:attack1c_smo_bids}}
\end{figure}

We also compare the flexibility bids of the SMOs before and after the attack in \cref{fig:attack1c_smo_bids}. As expected, the net load of the bids generally increases across all SMOs due to the loss of local DGs at their respective SMAs. However, by leveraging their SMA flexibilities, the SMOs are still able to offer some flexibility to the PM to help mitigate the attack.

\subsection{Mitigation of Attack 2a \label{sec:attack2a}}
\begin{figure}[htb]
\centering
\includegraphics[width=\columnwidth]{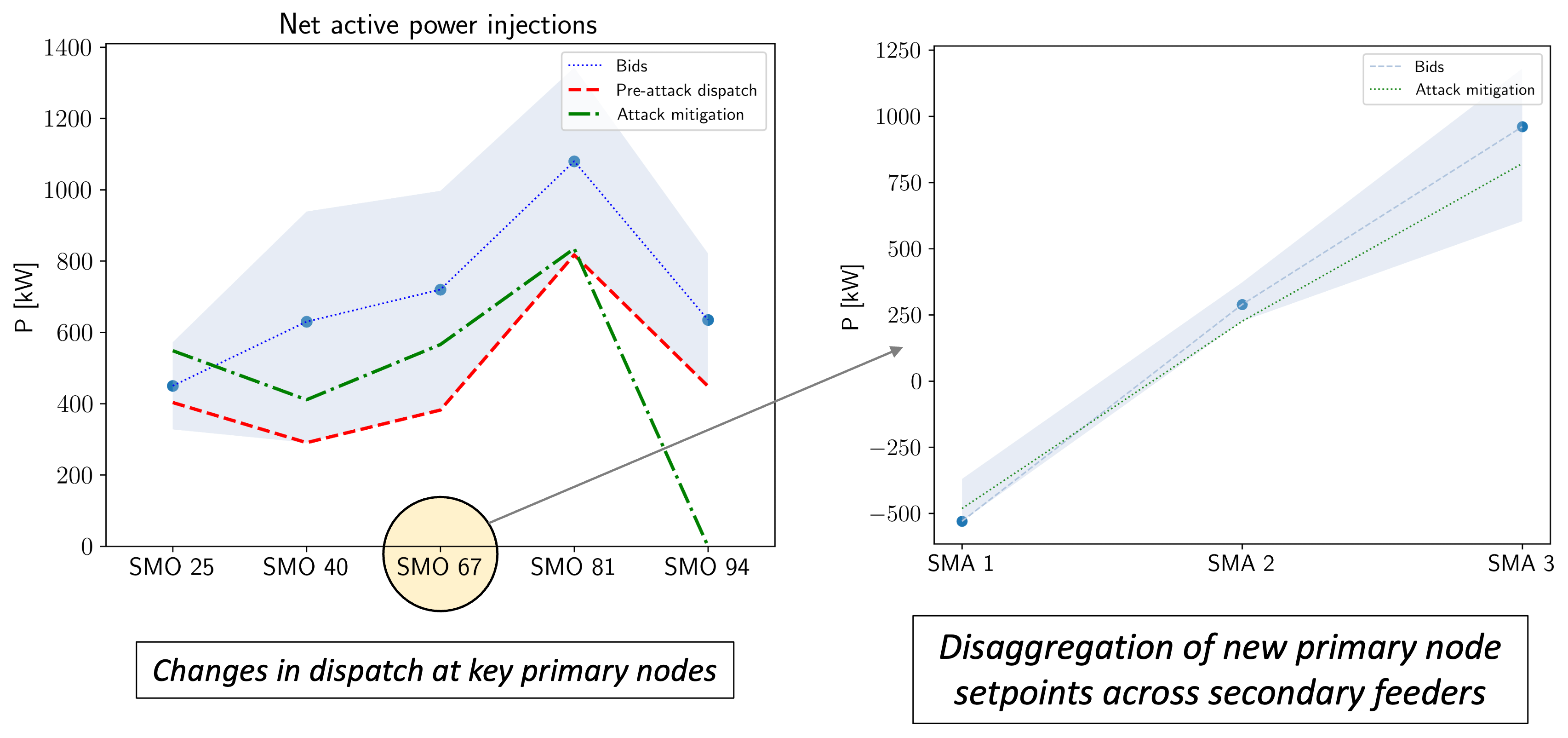}
\caption{Mitigation of small-scale attack 2a. \label{fig:attack_small}}
\end{figure}

This corresponds to a case where there are five large distributed generators in the modified IEEE-123 system, one of which (at SMO node 94) is taken offline. Here, we see that the remaining four SMO nodes (25, 40, 67, 81) have more than enough remaining generation capacity to meet the shortfall caused by the attack. Without mitigation, the attack would have resulted in an additional import of about 261 kW from the main grid. However, by utilizing the upward flexibility of remaining SMOs, we are able to fully resolve the attack and bring the total power imported back to pre-attack levels. The left figure in \cref{fig:attack_small} shows the results of the PM dispatch before the attack and after attack mitigation for the five key SMO nodes of interest. The plot also shows the SMO's bids into the PM, with the dashed blue line being the baseline injection bid and the blue-shaded region representing the upward/downward flexibility around it. The right figure shows the results of the SM re-dispatch after the attack mitigation and PM re-dispatch for SMO 67 as an example. It disaggregates the new setpoints among its three SMAs, with SMA 1 being a net load while SMAs 2 and 3 are net generators.

\subsection{Effects of Attack 2b on prices and economic implications\label{sec:attack_2b_details}}

\begin{figure}[htb]
\centering
\includegraphics[width=0.6\columnwidth]{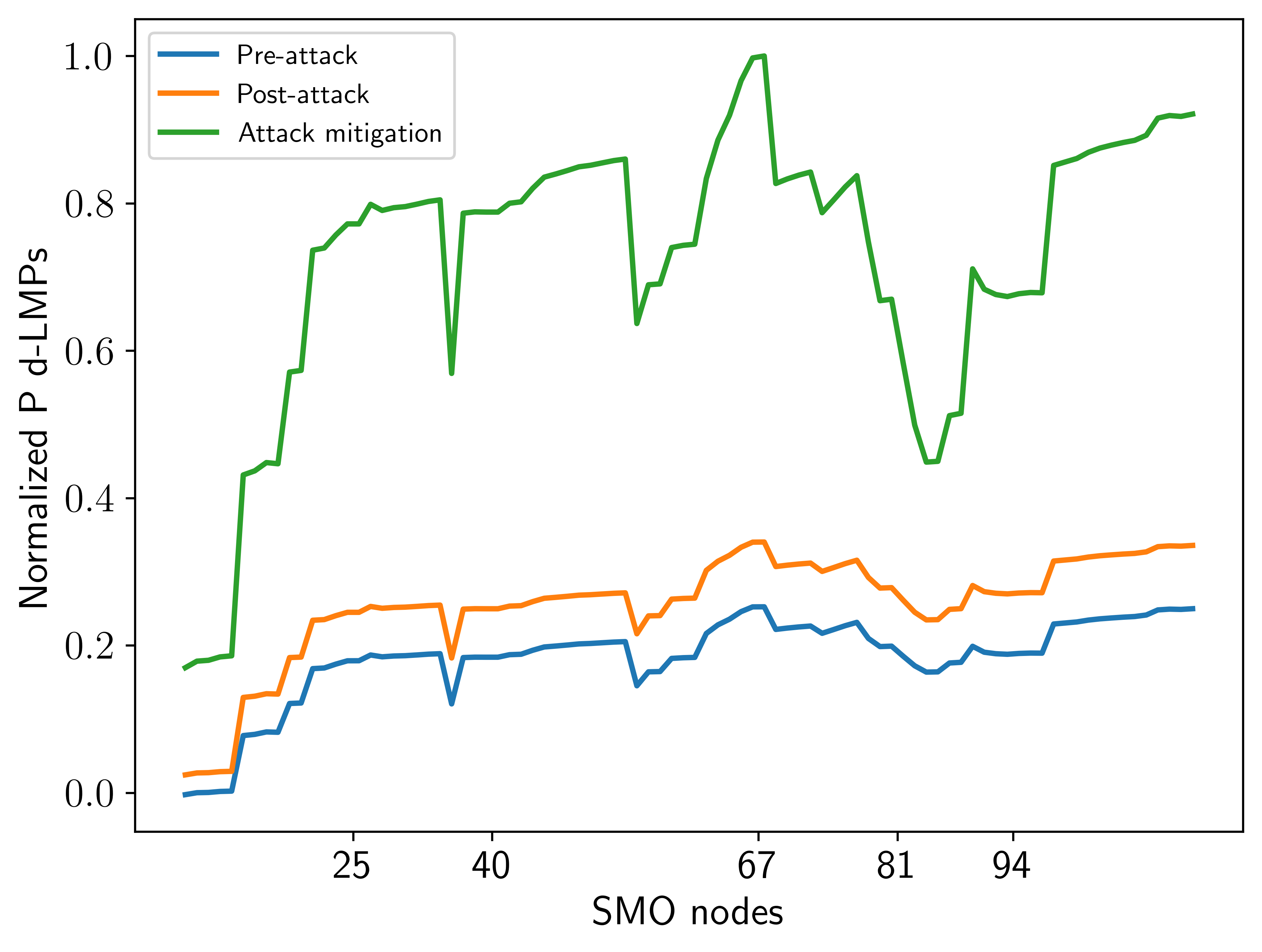}
\caption{Effects of large-scale attack and mitigation on nodal d-LMPs at SMO nodes.\label{fig:attack_large_prices}}
\end{figure}
We can obtain the electricity prices in the PM from the dual variables associated with the power balance constraints in \cref{eq:pmo}. We refer to these as distribution-LMPs (d-LMPs) at each node (with an SMO) in the primary feeder. We compared the normalized d-LMPs for active power before and after the attack, as well as post-attack mitigation, shown in \cref{fig:attack_large_prices}. As intuitively expected, we see that nodal prices increase throughout the grid after the attack and rise even further after implementing the attack mitigation steps, indicating that the loss of some local generation makes it more expensive to satisfy network constraints and results in sub-optimal solutions. The pre-attack and post-attack prices have nearly the same spatial profile across all the SMO nodes, with the post-attack values essentially being higher by an offset. This makes sense because the d-LMP variations between nodes are influenced by congestion on lines. In the attack case without mitigation, the shortfall caused by the attack would've been compensated for entirely by importing extra power from the grid, and thus the relative congestion variation over the rest of the network remains largely unchanged. The price trends after attack mitigation look more different since the changes in power flow and congestion (resulting from the PM re-dispatch) are not uniform throughout the network. Notably, we see that the prices are significantly more volatile, especially around the nodes affected by the attack. The price also peaks at node 67 - this makes sense since it has the highest increase in injection after attack mitigation, which in turn worsens congestion in lines connected to it.

Another important consideration is the impact of our mitigation approach on the different market participants, i.e., the SMOs and SMAs themselves. The objective function update rules from \cref{sec:costcoeff1,sec:costcoeff2} generally imply that these local resources will be compensated less per unit (kW or kVAR) of grid support they provide, either in terms of load flexibility or generation dispatch. It may also lead to significant load shifting and curtailment in order to meet grid objectives, which can reduce the overall utility of end-users. We also need to more carefully study the distributional impacts of such methods since they may end up disproportionately negatively impacting certain groups of customers or prosumers, which could in turn have important implications for energy affordability, equity, and fairness.

\section{FURTHER VALIDATION DETAILS OF ALL ATTACK SCENARIOS\label{sec:valid_other_attacks}}

\subsection{Validation of Attack 1a by PNNL using HELICS\label{sec:attack1a_pnnl}}

\begin{figure}[htb]
  \centering
 \includegraphics[width = \columnwidth]{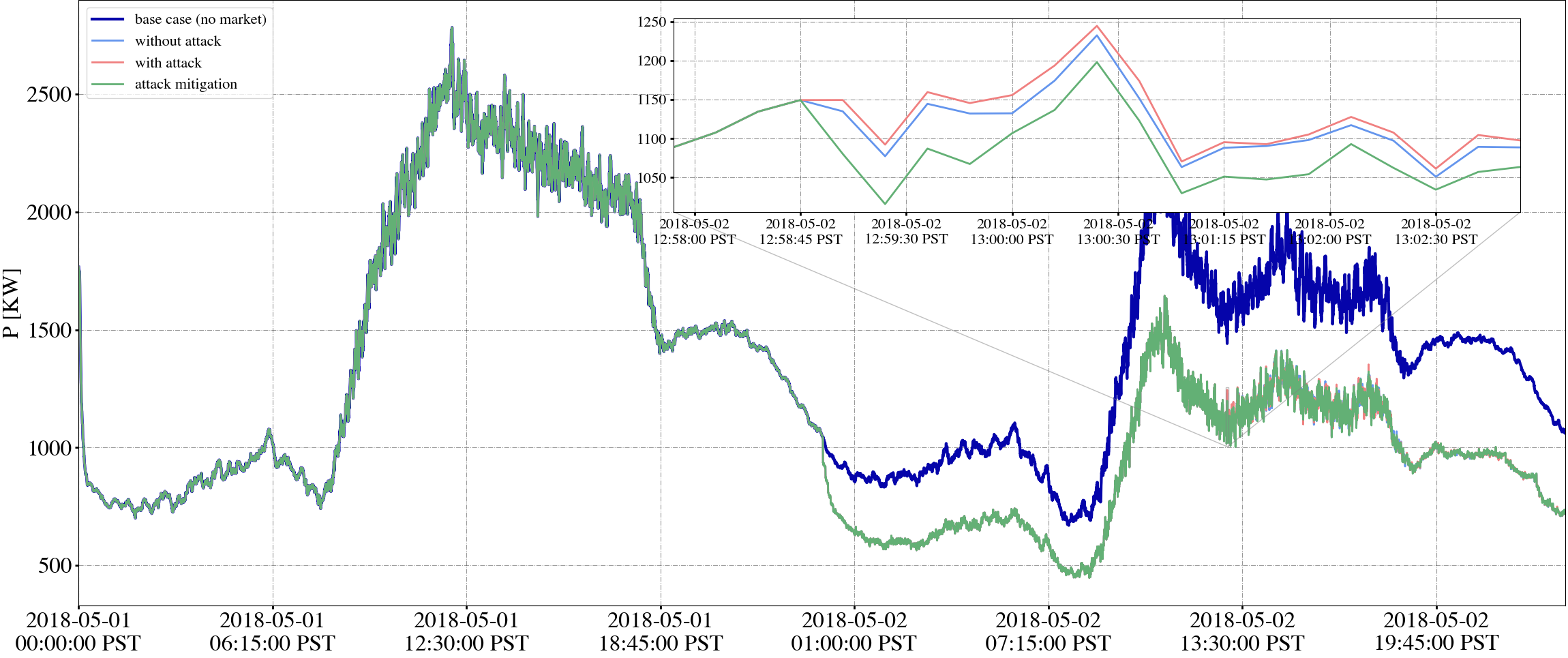}
  \caption{Validation of attack 1a mitigation effects of the EUREICA framework using HELICS, showing system load over 48 hours.\label{fig:attack2-total-load-48hours}}
\end{figure}

This attack artificially increases the load at several devices throughout the network. \cref{fig:attack2-total-load-48hours} shows the effects of attack 1a and mitigation on the total feeder load over the course of the 48-hour simulation. This was performed using the HELICS platform and a GridLAB-D model (see \cref{sec:val_pnnl} for details). Firstly, we notice that the application of the LEM during day 2 generally results in curtailment of net load by leveraging DER flexibility, relative to day 1 (when the market is not used). Secondly, upon zooming in on the attack period (around 13:00 PST), we see that the LA attack increases the total system load. However, attack mitigation is quickly able to reduce the system load using flexibility and help the system recover.

\subsection{Validation of Attack 1b by PNNL using HELICS\label{sec:attack1b_pnnl}}
\begin{figure}[htb]
  \centering
 \includegraphics[width = \columnwidth]{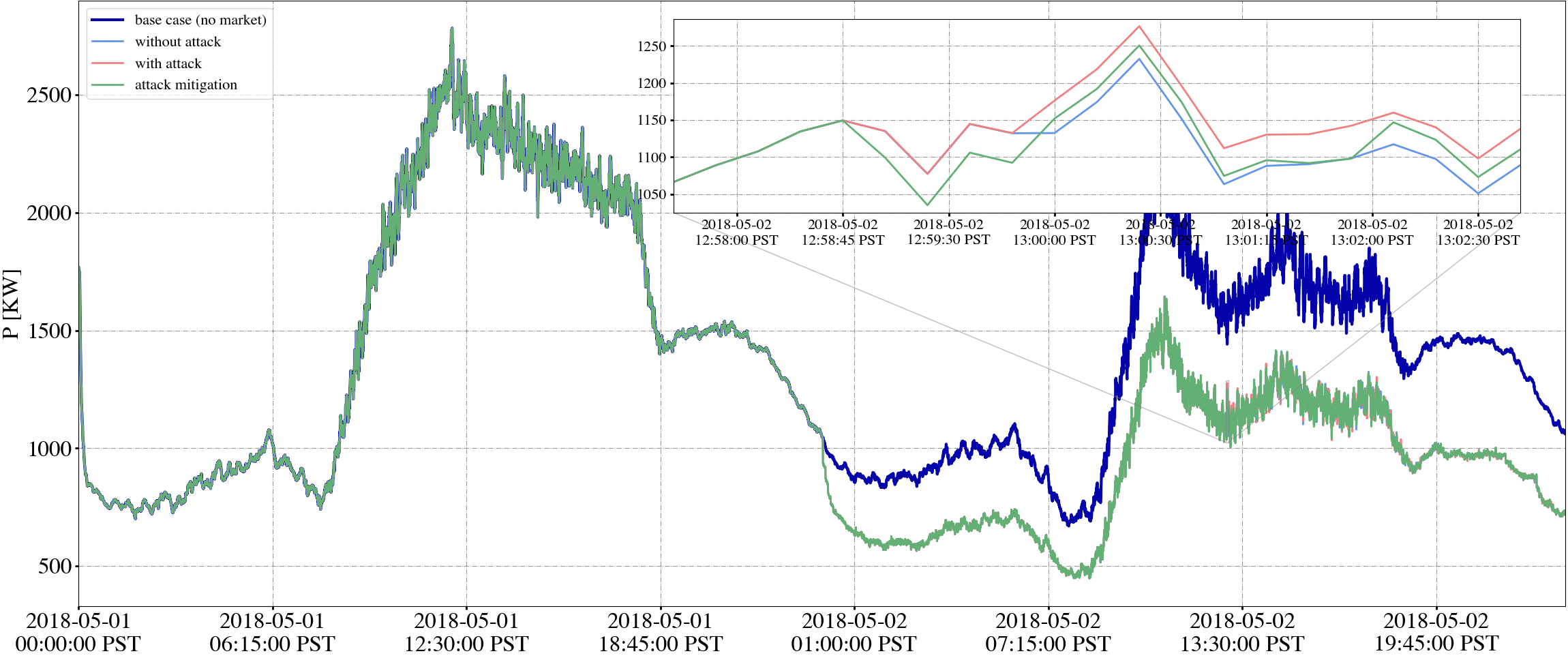}
  \caption{Validation of attack 1b mitigation effects of the EUREICA framework using HELICS, showing system load over 48 hours.\label{fig:attack1-total-load-48hours}}
\end{figure}

We utilized the HELICS-based co-simulation platform to simulate this use case in which several of the distributed generation resources are being disconnected leading to about a $44$ kW loss in generation. That is accomplished in the model simulation by taking offline the PVs at the buses as indicated in \cref{fig:attack1b_gen}. However, with the SA enabled through the market module, the SM agents are informed about how much they need to adjust their flexible assets, which results in an approximate $36$ kW load and local generation alteration after attack mitigation to counterbalance the distributed generation loss, as seen in \cref{fig:attack1b_curtail}.
The effect of the market integration on the total system load during the second 24-hour period of a 48-hour simulation is depicted in \cref{fig:attack1-total-load-48hours}. In particular, the window details the attack that happens around 13:00 on the second day and how the total flexible load is manipulated to mitigate the need for increased generation demand from the main grid. We see that the attack mitigation reduces the impact of the attack by lowering the total feeder load and bringing it back down closer to the values if there wasn't an attack. However, we note that even after mitigation, the load is still slightly higher than the `without attack' case for some periods but much lower than the `with attack' case.

\subsection{Validation of Attack 3 by PNNL using HELICS\label{sec:attack3_pnnl}}

The enhanced EUREICA IEEE 123-node feeder is covered partially by the local distributed generation resources, that is the PVSs and BESSs, and mainly from the main transmission and generation grid through a connection at node 150, as shown in \cref{fig:attack4-diagram}. Moreover, the system has available 2 large diesel generators at buses 48 (150 kVA rated capacity) and 66 (1 MVA rated capacity), respectively, that could be called upon to serve loads in case of adversarial events. Also, a set of switches between certain nodes of the system configures it into 7 areas that could be isolated in certain scenarios to be able to serve critical loads, as in \cref{fig:attack4-diagram}. The initial configuration of the switches is given in \cref{tbl:ieee-123-node-switches}.

\begin{table}[htb]
  \caption{Original switch configuration in the EUREICA IEEE 123-node test feeder.}
  \label{tbl:ieee-123-node-switches}
  \centering
  \begin{tabular}{lll}
    Node A & Node B & Switch status\\
    \hline \hline
    13 & 152 & CLOSED\\
    18 & 135 & CLOSED\\
    60 & 160 & CLOSED\\
    61 & 610 & CLOSED\\
    97 & 197 & CLOSED\\
    150 & 149 & CLOSED\\
    250 & 251 & OPEN\\
    450 & 451 & OPEN\\
    300 & 350 & OPEN\\
    95 & 195 & OPEN\\
    54 & 94 & OPEN\\
    151 & 300 & OPEN\\
    13 & 18 & CLOSED\\
    86 & 76 & CLOSED\\
    48 & 48\_dg & OPEN\\
    65 & 65\_dg & OPEN\\
  \end{tabular}
\end{table}

\begin{figure}[htb]
  \centering
  \includegraphics[width =\columnwidth]{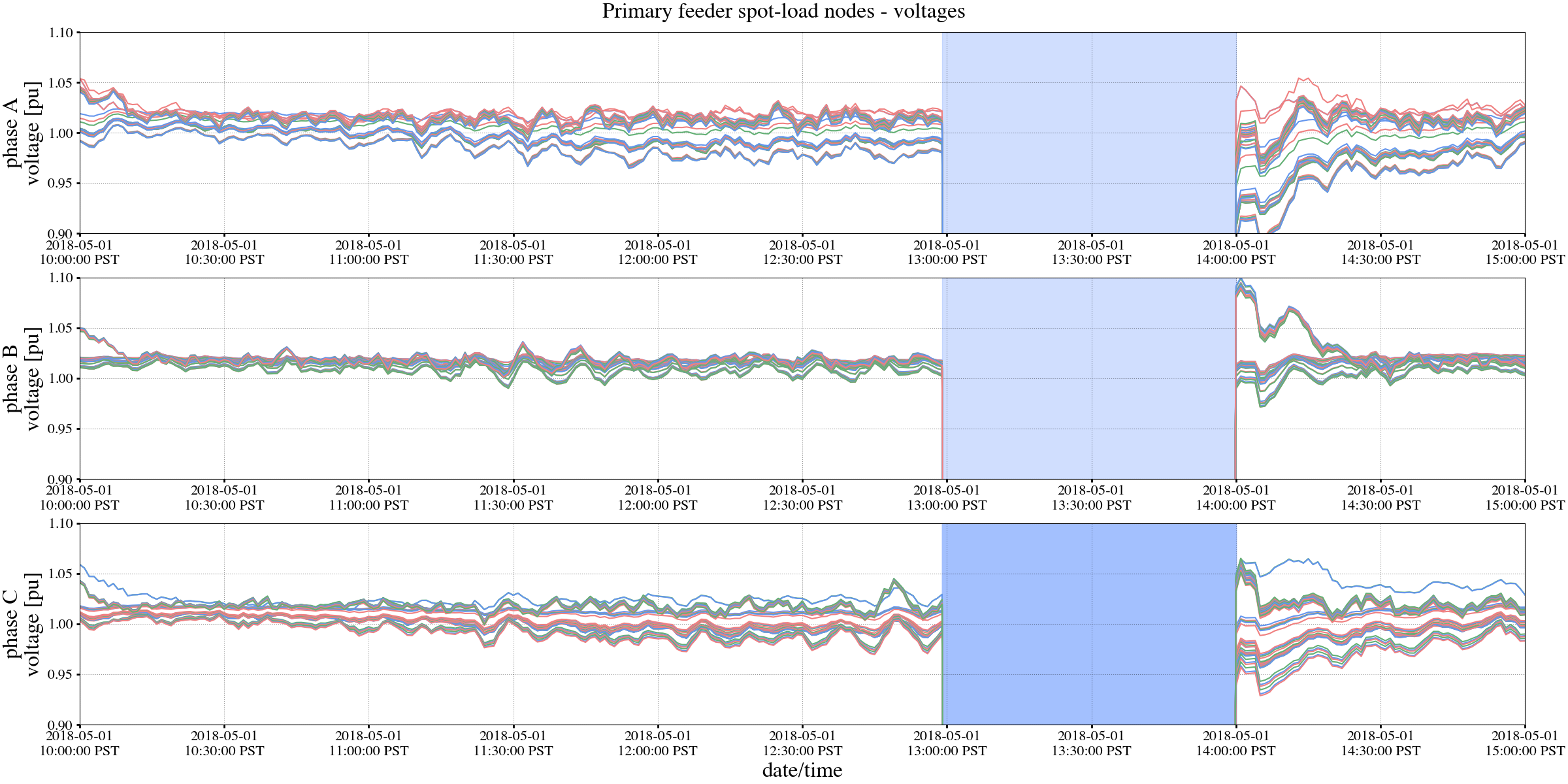}
  \caption{Black-out as a result of distribution system islanding in attack 3.}
  \label{fig:attack4-nodgnoreconf-voltages}
\end{figure}

The validation scenario assumes that due to an adversary event, either a cyber attack or a physical phenomenon, the distribution system gets islanded from the main grid, which is simulated by opening the switch between nodes 150 and 149 at 13:00. The system reconnection to the main grid is assumed to happen at 14:00. As expected, at 13:00 the system collapses, which is demonstrated by the sudden drop to 0 for all the spot-load bus voltages, as seen in \cref{fig:attack4-nodgnoreconf-voltages}.

\begin{figure}[htb]
  \centering
  \includegraphics[width =\columnwidth]{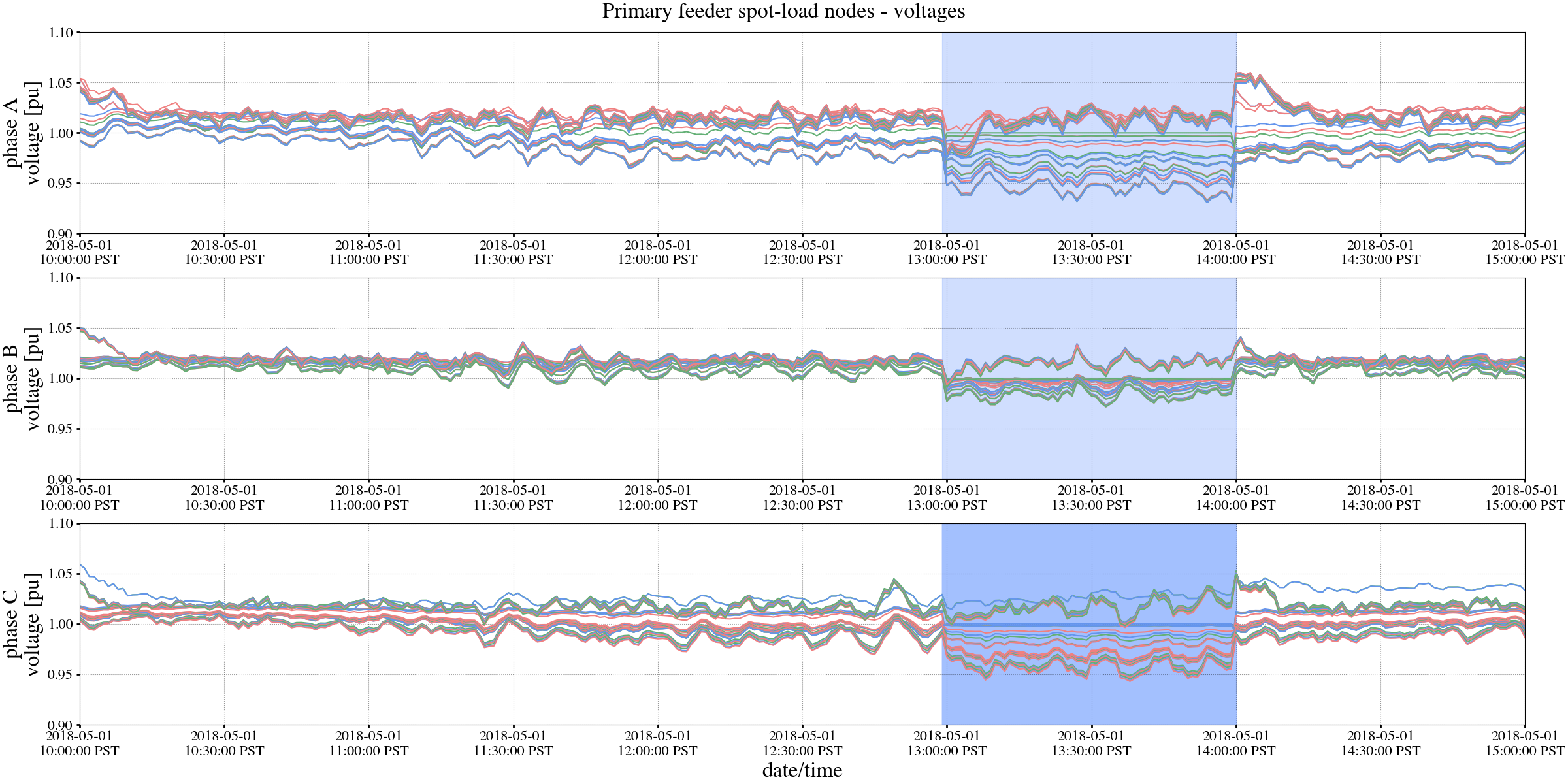}
  \caption{Voltage recovery after engaging diesel generators during attack 3.}
  \label{fig:attack4-dgnoreconf-voltages}
\end{figure}

The proposed reconfiguration and load shed approach addresses the situation created at 13:00 hours, creates situational awareness, and decides the switch statuses and loads that might need to be shed. If only the available diesel generators are brought online once the system is disconnected from the grid, by reconfiguring the corresponding switches status, the system would not be in a blackout, as seen in \cref{fig:attack4-dgnoreconf-voltages}.

\begin{figure}[htb]
  \centering
  \includegraphics[width = \columnwidth]{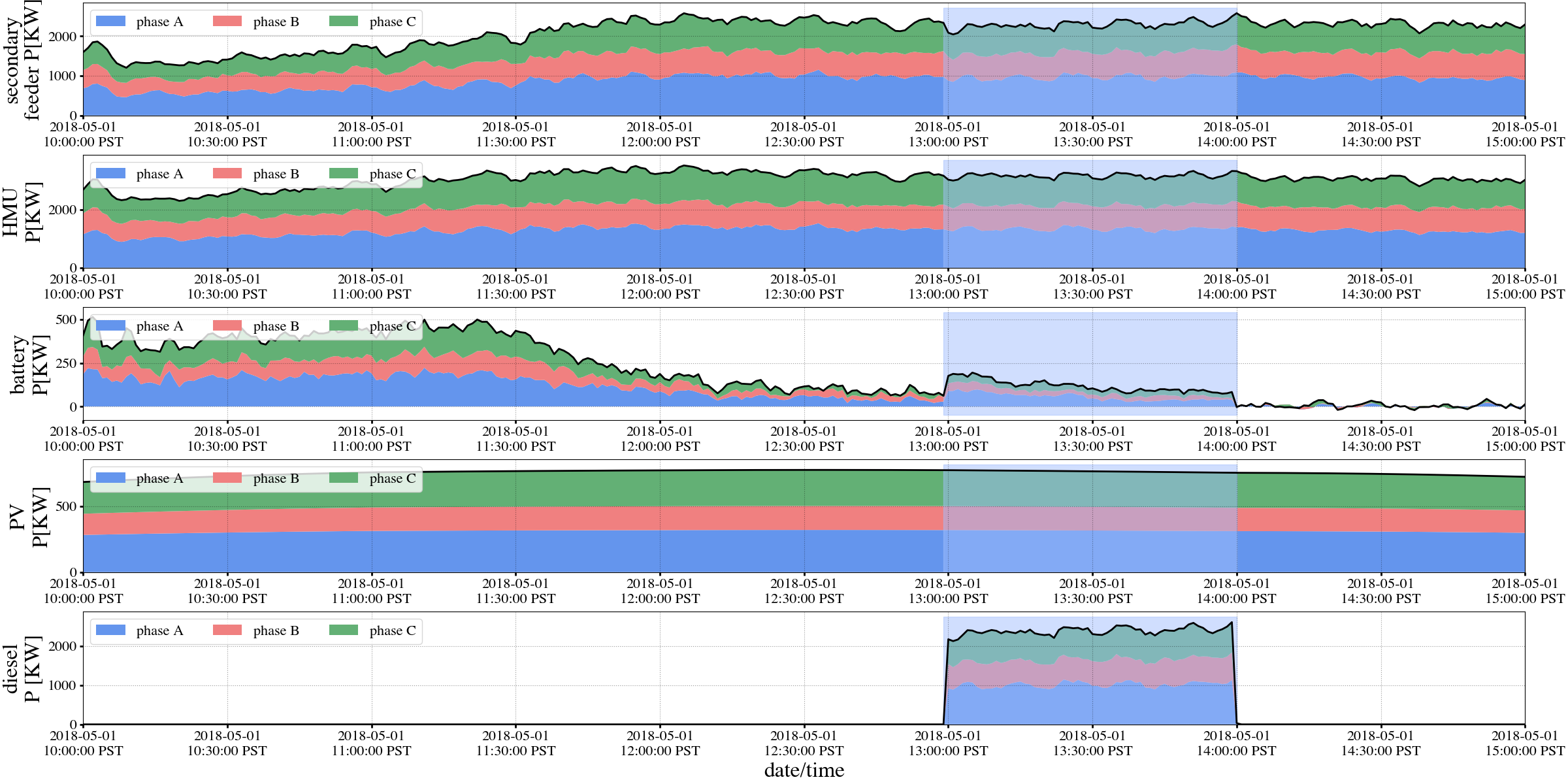}
  \caption{Demand and distributed generation without resilience-based reconfiguration during attack 3.}
  \label{fig:attack4-dgnoreconf-loads}
\end{figure}

However, as seen in \cref{fig:attack4-dgnoreconf-loads}, to supply the entire house population load (the total measurements from the house management units in the second graph from the top), even with the support of the PVs and batteries, the diesel generators would still need a total capacity of over 2 MW (as seen in bottom-most graph of \cref{fig:attack4-dgnoreconf-loads}), which is more than the maximum capacity of the model diesel generators.

Through the proposed resilience-based IoT load restoration with demand response optimization strategy, a feasible reconfiguration path is computed to open and/or close tie switches and shed either completely/partially grid edge loads to allow the available generation resources to cover the approximately $30\%$ critical load in the system, as identified in \cref{tab:distr-feeder-conf}. As seen in \cref{fig:attack4-dgreconf-loads-pnnl}, with the almost $70\%$ load shed (second graph from the top) between 13:00 and 14:00 hours, and batteries only allowed to discharge, if possible, to supply extra energy (third graph from the top), the burden on the diesel generators is significantly alleviated as they only need to ramp up to about $230$ KW.

\begin{figure}[htb]
  \centering
  \includegraphics[width=\columnwidth]{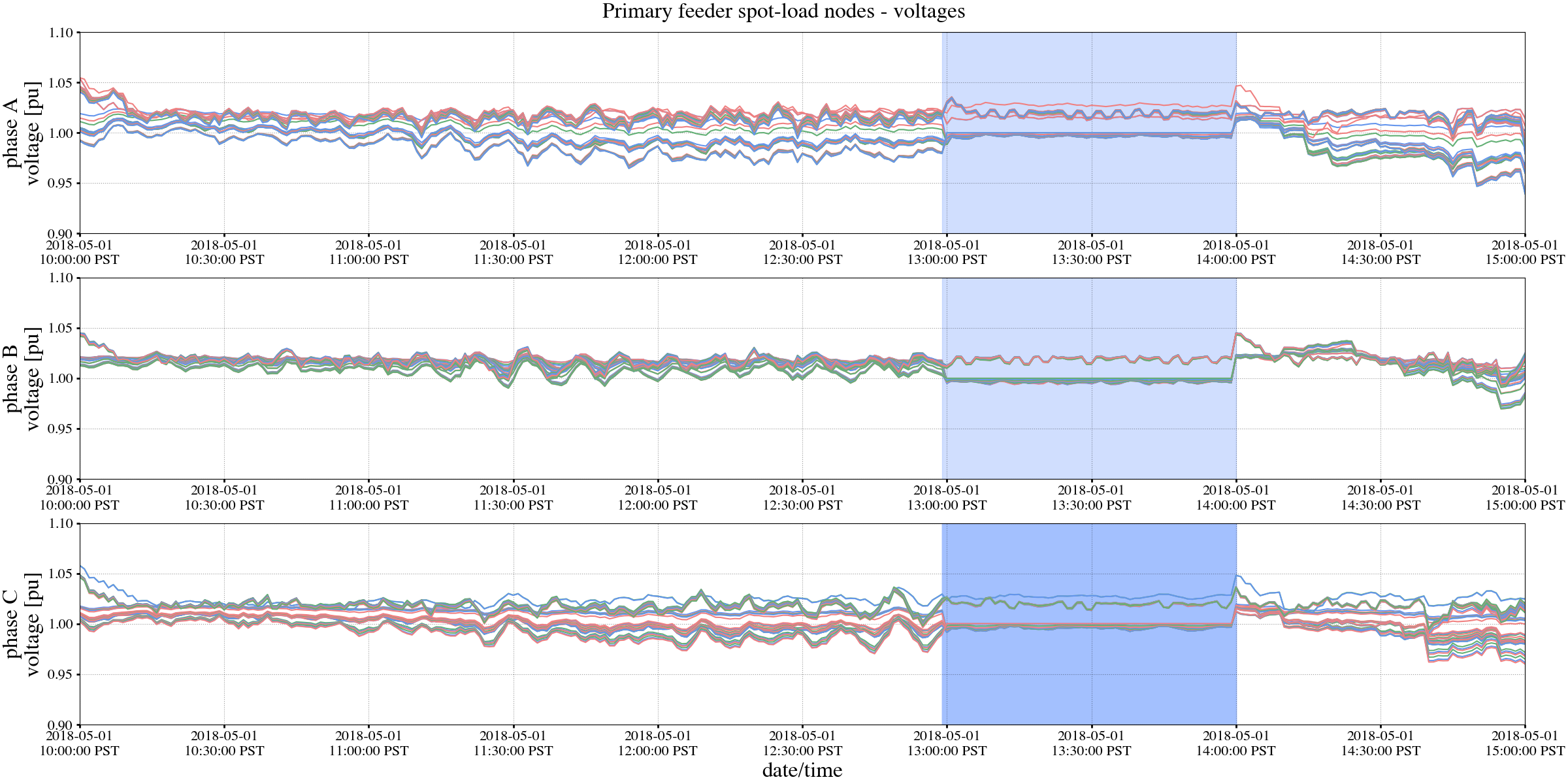}
  \caption{Voltage recovery after resilience-based reconfiguration during attack 3.}
  \label{fig:attack4-dgreconf-voltages-pnnl}
\end{figure}

For the islanded attack, the power flow redirection through switch reconfiguration as in SI \cref{fig:reconfig_alg} and load shed also helps with keeping the spot-load buses voltages within the admissible limits during the attack (\cref{fig:attack4-dgreconf-voltages-pnnl}). Moreover, by bringing the loads back online sequentially after system recovery, under-voltage problems due to load rebound are also avoided.

\subsection{Validation of Attack 1a by LTDES using DERIM-ADMS-DOTS\label{sec:attack1a_ltdes}}

\begin{figure}[htb]
  \centering
  \includegraphics[width=\columnwidth]{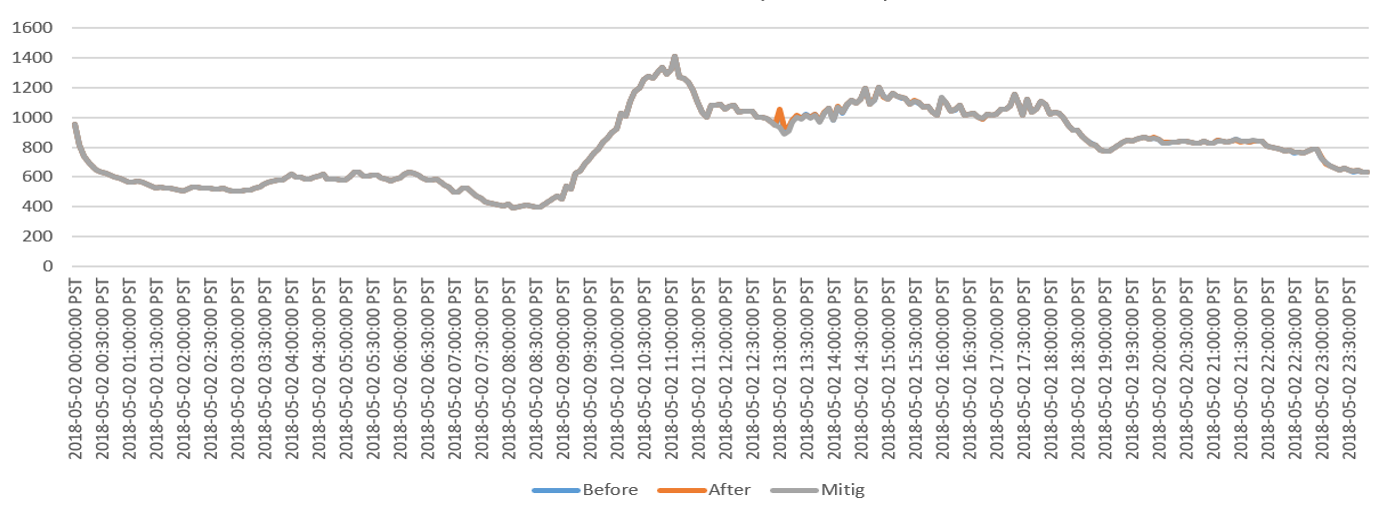}
  \caption{Effects of attack 1a on total load at feeder head over 24 h.}
  \label{fig:attack1a_ltdes_totalfeederload}
\end{figure}

\begin{figure}[htb]
  \centering
  \includegraphics[width=\columnwidth]{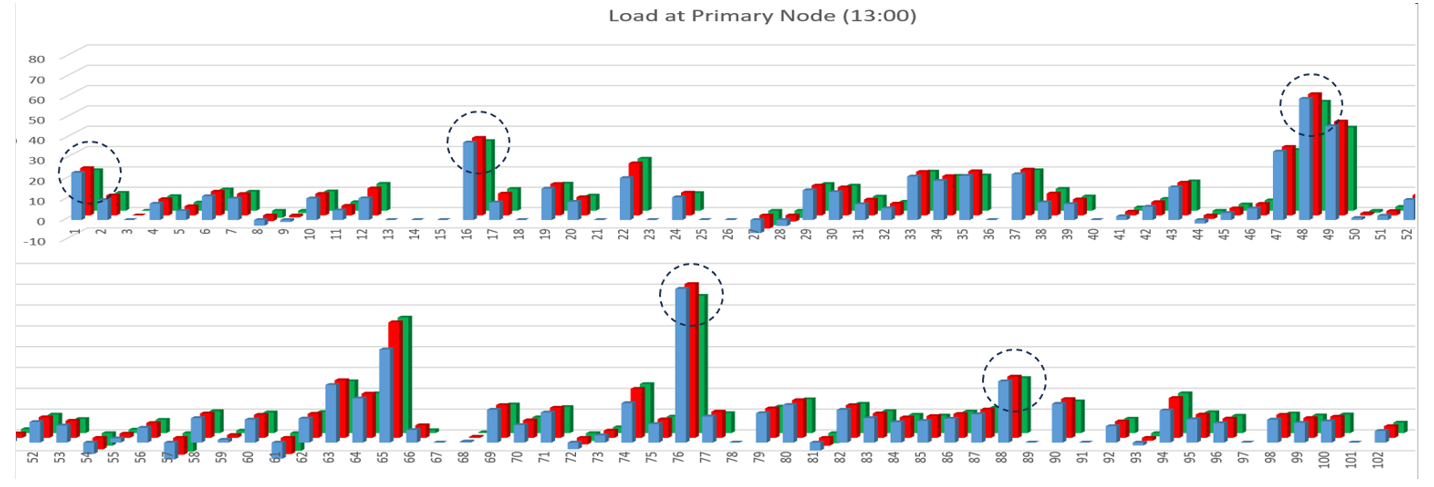}
  \caption{Load change at primary nodes during attack 1a. The values (i) without attack, (ii) with attack, and (iii) with attack mitigation are shown in the blue, red, and green bars, respectively. The SMO nodes providing the most flexibility are circled.}
  \label{fig:attack1a_ltdes_primarynodeloads}
\end{figure}

While \cref{figure18-ltdes} zooms in on the period around the attack timestep, \cref{fig:attack1a_ltdes_totalfeederload} shows the total feeder head load over the entire 24 h simulation horizon. We can clearly see the blip at 13:00 PST indicating the impact of the attack. \cref{fig:attack1a_ltdes_primarynodeloads} shows the effects of attack and mitigation on the net load at all the SMO primary nodes. This shows that the DERIM-ADMS-DOTS validation produces similar results to our market simulation in \cref{fig:attack1a_load} and \cref{fig:attack1a_curtail}. The attack increases the load at the following nodes: 12, 17, 21, 36, 65, 75, 95, 105, 112, and 113. The majority of load curtailment for mitigation is contributed by the larger loads at nodes 1, 16, 48, 76, and 88.

\subsection{Validation of Attack 1b by LTDES using DERIM and ADMS-DOTS\label{sec:attack1b_ltdes}}

\begin{figure}[htb]
  \centering
  \includegraphics[width=\columnwidth]{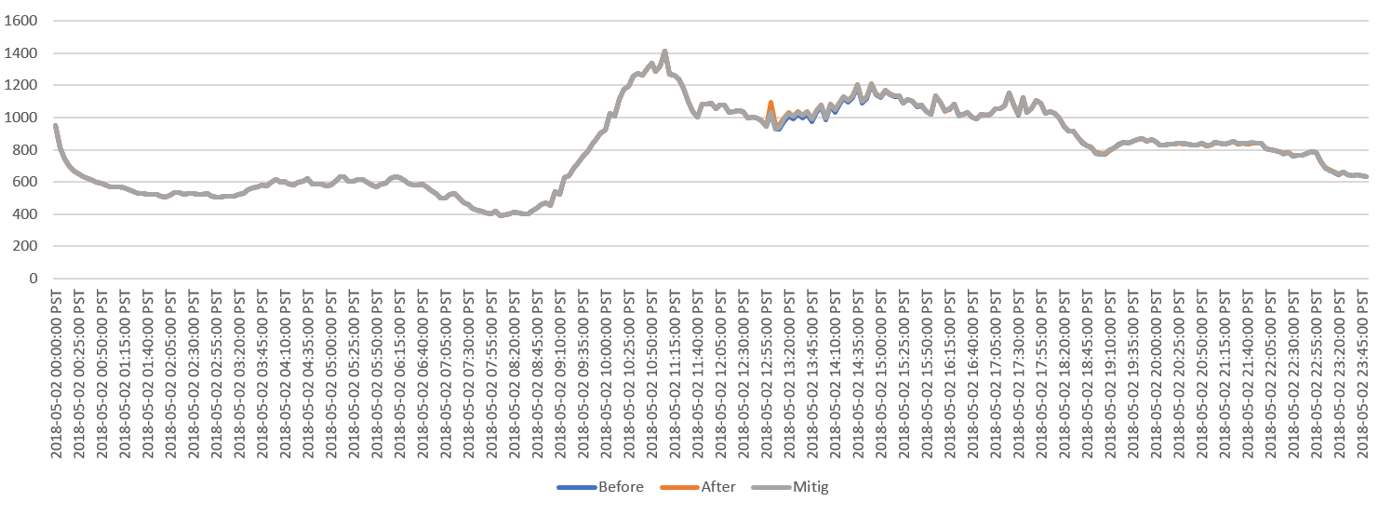}
  \caption{Effects of attack 1b on the total load at the feeder head over 24 h.}
  \label{fig:attack1b_ltdes_totalfeederload}
\end{figure}

\begin{figure}[htb]
\centering
\includegraphics[width=\columnwidth]{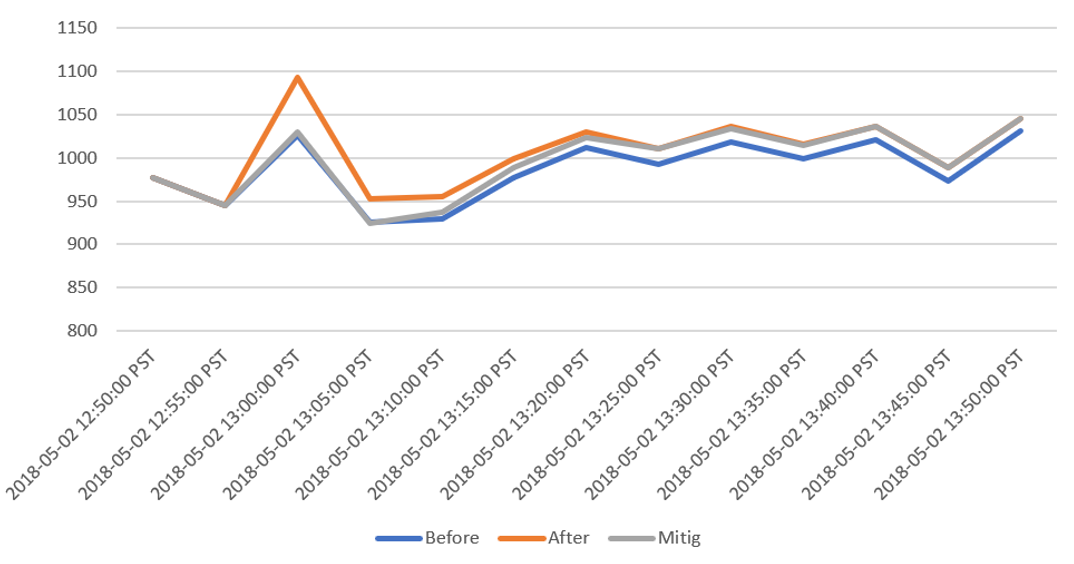}
\caption{LTDES validation of attack 1b in the DERIM-ADMS platform, showing total power import at the substation around the attack time at 13:00.} \label{fig:attack1b_ltdes_attacktime}
\end{figure}

\begin{figure}[htb]
  \centering
  \includegraphics[width=\columnwidth]{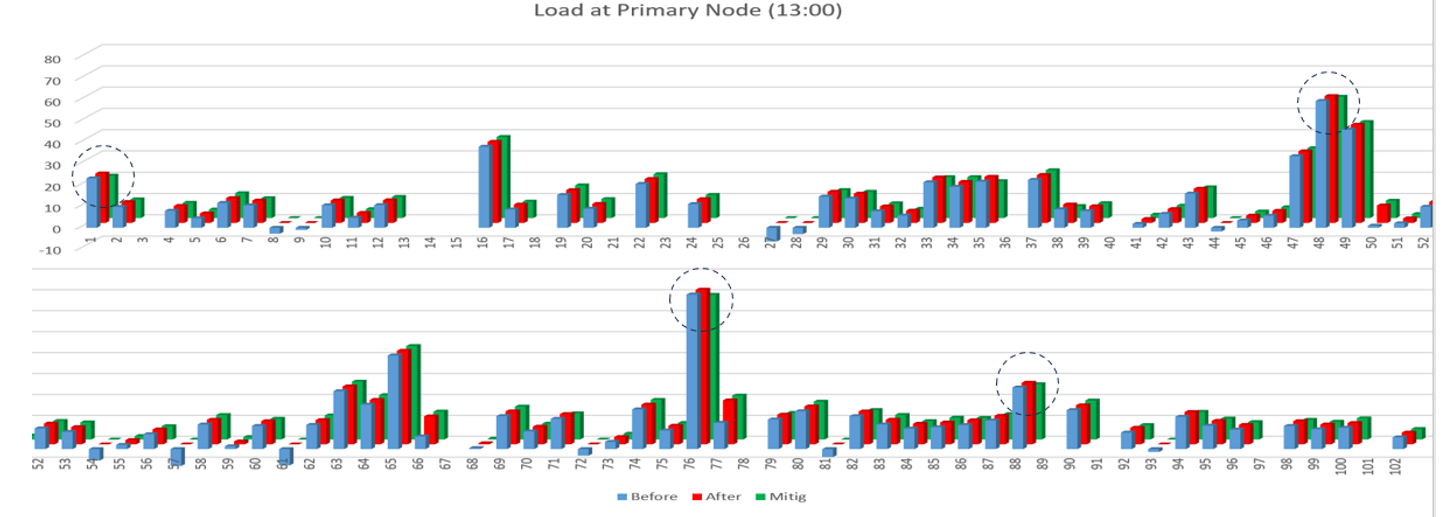}
  \caption{Load change at primary nodes during attack 1b. The values (i) without attack, (ii) with attack, and (iii) with attack mitigation are shown in the blue, red, and green bars, respectively. The SMO nodes providing the most flexibility are circled.}
  \label{fig:attack1b_ltdes_primarynodeloads}
\end{figure}

\cref{fig:attack1b_ltdes_totalfeederload} shows the effects of the DG attack 1b on the total system load over the full 24 h simulation while \cref{fig:attack1b_ltdes_attacktime} zooms in on the period around the attack. We see that without the market-based mitigation, the feeder demand would have jumped by 68 kW due to the attack. However, with mitigation, the attack impact is minimal since there's only a 4 kW increase in feeder demand. \cref{fig:attack1b_ltdes_primarynodeloads} shows the changes in net injections at all primary nodes during attack 1b. This essentially shows that we leverage flexibility from several primary nodes across the feeder, producing results similar to those shown in SI \cref{sec:attack1b} and \cref{sec:attack1b_rs_mit}. The attack causes the following DG nodes to lose power and go offline: 9, 28, 45, 55, 56, 58, 62, 73, 82, and 94. The following flexible load nodes contribute a majority of the curtailment needed to mitigate: 1, 48, 76, and 88.

\begin{figure}[htb]
\centering
\includegraphics[width=\columnwidth]{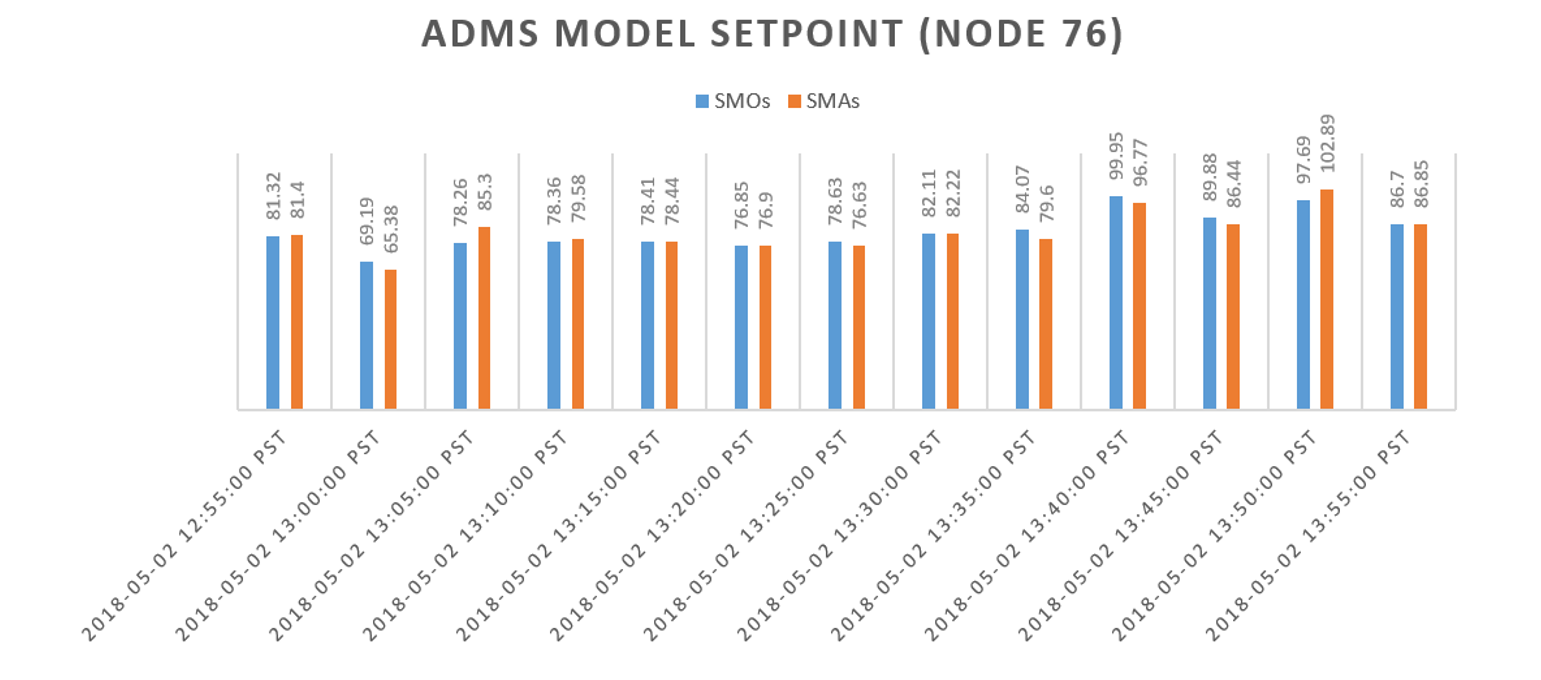}
\caption{Forecasted values of SMO and SMA setpoints at primary node 76 during attack 1b mitigation.\label{fig:attack1b_ltdes_smosma}}
\end{figure}

\cref{fig:attack1b_ltdes_smosma} compares the load setpoints at the SMO level (updated every 5 minutes) for node 76 versus the aggregated setpoints over all the SMAs at this node (cleared every 1 minute). Although these are largely similar, there are some slight differences between the two values. Thus, it may make more sense to utilize the more precise SMA setpoints directly for the ADMS simulation.

\begin{figure}[htb]
\centering
\includegraphics[width=\columnwidth]{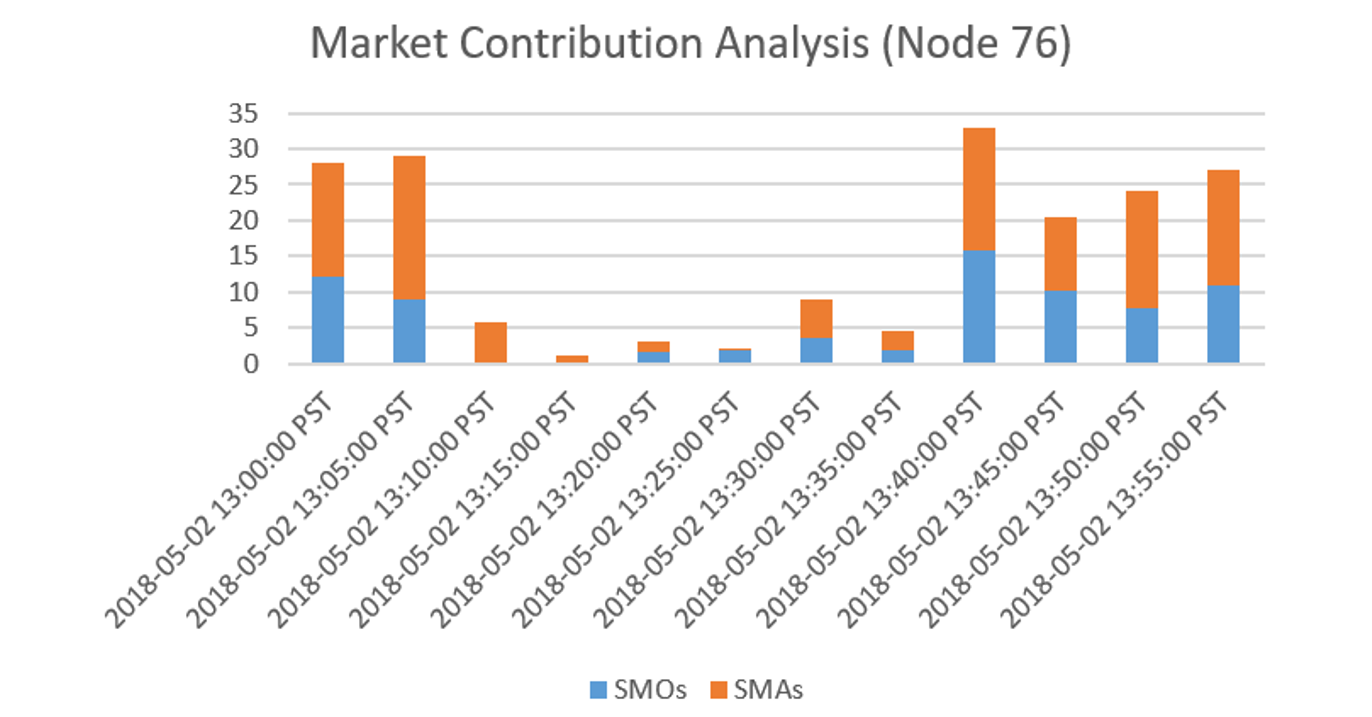}
\caption{Comparison of forecasted changes in SMO and SMA setpoints due to attack 1b mitigation.\label{fig:attack1b_ltdes_smpmcontrib}}
\end{figure}

We also did some further analysis on the role of the SM and PM in attack mitigation. \cref{fig:attack1b_ltdes_smpmcontrib} compares the contributions of the setpoints of the SMOs (5 minutes) and the SMAs (1 minute). The blue bar shows the 5-minute setpoint changes expected from the SMOs, while the orange bar shows the 1-minute setpoint changes at the SMA level. We see that the SM clearing every minute and the associated SMA setpoint changes contribute more toward the overall primary load adjustment when compared to the SMO-level changes alone.

\subsection{Validation of Attack 3 by LTDES using DERIM and ADMS-DOTS\label{sec:attack3_ltdes}}

\subsubsection{Case 1: Critical loads distributed across the feeder\label{sec:attack3_ltdes1}}

\begin{figure}[htb]
\centering
\includegraphics[width=\columnwidth]{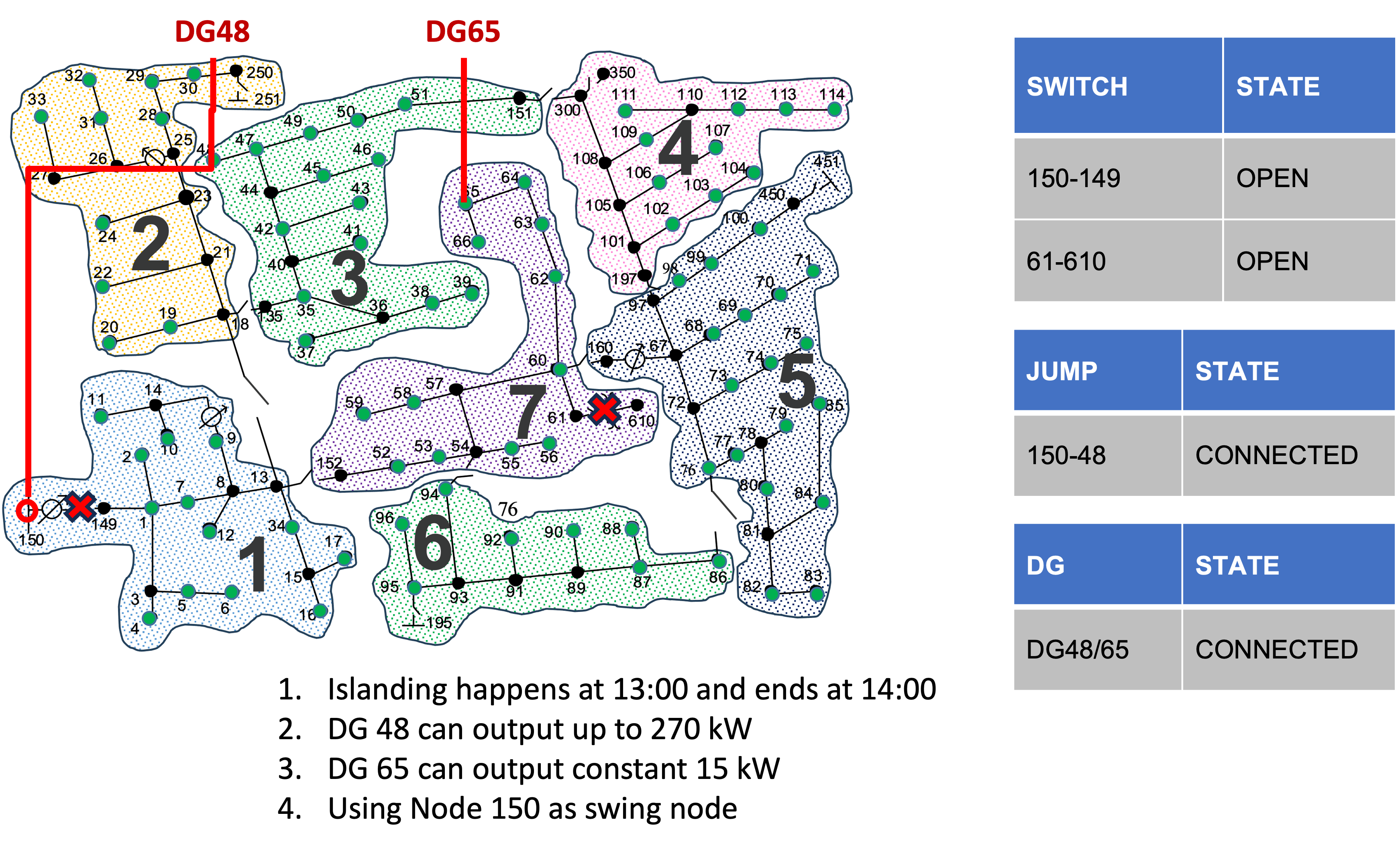}
\caption{Switch setting changes and network reconfiguration in the case when there are critical loads throughout the feeder.\label{fig:attack3_ltdes_islanded_diagram}}
\end{figure}

\cref{fig:attack3_ltdes_islanded_diagram} shows the new switch settings and updated topology after applying the resilience-based reconfiguration during attack 3. In this case, we assume that there are critical loads distributed throughout the feeder. The system is islanded from the main grid at 13:00 PST and islanding ends at 14:00.

\begin{figure}[htb]
\centering
\includegraphics[width=\columnwidth]{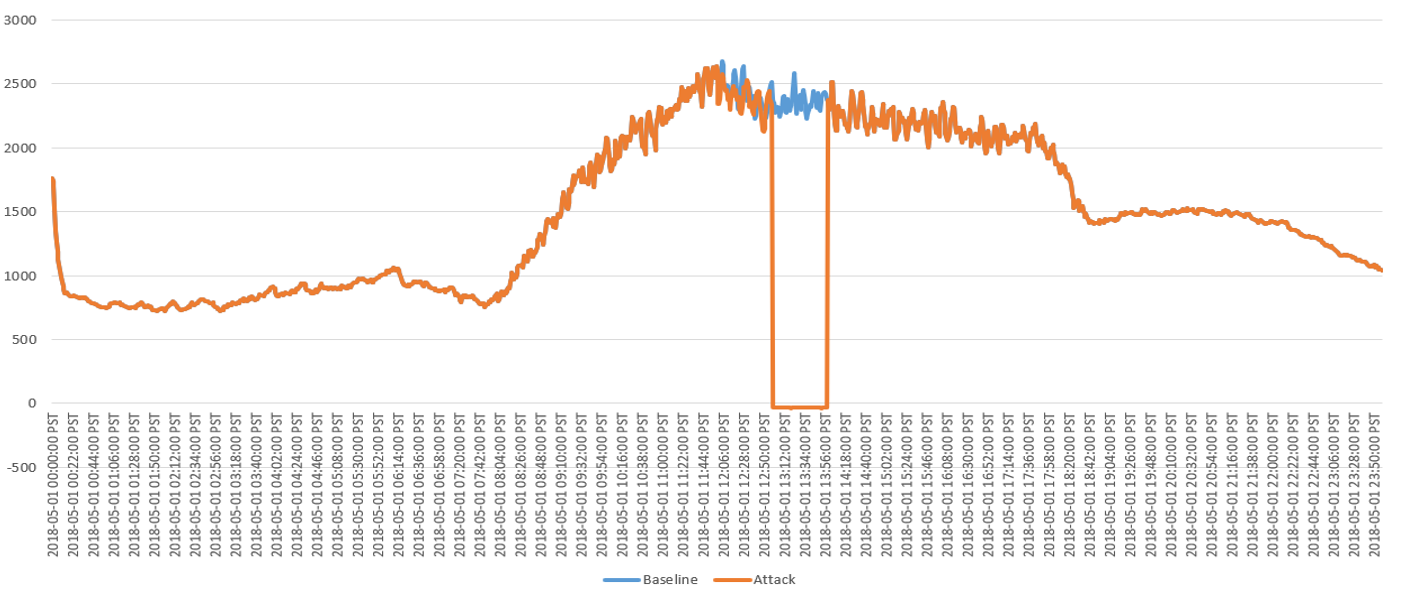}
\caption{Total feeder head load over 24 h simulation, when there are critical loads throughout the feeder.\label{fig:attack3_ltdes_islanded_feederload}}
\end{figure}

\begin{figure}[htb]
\centering
\includegraphics[width=\columnwidth]{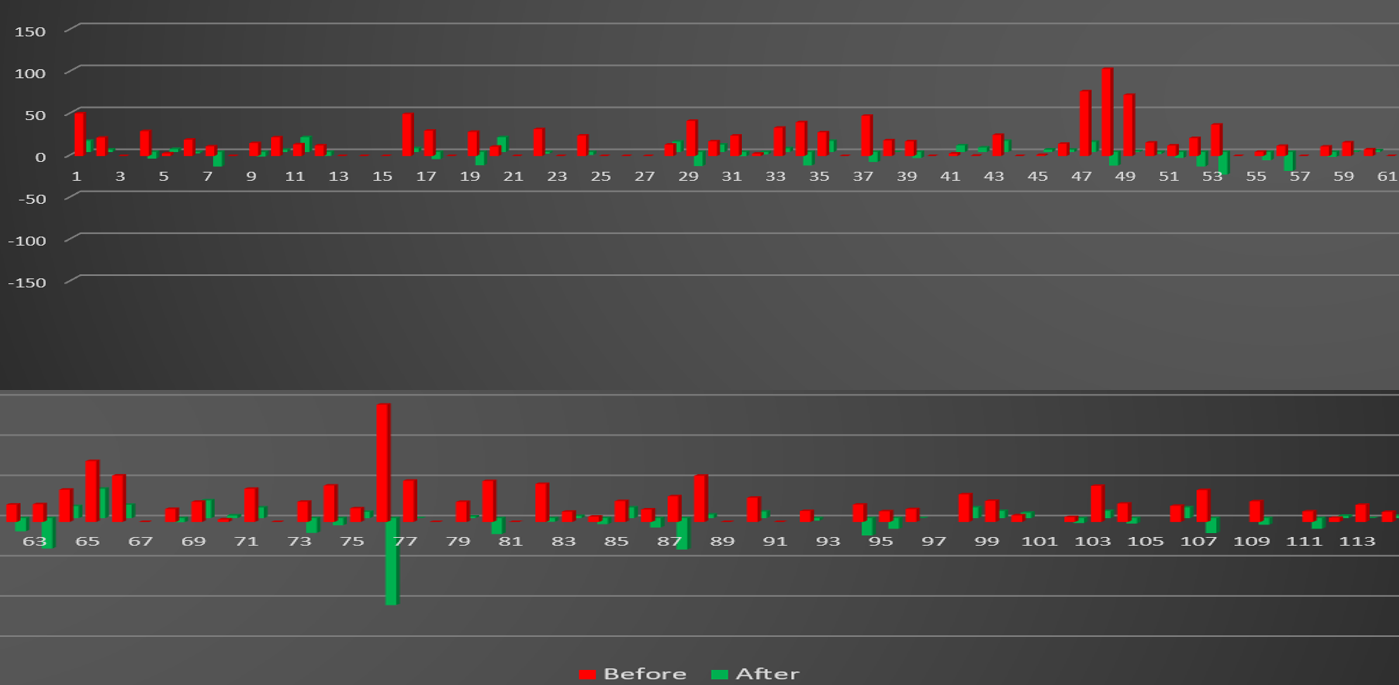}
\caption{Primary node load change during attack 3 between 12:59 and 13:00 PST, when there are critical loads throughout the feeder.\label{fig:attack3_ltdes_islanded_primarynodeload}}
\end{figure}

The DG at node 48 can output up to 270 kW while DG 65 outputs a constant 15 kW, and we use node 150 as the swing node (or slack bus) for the simulation. \cref{fig:attack3_ltdes_islanded_feederload} shows the impact of the attack on the total feeder load and \cref{fig:attack3_ltdes_islanded_primarynodeload} shows the changes in the net load at all primary nodes without the attack and with the attack (and associated reconfiguration). We see that the DGs at nodes 48 and 65 together pick up about 300 kW of the critical load, which represents about 20\% of the total baseline load. The remaining 80\% of the feeder load which is non-critical is shed (goes to zero after the attack in \cref{fig:attack3_ltdes_islanded_primarynodeload}) to maintain feasibility.

\subsubsection{Case 2: Critical loads aggregated in a single zone\label{sec:attack3_ltdes2}}

\begin{figure}[htb]
\centering
\includegraphics[width=\columnwidth]{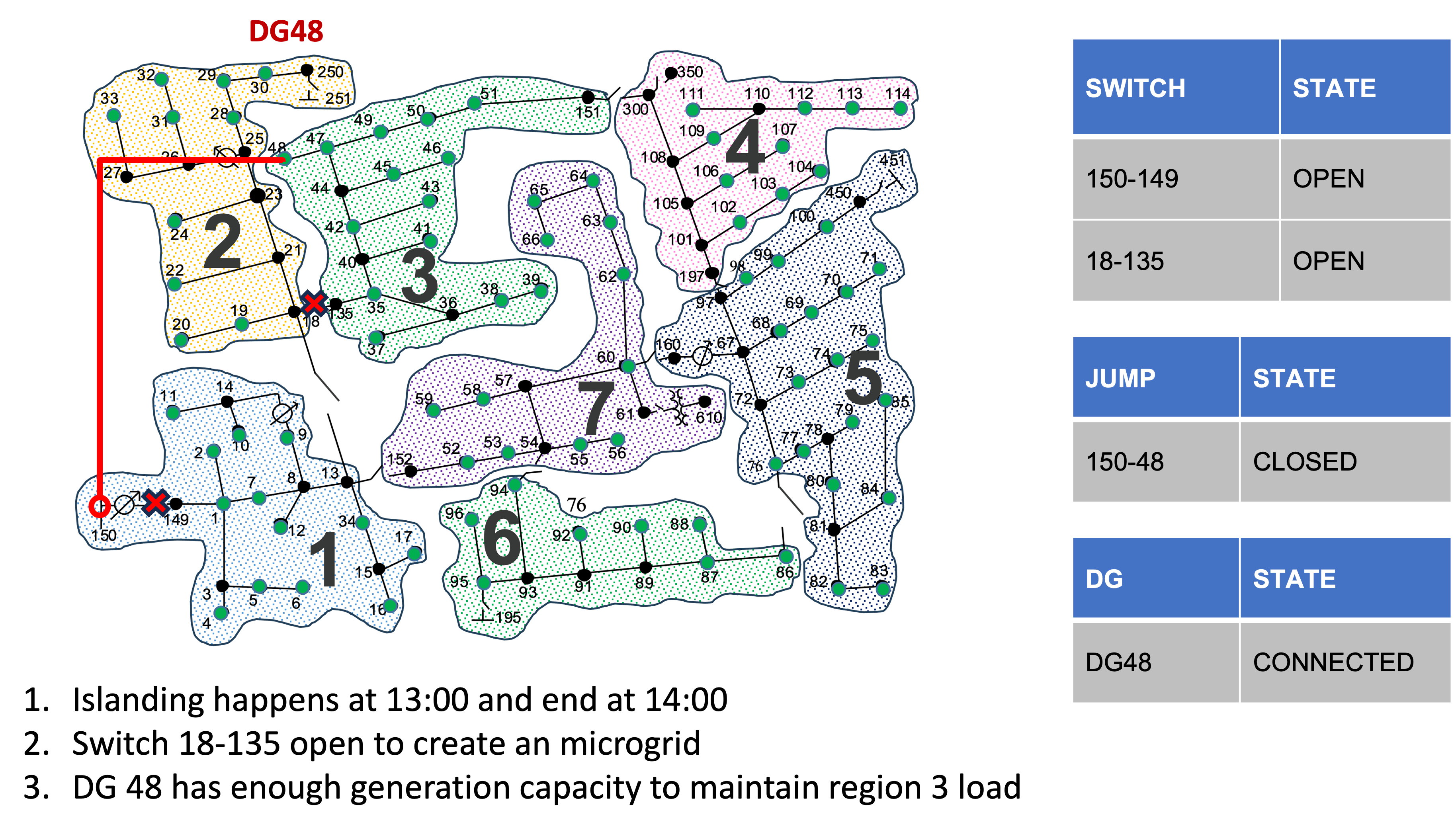}
\caption{Attack 3 case where critical loads are only located in zone 3 as a microgrid.\label{fig:attack3_ltdes_microgrid_diagram}}
\end{figure}

\cref{fig:attack3_ltdes_microgrid_diagram} shows the new switch settings and updated topology after applying the resilience-based reconfiguration during attack 3. In this case, we assume all the critical loads are concentrated only in zone 3. The system is islanded from the main grid at 13:00 PST and islanding ends at 14:00. During reconfiguration, switch 18-135 is opened so that cluster 3 becomes a microgrid. The DG at node 48 has sufficient capacity to meet all the load in zone 3 alone. Again, node 150 is used as the slack node for the simulation.

\begin{figure}[htb]
\centering
\includegraphics[width=\columnwidth]{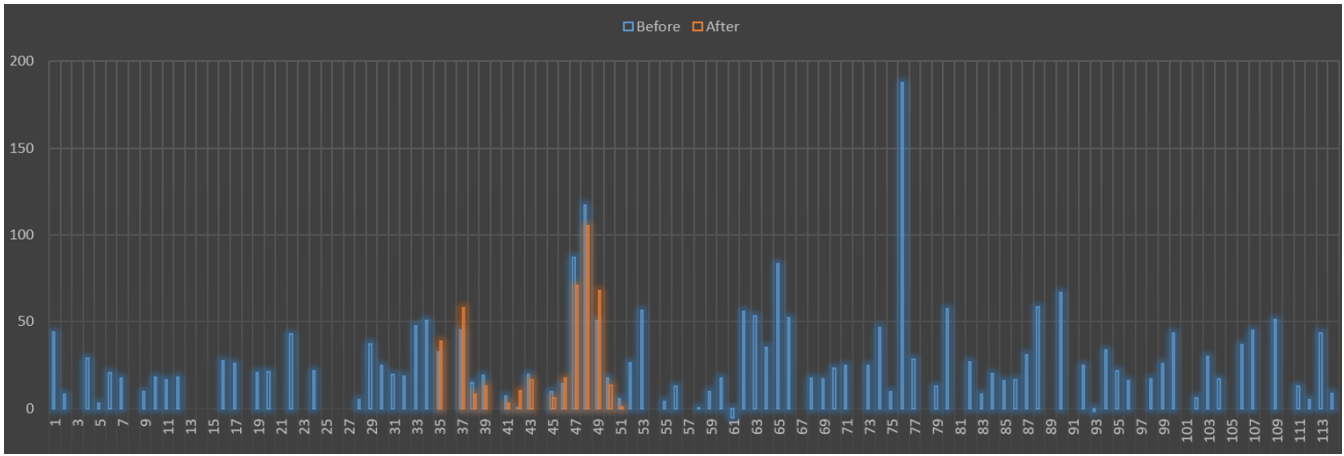}
\caption{Primary node load change during attack 3 between 12:59 and 13:00 PST, when critical loads are only located in zone 3 as a microgrid.\label{fig:attack3_ltdes_microgrid_primarynodeload}}
\end{figure}

\cref{fig:attack3_ltdes_microgrid_primarynodeload} shows the changes in the net load at all primary nodes without the attack and with the attack (and associated reconfiguration), as observed from the simulation results. We see that the DG at node 48 picks up all the expected load in zone 3 with 430 kW of generation output. 

Thus the main conclusions from the attack 3 validation using DERIM-ADMS-DOTS are as follows. Under case 1, the reconfiguration algorithm is able to restore all the critical loads throughout the islanded distribution circuit without relying on any power from the external transmission grid. In case 2, all the load in zone 3 (as a microgrid) was completely restored without any loss of load. In both cases, without the SA provided by the EUREICA framework, the control center operator at the substation would not have the necessary means to achieve restoration.

\subsection{Validation of Attack 1a by NREL using ARIES\label{sec:attack1a_nrel}}
The market structure is implemented using the same validation platform, with the primary feeders modeled on the RTDS, and secondary feeders, and below on Typhoon HIL and Raspberry Pis. In the implementation, the SMAs receive DER predictions from federated learning. The PMO and SMOs solve for primary and secondary market setpoints at each primary feeder node and secondary feeder, respectively, and then they are distributed to the SMAs, and ultimately to the IoT devices. Under nominal conditions (without an attack), the market operates with the objective of voltage regulation and minimization of power import from the main grid. 

\begin{figure}
    \centering
    \includegraphics[width=\columnwidth]{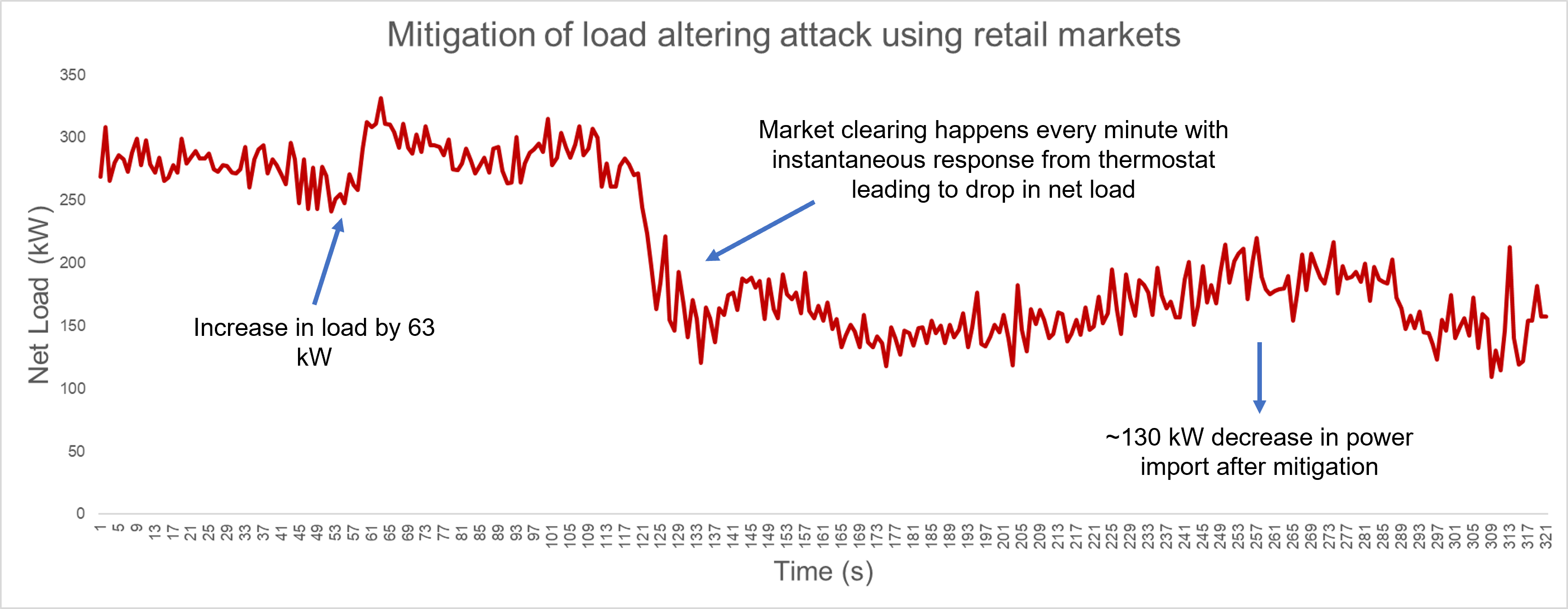}
    \caption{Implementation of market services to mitigate load increase in attack 1a.}
    \label{market_result}
\end{figure}

In the case of attack 1a, the secondary feeder load increases by 63 kW, which may be driven by various factors, such as weather-related load swings, or a coordinated cyber attack across IoT devices, such as the MadIoT attack. In this case, the mitigation is provided by using 30 flexible load nodes. Curtailment at the IoT device level ranges from a minimum of 0.2 kW reduction and a maximum of 0.5 kW reduction per primary feeder node. In total, approximately 130 kW of power import from the main grid decreases after mitigation. Market clearing happens every minute, and the drop in the load is shown in \cref{market_result}. The IoT device response, which is the thermostat in this case, has an instantaneous response, with an immediate drop in net load. 

\subsection{Validation of Attack 3 by NREL using ARIES\label{sec:attack3_nrel}}
\begin{figure}
    \centering
    \includegraphics[scale=0.52]{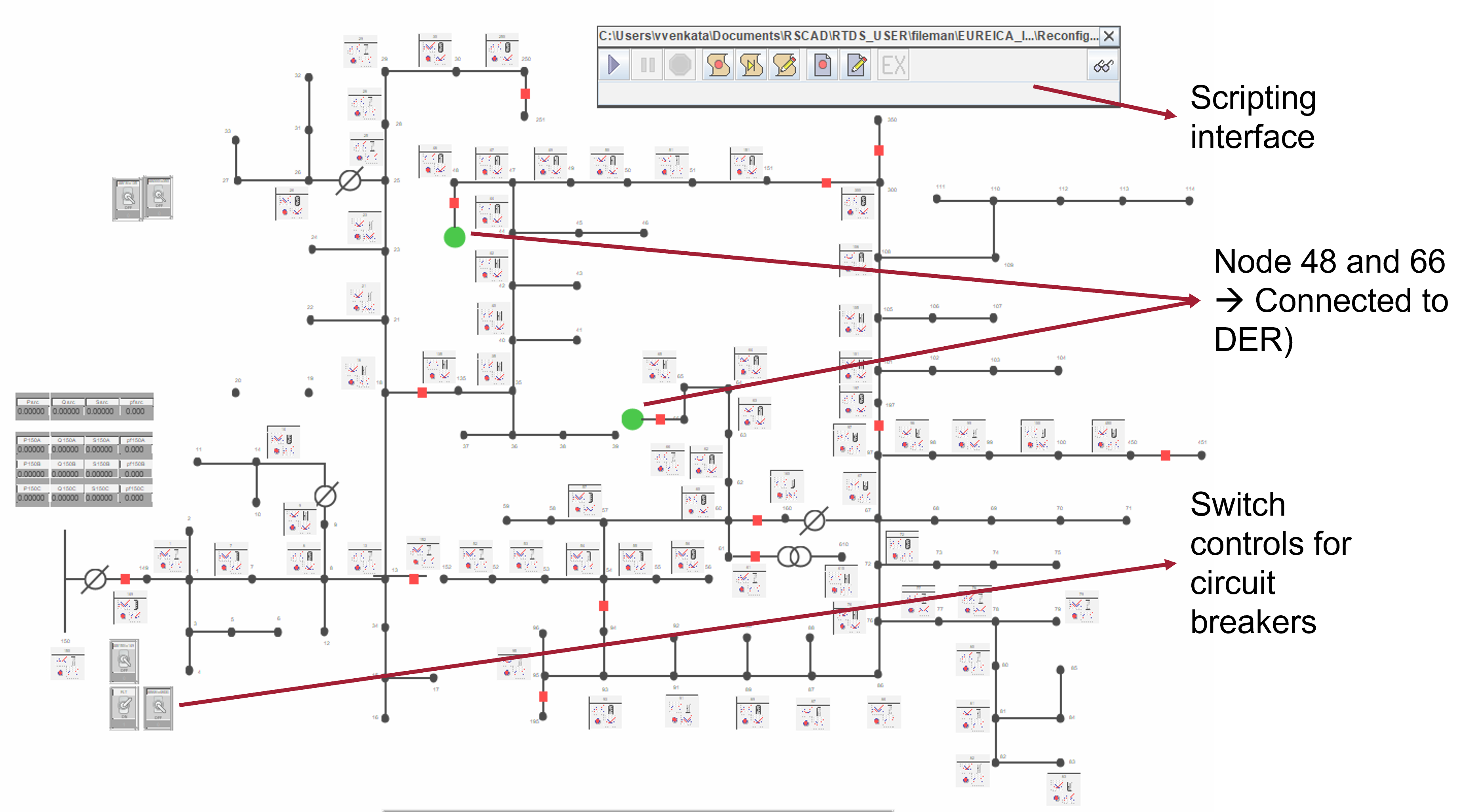}
    \caption{Implementation of reconfiguration algorithm in RTDS.}
    \label{rtds_reconfig}
\end{figure}

The mitigation of Attack 3 is validated using the RTDS at NREL-ARIES. The implementation of the reconfiguration algorithm in the RunTime environment of RTDS is shown in \cref{rtds_reconfig}.

\begin{figure}
    \centering
    \includegraphics[width=\columnwidth]{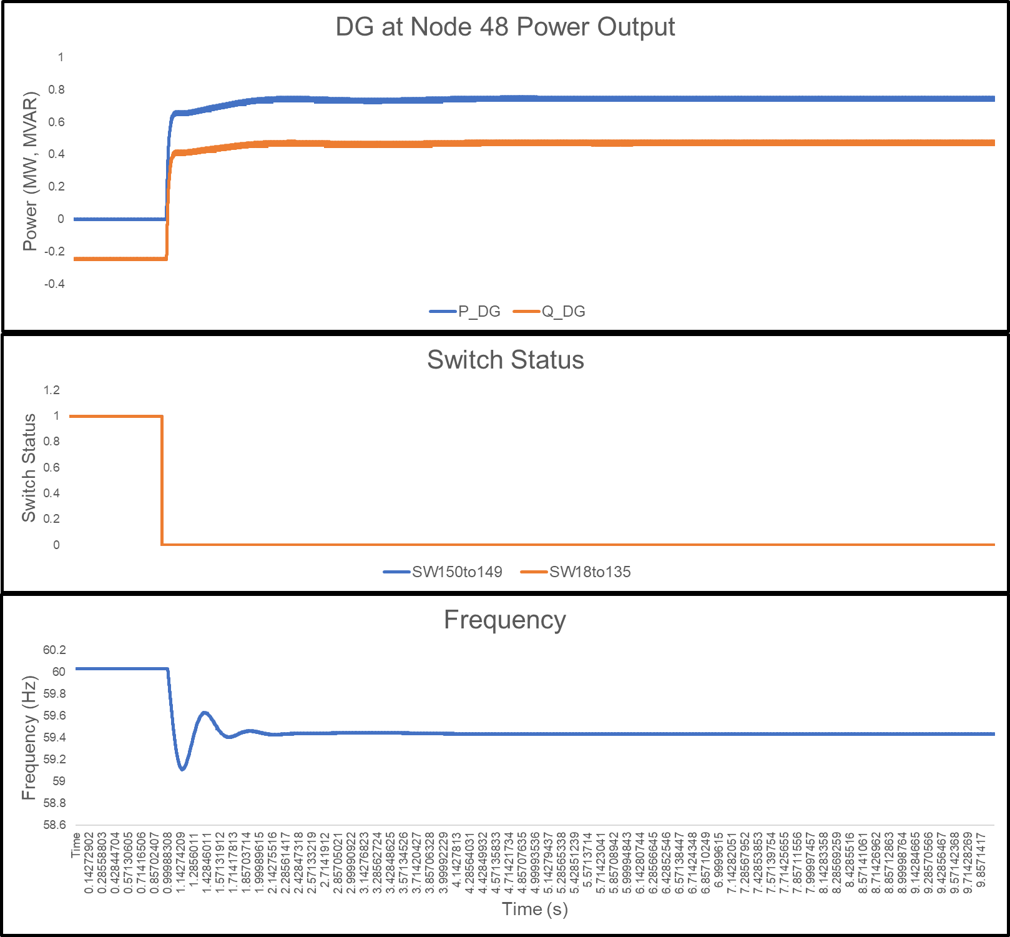}
    \caption{Distribution feeder broken into islands, with only Zone 3 load restored by DG at Node 48.}
    \label{reconfig_noeureica}
\end{figure}

In the case where the EUREICA framework is not used, the frequency of the system becomes unstable, and the distribution feeder is broken into islands and only the loads in Zone 3 are picked by the DG in Node 48. This plot is shown in \cref{reconfig_noeureica}.

The case with the EUREICA framework, with the contributions from various DGs and the military microgrid connected at Node 66 has already been demonstrated in \cref{mm_validation}. 

\bibliographystyleSI{pnas-new}
\bibliographySI{refs_manual}

\end{document}